\documentclass{jfm}
\usepackage{graphicx}
\usepackage{epstopdf, epsfig}
\pdfoutput=1

\usepackage{etoolbox}
\usepackage{xparse}
\usepackage{scalerel}

\usepackage{psfrag}
\usepackage{bm}
\usepackage{amsmath}
\usepackage{amssymb}
\usepackage{color}
\usepackage{rotating}
\usepackage{threeparttable}

\shorttitle{Shock-induced heating and transition to turbulence in hypersonic boundary layers}
\shortauthor{L. Fu et al.}

\title{{Shock-induced heating and transition\\ to turbulence in a hypersonic boundary layer}}

\author{Lin Fu\aff{1}, Michael Karp\aff{1}, Sanjeeb T. Bose\aff{2},\\[0.2ex] Parviz Moin\aff{1} \and Javier Urzay\aff{1}\corresp{\email{jurzay@stanford.edu}}}

\affiliation{\aff{1}{Center for Turbulence Research, Stanford University, Stanford CA 94305}\\
\aff{2}{Cascade Technologies Inc., Palo Alto, CA 94303}}

\begin{document}

\maketitle

\begin{abstract}

{The interaction between an incident shock wave and a Mach-6 undisturbed hypersonic laminar boundary layer over a cold wall is addressed using direct numerical simulations (DNS) and wall-modeled large-eddy simulations (WMLES) at different angles of incidence. At sufficiently high shock-incidence angles, the boundary layer transitions to turbulence via breakdown of near-wall streaks shortly downstream of the shock impingement, without the need of any inflow free-stream disturbances. The transition causes a localized significant increase in the Stanton number and skin-friction coefficient, with high incidence angles augmenting the peak thermomechanical loads in an approximately linear way. Statistical analyses of the boundary layer downstream of the interaction for each case are provided that quantify streamwise spatial variations of the Reynolds analogy factors and indicate a breakdown of the Morkovin's hypothesis near the wall, where velocity and temperature become correlated. A modified strong Reynolds analogy with a fixed turbulent Prandtl number is observed to perform best. Conventional transformations fail at collapsing the mean velocity profiles on the incompressible log law. The WMLES prompts transition and peak heating, delays separation, and advances reattachment, thereby shortening the separation bubble. When the shock leads to transition, WMLES provides predictions of DNS peak thermomechanical loads within $\pm 10\%$ at a computational cost lower than DNS by two orders of magnitude. Downstream of the interaction, in the turbulent boundary layer, WMLES agrees well with DNS results for the Reynolds analogy factor, the mean profiles of velocity and temperature, including the temperature peak, and the temperature/velocity correlation.}

\end{abstract}


\section{Introduction}\label{sect:introduction}
Airframes and propulsion systems of high-speed aerospace vehicles are subject to large wall heating rates and drag forces caused by viscous friction and shock waves \citep{Leyva2017,urzay2018supersonic,Candler2019}. However, the mechanisms responsible for these extra thermomechanical loads are complex and multi-scale.

{The model problem considered in the current study concerns the interaction between an oblique shock and an undisturbed hypersonic laminar boundary layer. In recent years, the related problem of interaction between shock waves and turbulent boundary layers has received considerable attention \citep{dupont2005space,dupont2006space,dupont2008investigation,dussauge2006unsteadiness,pirozzoli2006direct,loginov2006large,sandham2009numerical,touber2009large,gaitonde2013progress,bermejo2014confinement,adler2018dynamic}. In contrast, studies of the effects of incident shock waves on the transition of laminar boundary layers have remained comparatively more elusive. The basic triple-deck theory of weak shock waves interacting with laminar boundary layers was formulated first by \cite{Lighthill1950}, who quantified the upstream extent of the pressure disturbance on the wall surface. More recently, several efforts in characterizing shock waves interacting with transitional boundary layers have been undertaken \citep{Vanstone,sandham2014transitional,schuelein2014effects,Babinsky,Polivanov,willems2015experiments,Lash,currao2020hypersonic}. A recent review paper by \cite{Knight} summarizes important studies in this area. These studies have shed light upon realistic interaction cases under finite shock strength, including the overheating caused by transition of the post-interaction boundary layer.}


{Despite this progress, and similarly to other problems in high-speed aerodynamics involving transitional phenomena, it becomes difficult to computationally recreate the particular free-stream conditions in wind tunnels used for experiments, because they typically involve noise radiation that has a profound effect on the solution. Since it is currently challenging to provide complete measurements of the full structure of free-stream disturbances in wind tunnels, early simulations by \cite{sandham2014transitional} and \cite{yang2017aerodynamic} were conducted using a random perturbation field at the inflow of the computational domain, with the magnitude of the perturbations tuned to achieve a good match with the Stanton-number experimental measurements made by \cite{sandham2014transitional}, \cite{schuelein2014effects}, and \cite{willems2015experiments}.}

{The sensitivity of the transition process to the free-stream disturbances is greatly reduced as the shock incidence angle increases, in which case an absolute instability engendered in the separation bubble dominates the transition process \citep{hildebrand2018simulation}. Experiments at shock incidence angles higher than the ones considered in \cite{sandham2014transitional}, \cite{schuelein2014effects}, and \cite{willems2015experiments} have been recently addressed by \cite{currao2020hypersonic} in an experimental investigation performed concurrently with the present study. They studied the interaction between a Mach-5.8 laminar hypersonic boundary layer and a shock generated by a $10^\circ$ wedge. The measurements of the wall pressure and heat flux showed that the transition to turbulence is characterized by spanwise stationary fluctuations. \cite{currao2020hypersonic} proposed that these modulations were related to G{\"o}rtler-like streamwise vortices that grew exponentially along the concave streamlines above the post-interaction boundary layer near the interaction zone.}

{A relevant global stability analysis of shock waves interacting with laminar boundary layers was conducted by \cite{robinet2007bifurcations} in a study that employed a three-dimensional disturbance overlaid on a two-dimensional laminar boundary layer. It was found that for sufficiently strong shocks, the boundary layer became globally unstable to stationary disturbances with a finite spanwise wavenumber, in such a way that the eigenfunction had a purely exponential growth in time at each point in space without leading to any oscillations. The mechanism of instability was further analyzed by \cite{hildebrand2018simulation}, who showed that the interactions between streamwise vortices in the separation bubble created by an oblique shock impinging on a Mach-5.9 laminar boundary layer over an adiabatic wall are responsible for transition. Furthermore, the results in \cite{hildebrand2018simulation} indicated that, for shock incidence angles larger than the critical value $\beta=12.9^\circ$ (equivalent to a critical wedge angle $\alpha=4.5^\circ$), transition occurred due to round-off errors in the absence of any inflow disturbances, and that transition was accompanied by the formation of stationary streaky footprints in the wall heat flux. However, the exact value of the critical shock incidence angle is expected to generally depend on dimensionless flow parameters,  including the Reynolds and Mach numbers, and on the wall-to-free-stream temperature ratio, with additional thermochemical parameters being also required in regimes involving higher enthalpies.}



\begin{figure}
\begin{center}
\includegraphics[width=0.88\textwidth]{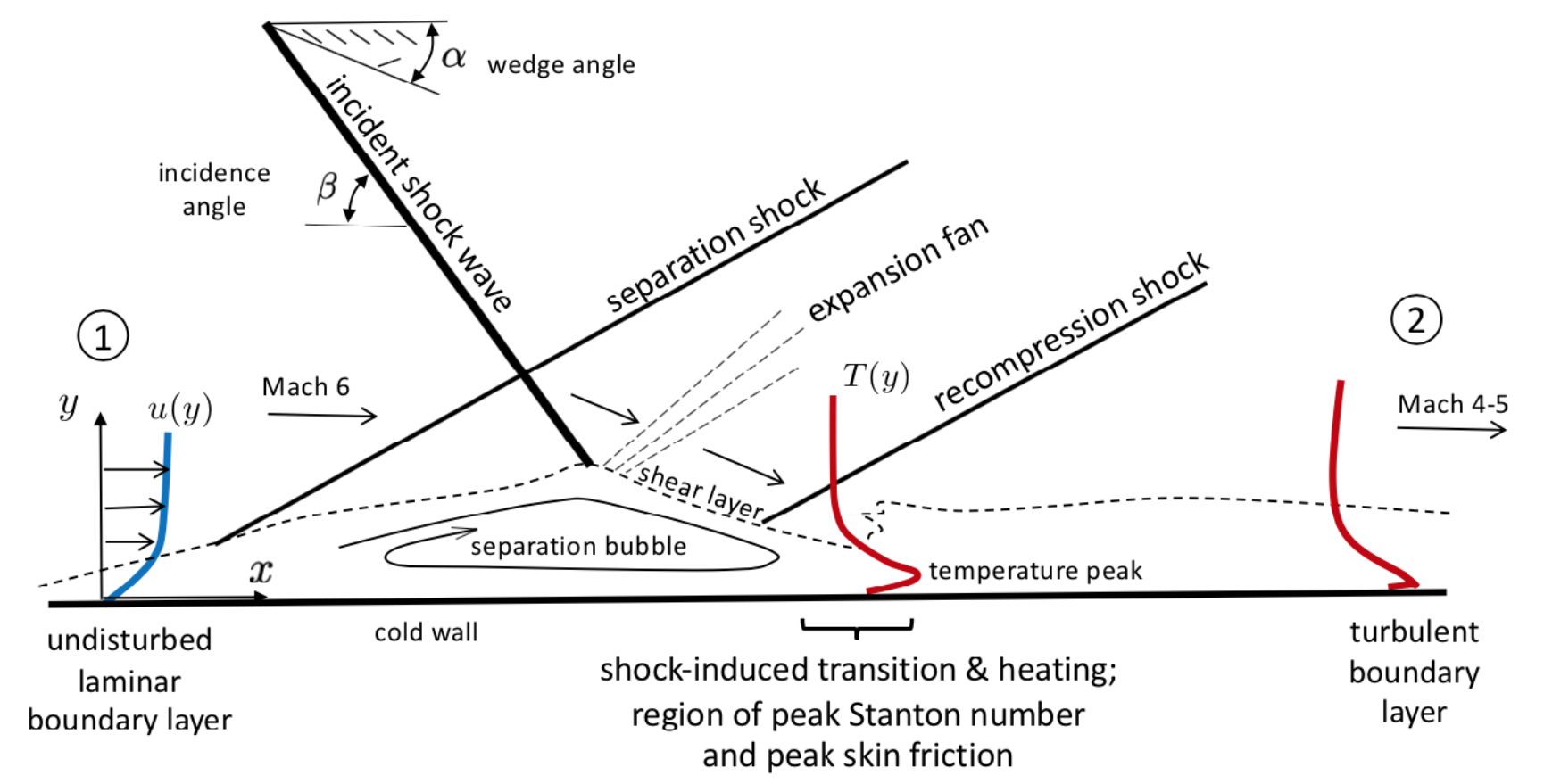}
\caption{{Schematics of the model problem: An oblique shock wave impinging on an undisturbed hypersonic laminar boundary layer.}}
\label{sketch1}
\end{center}
\end{figure}

{The focus of the present study is on the interaction of an incident oblique shock with a Mach-6 undisturbed laminar boundary layer overriding a cold isothermal flat plate. The main features of the flow are sketched in figure~\ref{sketch1}, and the setup resembles the experimental one outlined in \cite{sandham2014transitional}. However, in contrast to \cite{sandham2014transitional}, these simulations are concerned with shocks impinging at sufficiently high angles for transition to not rely on the presence of inflow disturbances. Specifically, the range of shock incidence angles considered here is $13.2^\circ\leq  \beta\leq  15.7^\circ$, which correspond to a range of wedge angles $5.0^\circ\leq \alpha\leq 8.0^\circ$. It will be shown below that, while transition is readily achieved near the upper end of this interval of wedge angles without the aid of free-stream disturbances, the transition process becomes utterly slow near the lower end, and does not lead to completion within the computational domain. Note however that the range of values of wedge angles tested here are smaller than the $\alpha=10^\circ$ wedge angle considered in the experimental investigation recently performed by \cite{currao2020hypersonic}. It should be stressed that increasing the wedge angle does not come at reduced computational cost. Specifically, as the incidence angle increases, the overshoot in the skin friction coefficient at transition increases, thus leading to an increasingly thinner viscous sublayer and consequently more stringent grid resolution requirements. Similarly, the larger the wedge angle is, the longer the separation bubble becomes upstream of the interaction region, thereby taxing the size of the computational domain.}

{In the present configuration, at sufficiently high incidence angles, a fully turbulent, highly supersonic boundary layer ensues downstream of the shock, as sketched in figure~\ref{sketch1}. Whereas compressible turbulent boundary layers are substantially more complicated than their incompressible counterparts, insight into their structure has been gained over the years by developing transformations that seek to convert velocity profiles from compressible turbulent boundary layer profiles into the well-known log law for incompressible turbulent boundary layers \citep{trettel2019transformations}. In addition, \cite{Morkovin1962} proposed that, for edge Mach numbers less than 5, any difference between compressible turbulent boundary layers and incompressible boundary layers can be accounted for by incorporating the variations of mean quantities, because flow dilatation plays a second-order effect. Many velocity transformations and scaling laws, which are verified by both experiments and DNS data \citep[e.g.,][]{Fernholz1980,guarini2000direct,pirozzoli2004direct,trettel2016mean}, have been developed on the basis of the Morkovin hypothesis, including the van Driest transformation \citep{vandriest} for adiabatic boundary layers, which converts the compressible mean velocity profile into the incompressible log law. However, these theories do not appear to perform adequately in non-adiabatic compressible boundary layers, and most particularly, in the practical case of boundary layers overriding cold walls~\citep{duan2010direct}. Specifically, the colder the wall temperature is relative to the free-stream stagnation temperature, the stronger the gradients of temperature are in the boundary layer as a result of the competition between the aerodynamic heating caused by the recovery of thermal energy, and the flow cooling induced by the wall. This well-known phenomenon leads to a non-monotonic temperature profile, whose maximum is observed in the present simulations to be located near or below the buffer layer, thereby leading to relatively large density gradients near the wall.}

{Beyond fundamental investigations of the problem, a relevant engineering question that often arises is whether the aforementioned physical processes, which are all concealed in the boundary layer, can be predicted with reasonable accuracy without incurring an exceedingly high computational cost. This question becomes particularly relevant when attempting to simulate high-speed flows around entire flight systems, since their resolution often renders impractical the utilization of direct numerical simulations (DNS). Typical strategies involve  utilization of coarser grids while relying on reduce-order models to partially account for the effects of the near-wall turbulence. Recent advances in numerical algorithms, computer hardware, and the related computer science have led to successful  predictions of complex multi-physics turbulent flows in aerospace applications by using wall-modeled large-eddy simulations (WMLES), but most of these breakthroughs have been limited to systems operating at subsonic and low-supersonic speeds \citep{bose2018wall}. While notable attempts to employ WMLES have been recently made in supersonic and hypersonic flows \citep{kawai2012wall,bermejo2014confinement,Johan,Subbareddy,Komives,Malik}, this research area is still in its infancy, particularly in relation to aspects connected with hypersonic transitional phenomena \citep{yang2017aerodynamic} and thermochemical effects \citep{Renzo}. The present study contributes to this progress by utilizing a relatively simple, yet challenging configuration for benchmarking wall models in hypersonic flows.}

{In this study, the equilibrium wall model described in \cite{yang2017aerodynamic} [see also \cite{kawai2012wall}] is employed with the goal of predicting the DNS results at reasonable cost. The comparisons between WMLES and DNS include metrics such as the location of transition and peak thermomechanical loads, the spatial extent of the separation bubble resulting from the adverse pressure gradient imposed by the shock, the first- and second-order flow statistics near the wall in the transitional and turbulent zones, and the physical processes responsible for the intense friction and overheating of the wall near the shock-impingement region.}




{The main research questions addressed by this study are: (a)~What are the physical mechanisms responsible for heat and friction augmentation near the shock-impingement region? (b)~do the classic Reynolds analogies, the Morkovin hypothesis, and the velocity log law hold in DNS and WMLES despite the high Mach numbers and cold wall temperatures? and (c) can WMLES predict the thermomechanical overloads at transition and the structure of the ensuing turbulent boundary layer?. The configuration analyzed in this study differs fundamentally from those in the literature of fully turbulent boundary layers in that it allows probing relevant quantities along the streamwise direction through very dissimilar flow environments ranging from laminar, to shock-induced transitional, and to fully turbulent farther downstream.}


The remainder of the paper is organized as follows. The computational setup is outlined in \S\ref{setup}, including the numerical method, boundary conditions, and grid resolutions employed in the simulations, along with a brief summary of the equilibrium wall model. Simulation results are described in \S\ref{results}, including predictions of  boundary-layer statistics in the transitional and turbulent zones. Conclusions are provided in \S\ref{conc}. {Additionally, four appendices are included that provide code verification and validation exercises (appendix~A), wall model formulation (appendix~B), a discussion of the performance of the wall model in the laminar portion of the boundary layer (appendix~C), and a supplementary grid-resolution study for WMLES (appendix~D).}

\section{Computational setup}\label{setup}

{This section focuses on a description of the computational setup. A sketch of the computational domain is provided in figure~\ref{sketch2} that supplements the discussion. Details are outlined below about numerical solver, boundary conditions, computational grids, and wall-model parameters employed in the simulations.}

\subsection{Numerical solver and boundary conditions}

{The simulations presented in this study are conducted using the finite-volume compressible solver charLES, which computes the solution on arbitrary polyhedral meshes. Specifically, charLES utilizes a low-dissipation spatial discretization based on principles of discrete entropy preservation \citep{tadmor2003,chandrashekar2013}, in which the fluxes are constructed to globally conserve entropy in inviscid shock-free flows, and to conserve the kinetic energy in inviscid low-Mach-number flows. Artificial diffusivity is employed in order to suppress oscillations in the vicinity of shock waves. Conserved quantities (i.e., mass, momentum, and total energy) are explicitly integrated in time using a three-stage strong-stability-preserving (SSP) Runge-Kutta scheme \citep{gottlieb2001}. The spatial and temporal schemes converge to second- and third-order with respect to the nominal mesh spacing and time step, respectively. Additional discussions regarding the solver discretization and its capabilities can be found in \cite{lozano2020}, \cite{lakebrink2019}, \cite{bres2018large}, and \cite{lehmkuhl2018}.} {A set of validation and verification exercises for charLES is provided in appendix~A that includes hypersonic laminar boundary layers, evolution of small amplitude disturbances in a high Mach number channel flow, along with a hypersonic flow around the boundary-layer transition (BOLT) subscale vehicle geometry.}

The formulation of the problem is described in \cite{yang2017aerodynamic}. Briefly, the charLES code integrates the conservation equations of mass, momentum, and total energy. {Favre-filtered} versions of these equations are employed for LES cases, with the subgrid-scale (SGS) tensor and SGS energy flux being modeled using the constant-coefficient Vreman model \citep{vreman2004}, with model constant $0.07$, along with a constant subgrid-scale turbulent Prandtl number $Pr_{sgs}=0.90$. The conservation equations are supplemented with Sutherland's law for the dynamic viscosity under a constant molecular Prandtl number $Pr=0.72$ (with Sutherland's model constants satisfying $T_{\textrm{ref}} = T_{1}$ and $S/T_{1} = 1.69$, with {$T_1$ being the temperature of the inflow free stream}), the ideal gas equation of state, and the assumption of calorically-perfect gas with $\gamma=1.4$.

\begin{figure}
\begin{center}
\includegraphics[width=0.85\textwidth]{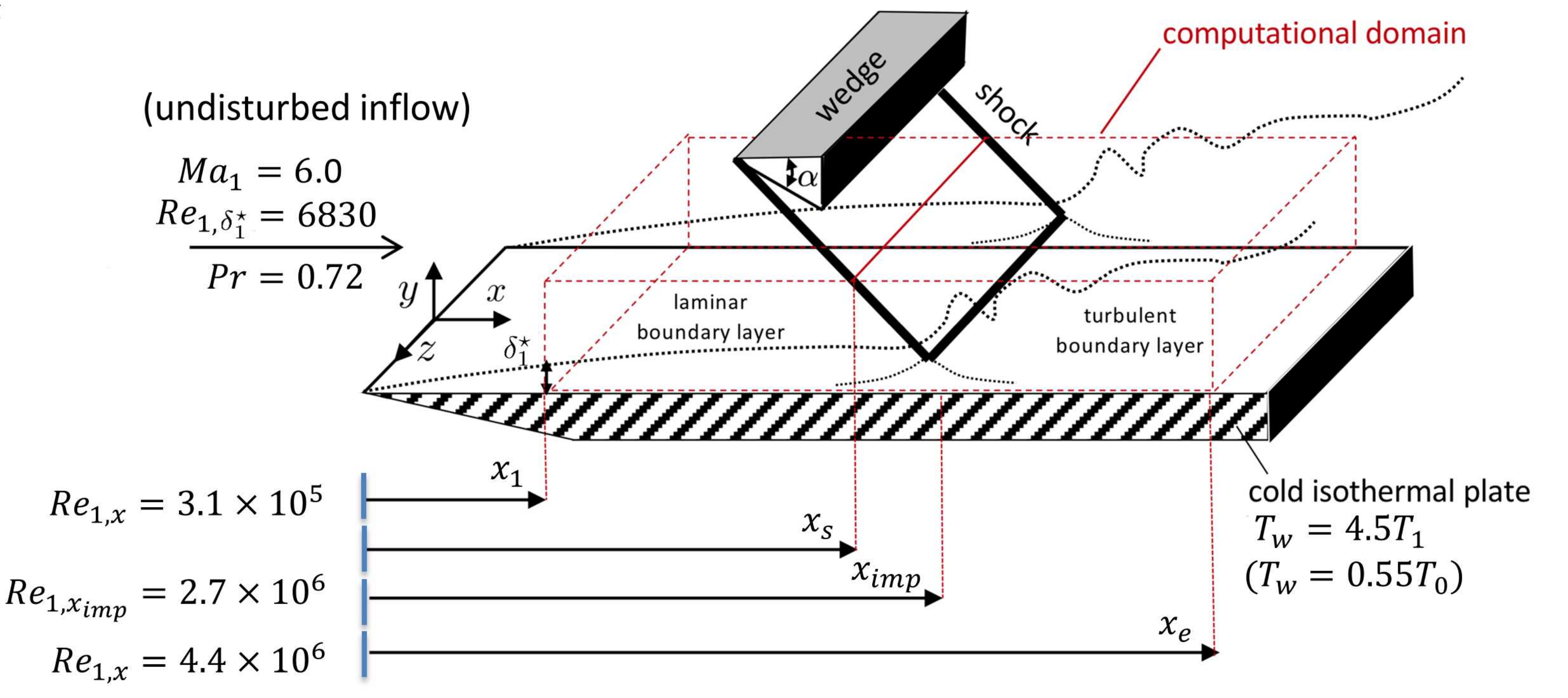}
\caption{Schematics of the computational domain.}
\label{sketch2}
\end{center}
\end{figure}

The geometry and operating conditions are explained in \cite{schuelein2014effects}, \cite{sandham2014transitional}, and \cite{willems2015experiments}. {Specifically, air at Mach {$Ma_{1}=U_1/a_1=6.0$}, based on the inflow free-stream velocity $U_1$ and speed of sound $a_1$, flows over an isothermal flat plate held at temperature {$T_w=4.5T_{1}$}, as schematically shown in figure~\ref{sketch2}} In these conditions, in which $T_w$ is smaller than the free-stream stagnation temperature {$T_{0}$} (i.e., $T_w/T_{0}=0.55$), the plate behaves as a cold one that receives heat from the flow. The resulting temperature profile in the wall-normal direction is non-monotonic, which is challenging to resolve with WMLES-like coarse grid resolution near the wall, as sketched in figure~\ref{sketch1}. A wedge held above the plate is responsible for generating the shock wave that impinges on the boundary layer.  In this work, four wedge angles $\alpha=5^\circ$, $6^\circ$, $7^\circ$, and $8^\circ$ are studied, while keeping all other parameters constant. {However, the wedge is not explicitly included in the computational domain, and therefore the expansion fan generated by its trailing edge is not considered.} Instead, the shock wave emanating from the leading edge of the wedge is imposed by appropriate jump boundary conditions, as described below.

{The Cartesian coordinate system $\{x,y,z\}$ used for the analysis is shown in figure~\ref{sketch2}, with $x=0$ corresponding to the leading-edge of the plate. At the inlet of the computational domain, the Reynolds number is $Re_{1,\delta_1^\star} =U_{1}\delta_1^\star/\nu_{1}= 6,830$ based on the inflow values of the displacement thickness $\delta_1^\star$ and of the free-stream velocity $U_1$ and kinematic viscosity $\nu_1$. The Reynolds number based on the distance $x_1=46\delta_1^\star$ from the leading edge of the plate to the inlet plane is  $Re_{1,x_1} =U_1 x_1/\nu_1= 314,252$. Correspondingly, the similarity solution for compressible laminar boundary layers is imposed at the inlet. In addition, periodic boundary conditions are used in the spanwise direction, while a characteristic non-reflecting boundary condition, with reference pressure chosen equal to the free-stream pressure, is applied at the outlet at a downstream distance $x_e$ such that $(x_e-x_1)/\delta_1^\star=600$, where the Reynolds number based on the inflow free-stream conditions is $Re_{1,x_e} =U_1 x_e/\nu_1= 4,410,887$. Note that the dimensionless streamwise distance from the edge of the plate, $(x-x_1)/\delta_1^\star$, and the Reynolds number based on the streamwise coordinate, $Re_{1,x}=U_1x/\nu_1$ can be used interchangeably for quantifying the streamwise distance in the plots below by using the relation
\begin{equation}
    Re_{1,x}=\left(\frac{x-x_1}{\delta_1^\star}\right)Re_{1,\delta_1^\star}+Re_{1,x_1}.\label{Rex}
\end{equation}
Different free-stream conditions emerge downstream of the recompression shock, denoted below by the subscript ``$2$'', as in $U_2$, $\rho_2$, $T_2$, $a_2$, and $\nu_2$. These quantities are useful, for instance, when examining the turbulent boundary layer ensuing downstream of the interaction, and they are utilized later in the text for defining the post-interaction values of the Reynolds number $Re_{2,x}=U_2x/\nu_2$ and Mach number $Ma_2=U_2/a_2$.


For a given wedge angle $\alpha$, the shock is made to emanate downwards from the top boundary of the domain at a streamwise position $x_{s}$ such that the point of inviscid intersection between the shock and the plate is located at a streamwise distance, $x_{\textrm{imp}}$, is given by $(x_{\textrm{imp}}-x_1)/\delta_1^\star=350$ in all cases, where the Reynolds number is $Re_{1,x_{\textrm{imp}}}=U_{1}x_{\textrm{imp}}/\nu_{1}=2,704,752$, as indicated in figure~\ref{sketch2}.} For $x>x_{s}$, an oblique flow entering the domain is prescribed at the top boundary using the Rankine-Hugoniot jump conditions for pressure, density, and velocities at the corresponding shock strength determined by the wedge angle $\alpha$, while the discretized fluxes at the boundary cell faces are obtained by solving a Riemann problem with a Harten-Lax van-Leer-Contact (HLLC) solver. The similarity solution for the compressible laminar boundary layer is imposed at the top boundary for $x<x_{s}$, including the vertical displacement velocity.

The simulations were initialized using the similarity solution for the laminar compressible boundary layer in the absence of an incident shock, and were evolved for 50 flow-through times. Cumulative statistics were calculated based on an on-the-fly analysis of the solution at every time step during 6 and 8  flow-through times in DNS and WMLES, respectively. {In the notation below, $\overline{f}$ and $\widetilde{f}$ denote, respectively, Reynolds and Favre averages of $f$, whereas $f'=f-\overline{f}$ and  $f^{''}=f-\widetilde{f}$ are the corresponding fluctuations.}

\subsection{Computational grids}

The dimensions of the computational domain are $600\delta_1^\star \times 75\delta_1^\star \times 45\delta_1^\star$ in the streamwise, wall-normal, and spanwise directions, respectively. The Cartesian grid used for DNS is  $6000 \times 600 \times 400$ (1440~M cells) and is stretched in the wall normal direction using a hyperbolic tangent clustering with a ratio of $\Delta y_{top}/\Delta y_w=10$. {The resolution of the DNS grid utilized here is comparable to the grid resolution employed in other studies on spatially evolving compressible turbulent boundary layers, including \cite{sandham2014transitional}, \cite{adams2000direct}, \cite{volpiani2018effects}, \cite{pirozzoli2010direct} and \cite{pirozzoli_bernardini_2011}}. Additionally, the DNS grid resolution employed here leads to reasonable agreement of statistical quantities such as the skin friction and the velocity-temperature relation with well established correlations.

Two different uniform Cartesian meshes are used for the WMLES to study the effects of grid resolution in the main text. The baseline WMLES grid is  $1024 \times 270 \times 144$ (40~M cells), whereas the coarse WMLES grid employs a coarser resolution in the wall-normal direction and is $1024 \times 192 \times 144$ (28~M cells) in order to assess the effects of varying the matching location between the wall model and the outer LES. The near-wall resolution in viscous units is listed in table~\ref{tableres} for all cases. In the baseline WMLES cases, the boundary layer was resolved with 5 points across the inlet plane (4 points in the coarse WMLES), 27 points across the outlet plane (9 points in the coarse WMLES), and 11 points across the wall-normal plane intersecting the streamwise location of maximum wall heat flux (7 points in the coarse WMLES), or equivalently, at the streamwise location of maximum Stanton number $St$, the latter being formally defined below in \S\ref{results}.

\begin{table}
\centering
\vskip 0.1in
\begin{tabular}{cc|ccc|ccc|ccc}
wedge angle & & & DNS & & & WMLES & & & WMLES coarse \\
$\alpha$ [deg] & & & $\Delta x^{+}\times \Delta y^{+}\times \Delta z^{+}$ [--] & & & $\Delta x^{+}\times \Delta y^{+}\times \Delta z^{+}$ [--] & & & $\Delta x^{+}\times \Delta y^{+}\times \Delta z^{+}$ [--] \\
 \hline
 5 & & & $ 4.09 \times 0.98 \times 4.60 $ & & & $ 20.37 \times  9.70  \times 10.91 $ & & & $ 20.24 \times 13.49 \times 10.79 $\\
 6 & & & $ 5.63  \times 1.35  \times 6.33 $ & & & $ 23.27  \times 11.08  \times 12.47 $ & & & $ 24.45  \times 16.30  \times 13.04 $ \\
  7 & & & $ 6.51 \times 1.56   \times 7.32$ & & & $ 27.80  \times 13.24  \times 14.90 $ & & & $ 28.74  \times 19.16  \times 15.33 $\\
  8 & & & $  7.46 \times  1.79  \times 8.40 $ & & & $ 31.75 \times  15.12 \times 17.01 $ & & & $ 33.17  \times 22.11  \times 17.69 $ \\
  \hline
\end{tabular}
\caption{{Minimum grid spacing near the wall in viscous units $\overline{\nu_w}/u_\tau$ at the outlet of the computational domain. In this notation, $\overline{\nu_w}$ is the time- and spanwise-averaged kinematic viscosity at the wall and $u_\tau=\sqrt{\overline{\tau_w}/\overline{\rho_w}}$ is the friction velocity based on  time- and spanwise-averaged values of the wall shear stress $\overline{\tau_w}$ and density at the wall $\overline{\rho_w}$.}} \label{tableres}
\end{table}

\subsection{Wall-model parameters}

In the WMLES cases, the equilibrium wall model described in appendix~B [see also \cite{kawai2012wall} and \cite{yang2017aerodynamic}] is utilized within a wall-modeled layer adjacent to the wall. Briefly, the equilibrium wall model consists of localized, RANS-like, steady one-dimensional versions of the wall-parallel momentum equation and the stagnation energy equation for a calorically perfect gas, with eddy-viscosity closures for the turbulent transport of momentum and energy, the latter relying on the assumption of a constant turbulent Prandtl number of $0.90$. A van~Driest damping function with constant $A^{+}=17$ is employed to exponentially suppress the eddy viscosity for $y^{+}\lesssim A^{+}$ in favor of the molecular viscosity. {Friction scaling is employed for the van Driest damping function, since mean density variations introduced by semi-local scaling have little effect because of the moderate wall-cooling levels utilized here.}    Additionally, the ideal gas equation of state is utilized in the wall model to relate the density $\rho$ with the temperature $T$, in such a way that the pressure across the wall-modeled layer remains equal to the pressure at the matching location $y=h_{wm}$.

The equations of the wall model are subject to non-slip and isothermal ($T=T_w$) boundary conditions at the wall, and to the instantaneous filtered values of the wall-parallel velocity, temperature, and pressure at the matching location. The outputs of the wall model are the local values of the wall shear stress $\tau_w$ and wall heat flux $q_w$, which are employed as boundary conditions for the LES conservation equations of the bulk flow.

{The thickness of the wall-modeled layer $h_{wm}$ employed in these simulations is equivalent to a single cell of the WMLES grid. Whereas \cite{kawai2012wall} have shown that this choice may lead to a log-layer mismatch, the results in \cite{yang2017log} indicate that temporal filtering  alleviates this problem. In this work, the approach proposed by \cite{yang2017log} is used because of its simplicity of implementation in unstructured grid environments.}

\begin{figure}
\begin{center}
\includegraphics[width=0.55\textwidth]{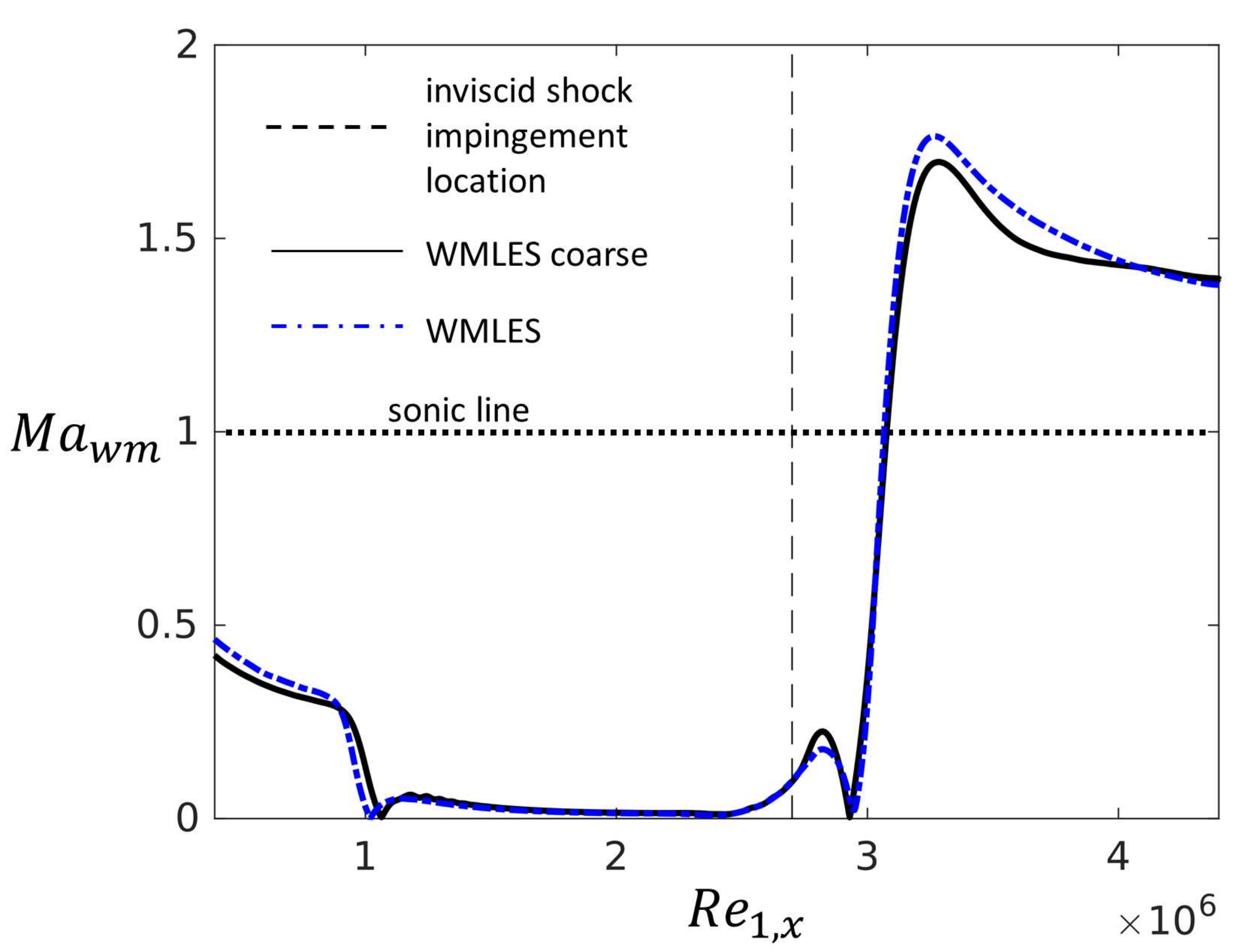}
\caption{{Distribution of the WMLES local Mach number $Ma_{wm} = [\overline{\rho} (\overline{u}^2+\overline{v}^2)/(\gamma \overline{P})]^{1/2}$ at the matching location $y=h_{wm}$, for the case $\alpha=7^\circ$, based on the time- and spanwise-averaged values of the local velocity, density, and pressure. The vertical dashed line denotes the inviscid shock-impingement location on the wall.}}
\label{mach}
\end{center}
\end{figure}

Since the wall model does not incorporate streamwise variations of any quantity, the upstream propagation of elliptic effects within the wall-modeled region -- for instance due to the shock-induced adverse pressure gradient -- can only occur through the boundary conditions applied at the matching location. {As shown in figure~\ref{mach}, for both WMLES resolutions, the Mach number $Ma_{wm}$ based on the time- and spanwise-averaged values of the streamwise velocity and local speed of sound at the matching location is everywhere less than 0.5 in the laminar portion for the case $\alpha=7^\circ$.} This is also the case for the other values of the wedge angle treated here. These considerations indicate that the wall-modeled layer is fully subsonic on average, and that the resolved field near the matching location is the one supporting the propagation of elliptic effects. Note that, had the wall-modeled layer been thick enough to bear the sonic line inside, no propagation of elliptic effects close to the wall would have been accounted for in the WMLES.

\begin{figure}
\begin{center}
\vskip 0.2in
\includegraphics[width=0.99\textwidth]{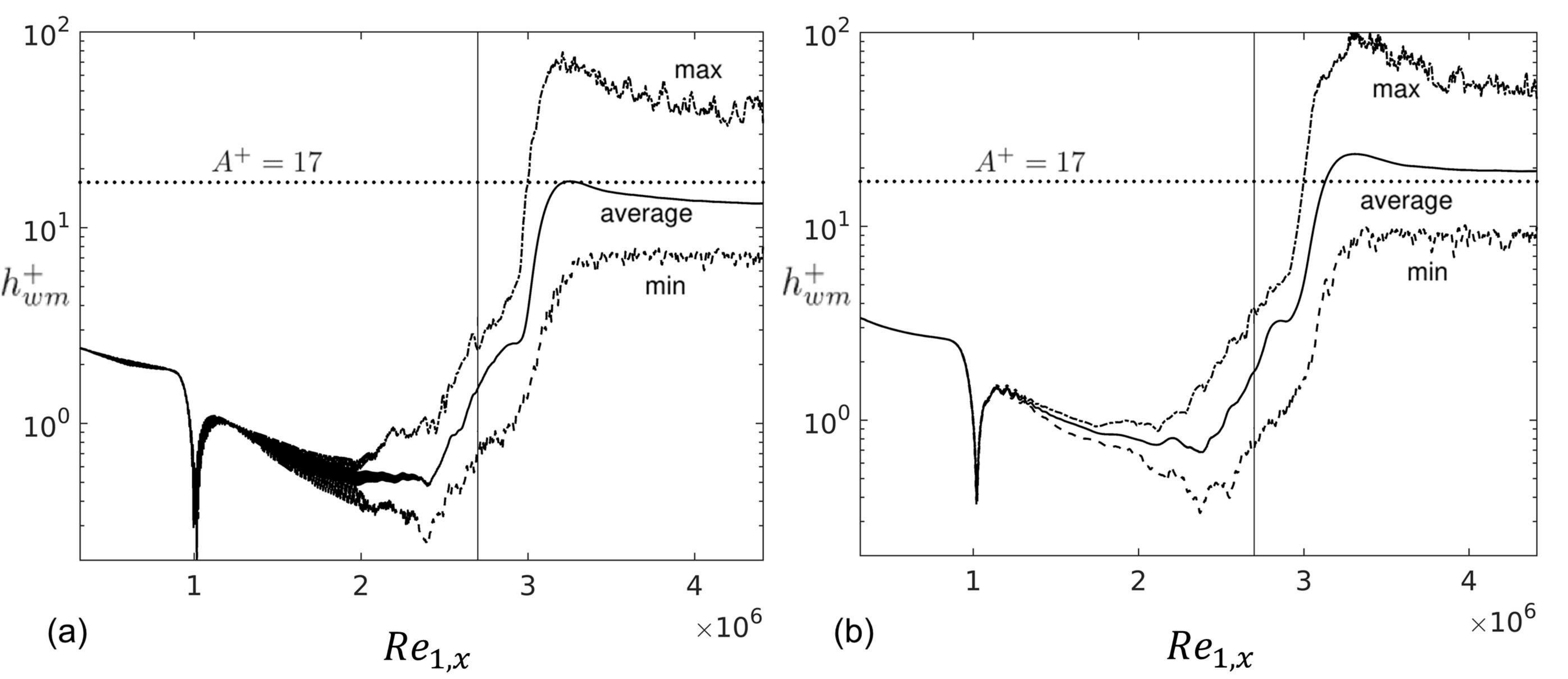}
\caption{{Distribution of the WMLES matching location $h^+_{wm}$ scaled in wall units for the case $\alpha=7^\circ$, including minimum (dot-dashed line), maximum (dashed line), along with the time- and spanwise-averaged value (solid line). The vertical dashed line denotes the inviscid shock-impingement location on the wall. Included are the data for (a)~baseline and (b)~coarse WMLES cases.}}
\label{yplusWMLES}
\end{center}
\end{figure}

{In figure~\ref{yplusWMLES}, the matching location expressed in viscous units, $h^{+}_{wm}$, plunges at {$Re_{1,x}\simeq 10^6$} for the case $\alpha=7^\circ$ because the flow separates there, and increases rapidly near the inviscid shock-impingement location {$Re_{1,x_{\textrm{imp}}}$} due to the sharp rise of the skin-friction coefficient, as shown below in \S\ref{results}. Whereas the time- and spanwise-averaged value of $h^{+}_{wm}$ remains everywhere around or below the damping constant $A^+$ in the baseline WMLES shown in figure~\ref{yplusWMLES}(a), its maximum value overtakes $A^+$ by a factor of 4. As a consequence,  based on the averaged $h^{+}_{wm}$, it may be tempting to disregard the effects of the eddy viscosity built in the wall model in the baseline WMLES. Nonetheless, it is shown in \S\ref{results} that the baseline WMLES without eddy viscosity in the wall-model equations (i.e., $\mu_{t,wm}=0$) does not lead to satisfactory results neither in the transitional nor in the turbulent portions of the boundary layer. Despite the fact that the eddy-viscosity hypothesis is questionable in transitional scenarios, these considerations highlight its dynamical relevance in regions where local overshoots in $h^{+}_{wm}$ occur. }

{ In both baseline and coarse WMLES cases, the equilibrium wall model is applied everywhere along the surface of the plate, including the laminar portion of the boundary layer. Two important aspects are worth remarking with regards to this choice that are discussed in the remainder of this section.}

It is shown in appendix~C that the WMLES adequately captures the velocity and temperature profiles in the laminar boundary layer, which remains mostly steady and two-dimensional until it becomes highly disturbed in the interaction region. The wall model performs correctly there because its conservation equations are equivalent to the steady laminar boundary-layer equations very close to the wall, where advection is negligible. This can be understood by examining the distribution of $h^{+}_{wm}$ in figure~\ref{yplusWMLES}. The values of $h^{+}_{wm}$ in both WMLES simulations remain much smaller than $A^+$ in the laminar region, thereby yielding negligible values of the eddy viscosity in the wall model. Since order-unity values of $h^{+}_{wm}$ in the laminar region are equivalent to very small values of $h_{wm}$ relative to the boundary-layer thickness, namely $h_{wm}/\delta^\star_1 =O\left(Re_{1,x_1}^{-7/4}\right)\ll 1$, the constant molecular stress predicted by the wall model in the first approximation for $y^{+}_{wm}/A^+\ll 1$ [i.e., see equation~\eqref{mom} in appendix~B] is equivalent to the $y/\delta^\star\rightarrow 0$ limit of the steady laminar boundary-layer equations in the absence of streamwise pressure gradient.

{ That the wall model performs correctly in the laminar portion of this flow can also be understood by noticing that the scenarios sought for transition in this study are not the classical ones in which unstable eigenmodes grow relatively slowly along the entire portion of the laminar boundary layer, eventually producing transition far downstream in a way that is rather well understood, at least for calorically perfect gases flowing over smooth flat surfaces in the absence of incident shocks \citep{mack1984boundary}. Instead, the physical processes leading to transition in the present study are spatially localized downstream of the shock on the leeward side of the separation bubble, and are triggered by the absolute instability of the separation bubble without participation of any intentional disturbances at the inflow \citep{hildebrand2018simulation}. As a result, in the present study there are lesser consequences derived from the fact that neither the coarse grid resolution in WMLES nor the equilibrium wall model itself can appropriately support the growth of eigenmodes along the lengthy laminar portion of the boundary layer upstream of the shock. The task of the wall model there is limited to providing the velocity and temperature profiles within the fully-viscous wall-modeled layer.}

In this work, the computational cost of using WMLES was about 150 times less than DNS. Specifically, typical DNS cases took 25~M core hours at Argonne's Mira supercomputer, whereas only 150~K core hours were required on average for each WMLES case on the same machine. Furthermore, it is also shown in \S\ref{results} that non-wall-modeled LES at the resolutions listed in table~\ref{tableres} provide completely wrong predictions in the transitional and fully turbulent zones of the boundary layer, which underscores the positive role of the wall model in warranting acceptable predictions.

\section{Numerical Results}\label{results}

{In this section, the analysis begins by a quantification of the effect of the shock incidence angle on the peak thermomechanical loads. Next, a detailed analysis of the DNS flow field is conducted, followed by comparisons between DNS and WMLES, particularly near the shock-impingement region. This section concludes with a description of the DNS statistics in the turbulent boundary layer ensuing downstream of the reattachment zone along with associated comparisons with WMLES.}

{ A number of considerations in this section are based on the skin friction coefficient $C_f$ and the Stanton number $St$ as main figures of merit. These two parameters require information about the inviscid free stream flowing above the boundary layer. However, in the present problem, the aerothermodynamic state of the inflow free stream is different from that of the free stream found downstream of the recompression shock. These changes imperil a proper simultaneous scaling of $C_f$ and $St$ in both the laminar (i.e., pre-interaction) and turbulent (i.e., post-interaction) boundary layers. As a result, two different definitions of the skin friction coefficient and Stanton number are used depending on where the free-stream conditions are based, namely
\begin{equation}
C_{f,1} = \frac{2\overline{\tau_w}}{\rho_1 U_{1}^2}\label{Cf1}
\end{equation}
and
\begin{equation}
St_1= \frac{\overline{q_w}}{\rho_{1}U_{1}c_p\left(T_{aw,1}-T_w\right)}\label{St1}
\end{equation}
for conditions based on the inflow free-stream, and
\begin{equation}
C_{f,2} = \frac{2\overline{\tau_w}}{\rho_2 U_{2}^2}\label{Cf2}
\end{equation}
and
\begin{equation}
St_2= \frac{\overline{q_w}}{\rho_{2}U_{2}c_p\left(T_{aw,2}-T_w\right)}\label{St2}
\end{equation}
for conditions based on the free stream found downstream of the recompression shock. In equation \eqref{St1},  $T_{aw,1}=T_{1}[1+r_1(\gamma-1)Ma_{1}^{2}/2]$ is the adiabatic wall temperature based on a recovery factor $r_1= Pr^{1/2}=0.85$ corresponding to laminar boundary layers \citep{vandriest}. Instead, in equation \eqref{St2}, $T_{aw,2}=T_{2}[1+r_2(\gamma-1)Ma_{2}^{2}/2]$ is the adiabatic wall temperature based on a recovery factor $r_2=Pr^{1/3}=0.90$ appropriate for turbulent boundary layers~\citep{volpiani2018effects}. In all expressions, $c_p$ is the constant-pressure specific heat of the gas, whereas $\overline{\tau_w}$ and $\overline{q_w}$ are time- and spanwise-averaged values of the wall shear stress $\tau_w=\mu_w(\partial u/\partial y)_w$ and the wall heat flux $q_w=\lambda_w(\partial T/\partial y)_w$, respectively, where $T$ is the temperature, $\lambda_w$ is the thermal conductivity evaluated at the wall temperature, $u$ is the streamwise velocity, and $\mu_w$ is the dynamic viscosity evaluated at the wall temperature.}


\begin{figure}
\begin{center}
\includegraphics[width=1\textwidth]{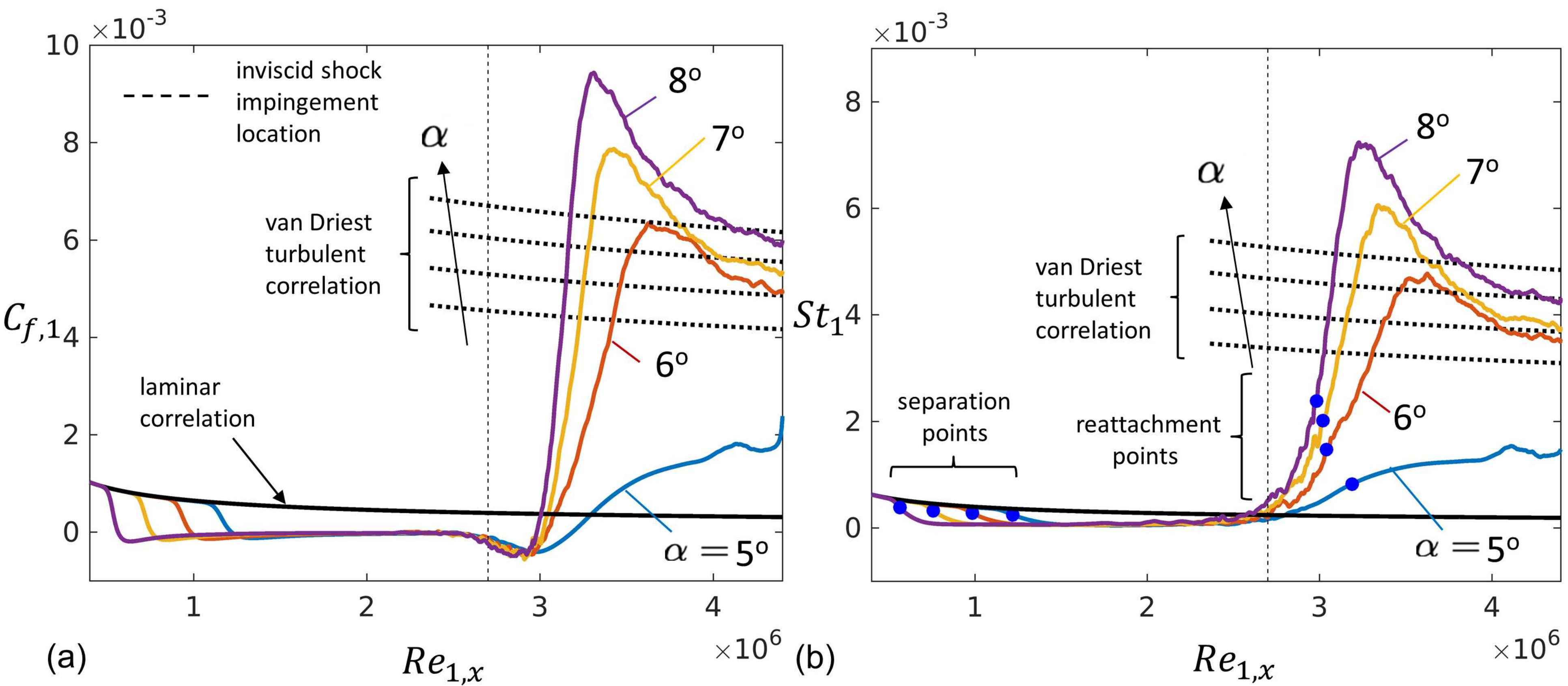}
\caption{{DNS results of (a)~skin friction coefficient and (b)~Stanton number as a function of the local Reynolds number $Re_{1,x}$ and the wedge angle $\alpha$. In this figure, the van Driest turbulent correlation for the skin friction coefficient $C_{f,2}$ is calculated based on post-interaction free-stream conditions, with a virtual origin equated to the leading edge of the plate, and is then re-scaled by a factor of $\rho_2 U_2^2/\rho_1U_1^2$ obtained by the DNS solution to refer the skin-friction coefficient to the pre-interaction free stream, $C_{f,1}$. Similarly, the van Driest turbulent correlation for the Stanton number $St_2$ is calculated from $C_{f,2}$ using the Reynolds analogy factor $2St_2/C_{f,2}=Pr^{-2/3}$, and is then re-scaled by a factor of $\rho_2 U_2 (T_{aw,2}-T_w)/[\rho_1 U_1 (T_{aw,1}-T_w)]$ obtained by the DNS solution to refer the Stanton number to the pre-interaction free stream, $St_1$.}}
\label{Cf_St_all_angles}
\end{center}
\end{figure}

\subsection{Effects of the shock incidence angle on peak thermomechanical loads} \label{angle}

{The DNS distributions of $C_{f,1}$ and $St_1$ as a function of the streamwise Reynolds number $Re_{1,x}$ are provided in figure~\ref{Cf_St_all_angles} for the wedge angles considered here. Initially all the curves collapse on the laminar correlation obtained from the similarity solution, as expected by the scaling with the pre-interaction free-stream values used in equations~\eqref{Cf1} and \eqref{St1}. The characteristic shapes of $C_{f,1}$ and $St_1$ include an early drop in the laminar zone due to  boundary-layer separation and a sudden overshoot downstream of the shock-impingement region because of transition. The separation of the laminar boundary layer causes a change of sign in $C_{f,1}$ due to the flow reversal and a decrease in $St_1$ due to the resulting weaker temperature gradient at the wall. In contrast, transition to turbulence leads to large spikes in $C_{f,1}$ and $St_1$, whose magnitude increase with the wedge angle. An additional discussion of this important phenomenon is provided in \S\ref{structures} upon examining flow structures participating in the augmentation of the local thermomechanical loads.}


{The case $\alpha=5^\circ$ behaves distinctly from the others. While overshoots are observed for higher wedge angles, the $\alpha=5^\circ$ case is characterized by a modest rise in $C_{f,1}$ and $St_1$, both of which stay far below the other cases. This is attributed to the fact that transition did not occur within the computational domain in the DNS of the $5^\circ$ case. Instead, the slight increments in $C_{f,1}$ and $St_{1}$ downstream of the shock are mainly produced by the variation of the free-stream aerothermodynamic state across the shock and its impact on $\overline{\tau_w}$ and $\overline{q_w}$, whereas the normalization used for $C_{f,1}$ and $St_1$ involves only the pre-interaction free-stream aerothermodynamic state, as indicated above.}

\begin{figure}
\begin{center}
\includegraphics[width=1\textwidth]{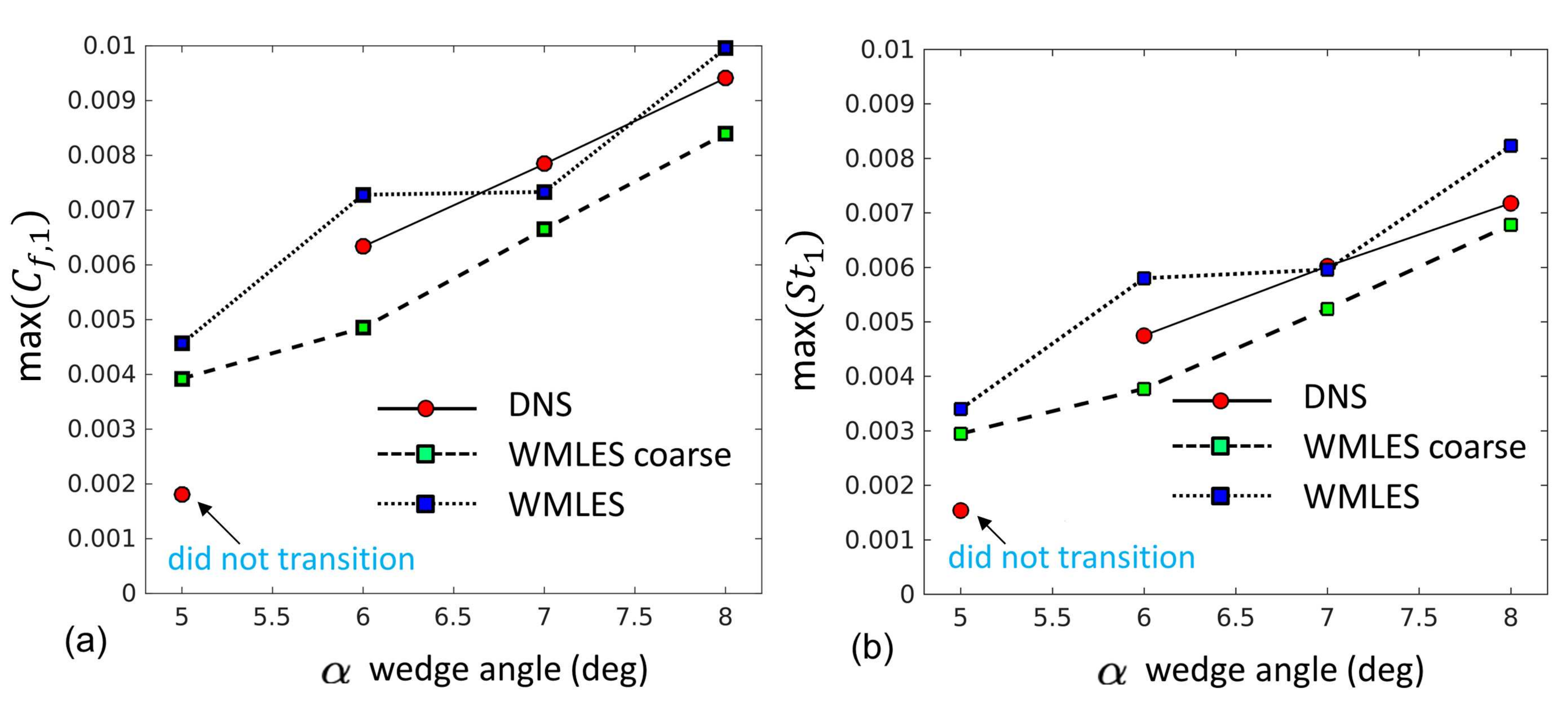}
\caption{{DNS (solid lines), baseline WMLES (dotted lines), and coarse WMLES (dashed lines) peak values of the (a)~skin friction coefficient and (b)~Stanton number as a function of the wedge angle $\alpha$. Since the boundary layer did not transition in the DNS of the $\alpha=5^\circ$ case, its data points are purposely disconnected from the DNS data points corresponding to the transitioning cases $\alpha=6^\circ$, $7^\circ$, and $8^\circ$.}}
\label{Cf_St_peaks}
\end{center}
\end{figure}

{Based on the above considerations, figure~\ref{Cf_St_all_angles} suggests that the critical wedge angle for the onset of shock-induced transition is somewhere between $5^\circ$ and $6^\circ$. For the transitioning cases $\alpha=6^\circ$, $7^\circ$ and $8^\circ$, the values of $C_{f,1}$ and $St_1$ downstream of transition do not agree well with the turbulent correlation of van Driest as expected, since both $C_{f,1}$ and $St_1$ are based on the pre-interaction values of the free stream, as mentioned above. In addition, the van Driest turbulent correlation for the Stanton number makes use of the Reynolds analogy factor $Pr^{-2/3}=1.24$ traditionally used to approximate a Reynolds analogy for boundary layers with non-unity Prandtl numbers. It is shown in \S\ref{Turbulence_statistics} that agreement with the van Driest turbulent correlations for the skin friction coefficient and Stanton number is obtained using the definitions (\ref{Cf2}) and (\ref{St2}) along with the modified Reynolds analogy factor of 1.16 proposed by \cite{chi1966influence}.}



{Qualitative comparisons between the DNS Stanton numbers for $\alpha=6^\circ$, $7^\circ$, and $8^\circ$ in figure~\ref{Cf_St_all_angles} with the experimental measurements by \cite{currao2020hypersonic} for $\alpha=10^\circ$ show that (i)~the minimum value of $St_1$ is in the separated region in both DNS and experiments, and (ii)~a monotonic increase of $St_1$ occurs near the reattachment in both DNS and experiments, after which transition of the boundary layer takes place simultaneously with an overshoot in $St_1$. Downstream of the transition zone, $St_1$ decays in both DNS and experiments, although the decay in the latter is much more substantial because of the expansion fan emanating from the trailing edge of the wedge.}

Small changes in the wedge angle have profound consequences on the flow field. In particular, the DNS results for $C_{f,1}$ and $St_1$ in figure~\ref{Cf_St_all_angles} indicate that increasing the wedge angle leads to earlier boundary-layer separation, longer separation bubbles, and higher overshoots of $C_{f,1}$ and $St_1$ near the shock-impingement region as a result of earlier transition. The dependency of the peak values $C_{f,1}$ and $St_1$ on the wedge angle $\alpha$ is shown in figure~\ref{Cf_St_peaks}. The trend in the DNS results is nearly linear, such that a $1^\circ$ increase in $\alpha$ causes approximately a 30\% increase in the average peak thermomechanical load acting on the plate. Comparisons between DNS and WMLES predictions of peak values of $C_{f,1}$ and $St_1$ in figure~\ref{Cf_St_peaks} are deferred to \S\ref{WMLES1}.

\begin{figure}
\begin{center}
\includegraphics[width=0.99\textwidth]{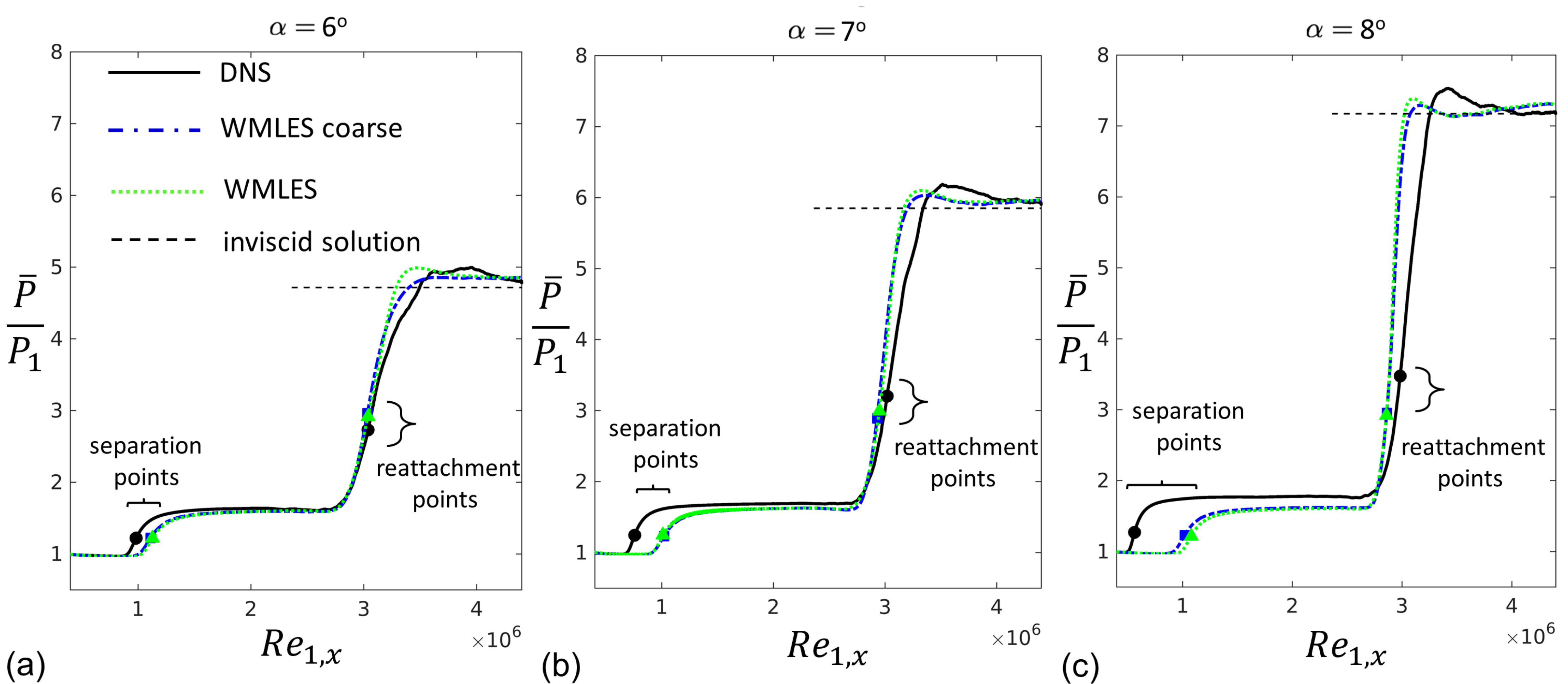}
\caption{{DNS (black solid lines), baseline WMLES (green dotted lines), and coarse WMLES (blue dot-dashed lines) results for time- and spanwise-averaged profiles of the wall pressure for wedge angles of (a)~$\alpha=6^\circ$, (b)~$\alpha=7^\circ$, and  (c)~$\alpha=8^\circ$. The black dashed lines indicate the dimensionless post-interaction static pressure $P_2/P_1$ calculated assuming inviscid flow.} }
\label{DNS_7_pressure}
\end{center}
\end{figure}

\begin{table}
\begin{center}
\begin{tabular}{cccl|ccc|ccc|ccc}
Wedge angle $\alpha$ [deg] & & &  & & & $U_2/U_1$ & & & $P_2/P_1$ & & & $T_2/T_1$\\
 \hline
6 & & & DNS              & & & 0.956 & & & 4.850 & & & 1.642\\
 & & & WMLES            & & & 0.958 & & & 4.845 & & & 1.650\\
 & & & WMLES coarse     & & & 0.957 & & & 4.844 & & & 1.634\\
 & & & inviscid theory  & & & 0.957 & & & 4.714 & & & 1.600\\

\hline

7 & & & DNS              & & & 0.945 & & & 6.050 & & & 1.807\\
& & &  WMLES            & & & 0.944 & & & 5.936 & & & 1.814\\
& & & WMLES coarse     & & & 0.946 & & & 5.932 & & & 1.796\\
& & & inviscid theory  & & & 0.948 & & & 5.848 & & & 1.724\\

\hline

8& & & DNS              & & & 0.933 & & & 7.253 & & & 1.958\\
 & & & WMLES            & & & 0.926 & & & 7.270 & & & 2.040\\
 & & & WMLES coarse     & & & 0.929 & & & 7.252 & & & 2.007\\
& & &  inviscid theory  & & & 0.939 & & & 7.172 & & & 1.856\\
 \hline
\end{tabular}
\caption{{Comparison of the free-stream velocity, pressure, and temperature ratios across the interaction zone for the transitioning cases.}}
\label{tab:jump_quantities}
\end{center}
\end{table}

\subsection{Flow field ensued by the incidence of the shock on the boundary layer} \label{structures}

{The separation of the laminar boundary layer upstream of the shock-impingement region is induced by the adverse pressure gradient created by the incident oblique shock wave, whose effect is communicated upstream along the subsonic flow close to the wall. The time- and spanwise-averaged profiles of static pressure on the wall showing the footprint of the shock wave are provided in figure~\ref{DNS_7_pressure} for the transitioning cases. The curves are composed of three plateaus (from left to right) that correspond, respectively, to the laminar zone, the separation bubble, and the turbulent zone downstream of the recompression shock. Note that the third plateau may not be present in experiments subjected to expansion effects from the trailing edge of the wedge~\citep{currao2020hypersonic}. In the peresent simulations, approximately five-, six-, and seven-fold overall increase in the pressure is observed across the interaction region for the cases $\alpha=6^\circ$, $7^\circ$, and $8^\circ$, respectively, mostly in agreement with the inviscid theory. Similar agreements between DNS and the inviscid theory are observed for velocity and temperature ratios in table~\ref{tab:jump_quantities}.}

\begin{figure}
\begin{center}
\includegraphics[width=1\textwidth]{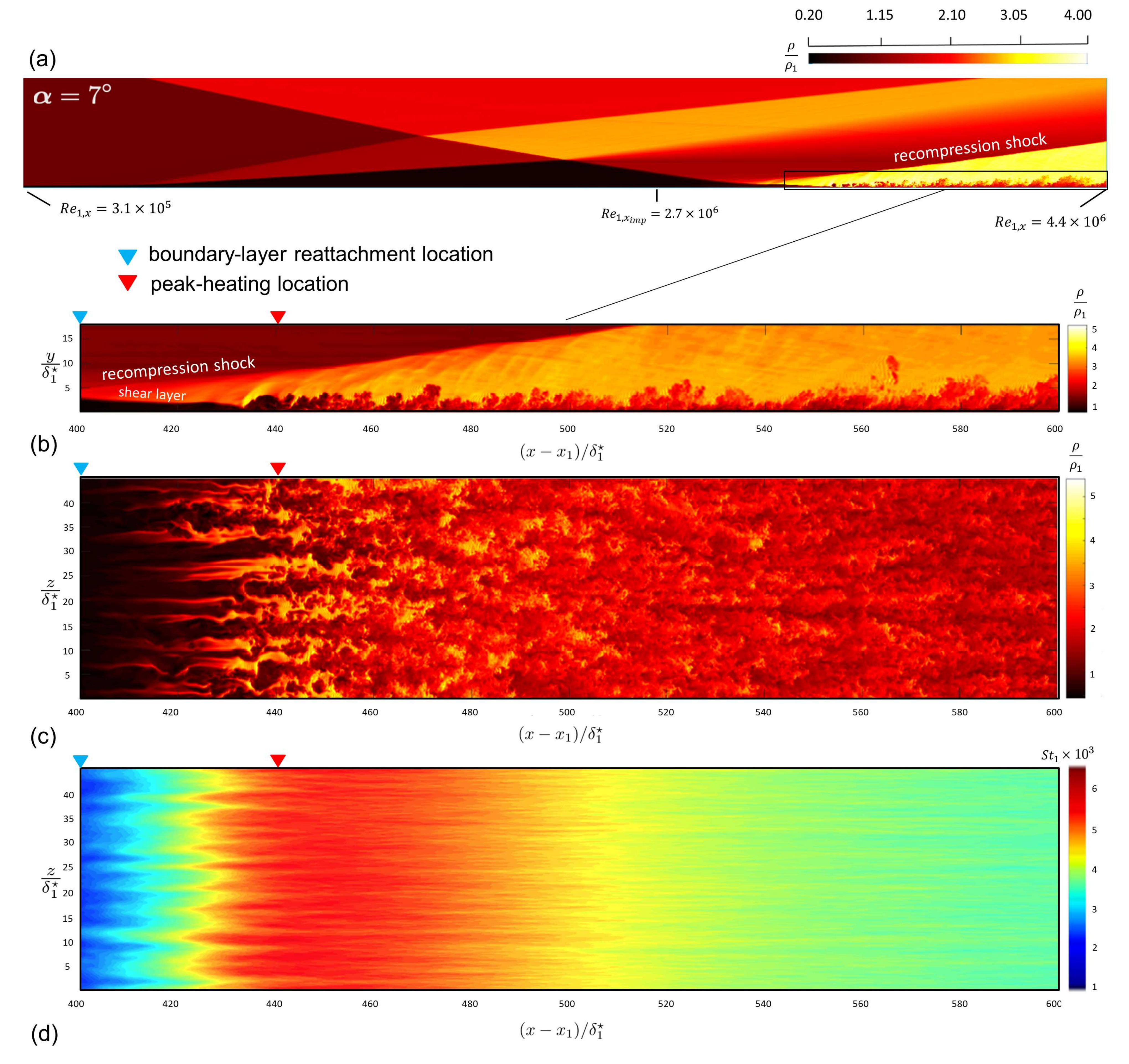}
\caption{DNS instantaneous contours for the case $\alpha=7^\circ$ including (a)~sideways view of the density field, (b)~zoomed view of the post-recompression-shock region, (c)~plane view of the same region at $y^{+}=100$ based on outflow conditions, along with (d)~time-averaged contours of the Stanton number {$St_1$} along the wall. The streamwise locations indicated by the triangles are averaged in time and along the spanwise coordinate.}
\label{DNS_777}
\end{center}
\end{figure}

{A side view of the resulting separation bubble for the case $\alpha=7^\circ$ can be approximately identified as the dark triangle-shaped region at the foot of the incident shock in figure~\ref{DNS_777}(a), where the gas heats up by the reversing flow deceleration and its density reaches small values. Despite of the high temperatures of the gas in the separation bubble, the wall heat flux is relatively small in this region, since the velocity gradients involved in the recirculating flow are small in comparison with those present in the laminar and turbulent portions of the boundary layer. Experimental flow visualizations by \cite{currao2020hypersonic} show flow features qualitatively similar to those revealed by the density field in figure~\ref{DNS_777}(a).}

{In addition to the separation bubble, figure~\ref{DNS_777}(a) indicates that the structure of the flow ensuing from the interaction consists of a separation shock emanating from the point of flow reversal, an expansion fan radiated from the crest of the separation bubble as the supersonic overriding flow turns downwards around it, and a recompression shock created at the point of reattachment.} The incident and separation shocks intersect along a horizontal line in the spanwise direction above the separation bubble, perpendicularly to the plane of figure~\ref{DNS_777}(a). The result is a regular reflection that shifts the effective interaction region downstream by approximately $75\delta_1^\star$ with respect to the inviscid shock-impingement location $x_{\textrm{imp}}$. There, the effective incidence angle $\beta$ of the shock impinging on the boundary layer is closer to $\beta\approx 11^\circ$ than to the theoretical value $\beta=14.8^\circ$ corresponding to the weak solution of an oblique shock created by an $\alpha=7^\circ$-wedge. {As a consequence, the effective incidence angle of the shock is always smaller than the theoretical one predicted by the inviscid solution unless the incident shock is sufficiently weak to prevent separation.}

\begin{figure}
\begin{center}
\includegraphics[width=1\textwidth]{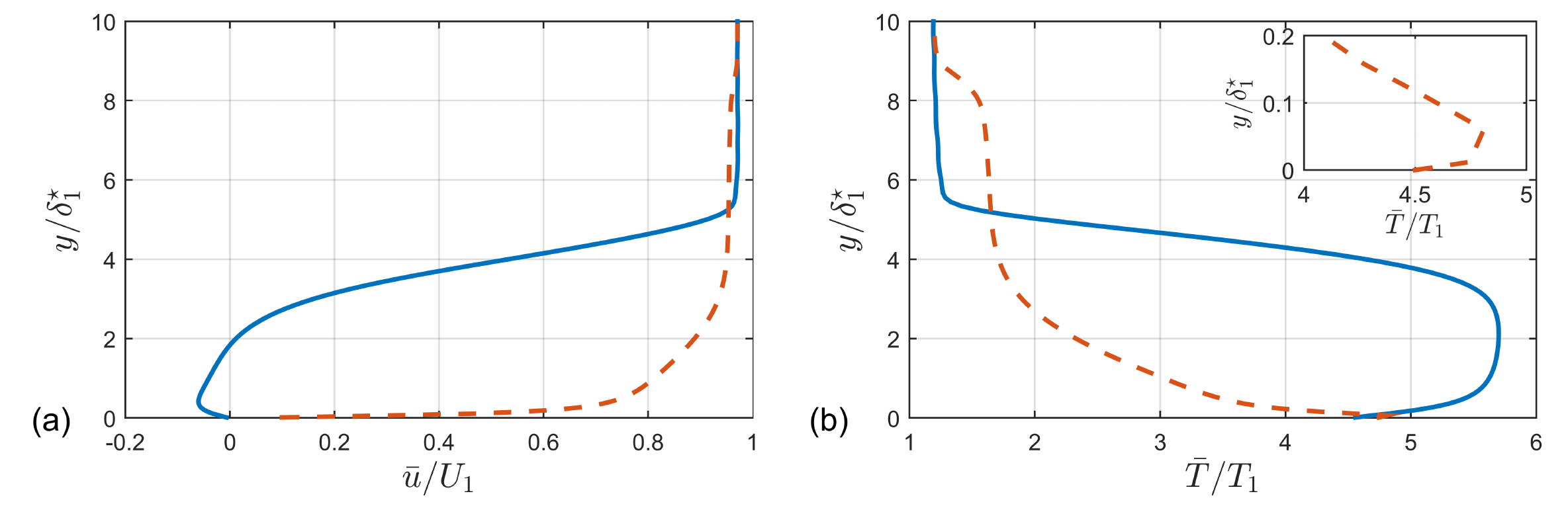}
\caption{{DNS results for the case $\alpha=7^\circ$ including time- and spanwise-averaged profiles of (a)~streamwise velocity and (b)~temperature. The profiles are extracted at the streamwise locations $(x-x_1)/\delta^\star_1=380$ (solid lines; location within the separation bubble) and $(x-x_1)/\delta^\star_1=445$ (dashed lines; location near peak heating). In panel~(b) a peak in the temperature profile at the station $(x-x_1)/\delta^\star_1=445$ develops very close to the wall, as shown in the inset.}}\label{UT_380_445}
\end{center}
\end{figure}

{In the cases $\alpha=6^\circ$, $7^\circ$, and $8^\circ$, the boundary layer transitions to turbulence on the leeward side of the separation bubble, shortly downstream of the time- and spanwise-averaged streamwise coordinate for reattachment.} A zoomed side view of this region is provided in figure~\ref{DNS_777}(b). Dynamic visualizations of the flow in this region show a persistent flapping motion of the shear layer formed between the low-speed recirculating flow within the separation bubble and the high-speed flow above. This flapping motion, in conjunction with early streaks generated shortly upstream of the reattachment point, lead to the onset of broadband turbulence at the same location where the maximum value of the Stanton number occurs, as observed in figure~\ref{DNS_777}(c) and further discussed below.

The DNS distribution of the time-averaged Stanton number shown in figure~\ref{DNS_777}(d) for the case $\alpha=7^\circ$ suggests the presence of quasi-stationary streaky thermal footprints of the flow onto the wall near the transition region. These structures have a spanwise wavelength of approximately $5\delta_1^\star$. {These structures do not vanish by increasing the averaging time interval, as corroborated by similar streaky thermal patterns observed experimentally using infrared thermography by \cite{currao2020hypersonic}. It should be noted that the variation of the time-averaged Stanton number along the spanwise direction is expected in the transitional region since this is the signature of the underlying instability mechanism, which is characterized by a non-zero spanwise wavenumber along with a purely exponential growth in time at each point in space \citep{hildebrand2018simulation}.}

The spanwise- and time-averaged velocity and temperature profiles in the transitional region at station $(x-x_1)/\delta^\star_1=380$ {within the separation bubble} are indicated by the solid lines in figure~\ref{UT_380_445}. The overall flow overriding the separation bubble corresponds to an inflectional shear layer. The temperature attains a maximum at $y/\delta^\star_1=2.15$ and attenuates towards the wall due to the cold-wall boundary condition.

\begin{figure}
\begin{center}
\includegraphics[width=1\textwidth]{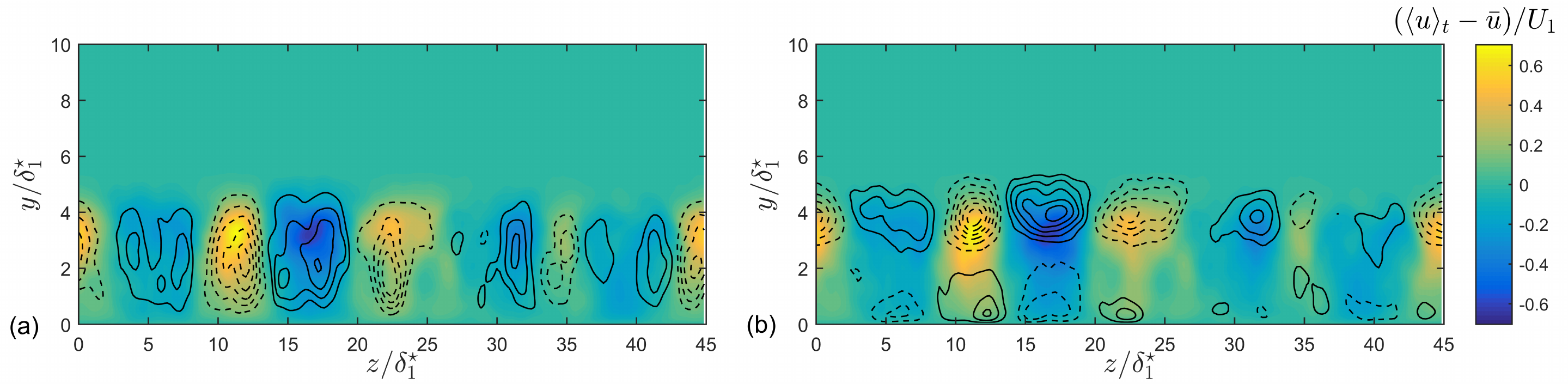}
\caption{{DNS results for the case $\alpha=7^\circ$ including solid contours of the time-averaged fluctuations of streamwise velocity (denoted by  $\langle u \rangle_t$) at $(x-x_1)/\delta^\star_1=380$ within the separation bubble. The solid (dashed) lines indicate positive (negative) time-averaged fluctuations of the wall-normal velocity with contour spacing of 0.03 in dimensionless units [in panel~(a)], as well as positive (negative) time-averaged temperature fluctuations with contour spacing of 0.05 in dimensionless units [in panel~(b)].}}\label{perturb_U_380}
\end{center}
\end{figure}

{The lack of monotonicity in the temperature profile in figure~\ref{UT_380_445}(b) in the transitional region has important consequences on the cross-correlations between velocity and temperature fluctuations in the boundary layer. To visualize this, consider the time-averaged spatial fluctuations of the streamwise and wall-normal velocities in the cross-stream plane shown in figure~\ref{perturb_U_380}(a). Similarly to the stationary spanwise structures of the Stanton number observed in figure~\ref{DNS_777}(d),  the spatial inhomogeneity of the velocity fluctuations in the spanwise direction is stationary, since both are signatures of the underlying instability mechanism. Four sets of high and low-speed streaks are observed in figure~\ref{perturb_U_380}(a), with maximum magnitudes in the region of strong shear ($3\lesssim y/\delta^\star_1\lesssim 5$). The time-averaged spatial fluctuations of the streamwise and wall-normal velocities are anti-correlated along the span, where the interaction between the mean shear and streamwise vortices results in streamwise velocity streaks. The time-averaged spatial fluctuations of the temperature and streamwise velocity on the cross-stream plane are shown in figure~\ref{perturb_U_380}(b). Two clearly distinguished regions are observed there: a first region above the wall-normal location of maximum mean temperature ($y/\delta^\star_1=2.15$), where the fluctuations of $u$ and $T$ are anti-correlated, and a second region below the aforementioned wall-normal location, where the fluctuations of $T$ flip their sign and become positively correlated with the fluctuations of $u$. This sign change is in agreement with the lift-up effect, since positive wall-normal velocity shifts hot gas away from the wall for the portion $y/\delta^\star_1>2.15$ of the temperature profile that has a negative gradient, whereas the opposite happens for the portion $y/\delta^\star_1<2.15$ of the temperature profile that has a positive gradient.}

\begin{figure}
\begin{center}
\includegraphics[width=1\textwidth]{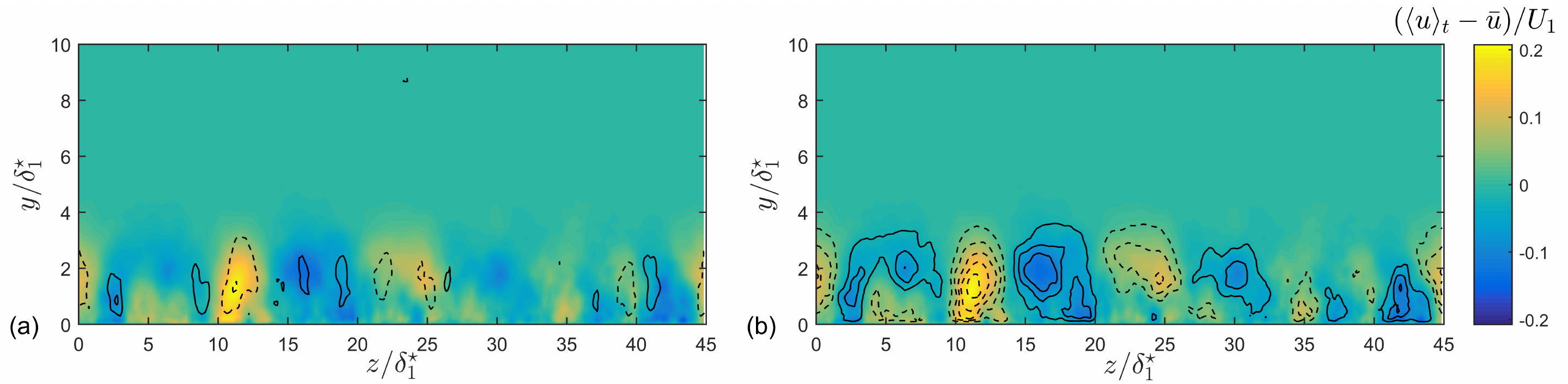}
\caption{{DNS results for the case $\alpha=7^\circ$ including solid contours of the time-averaged fluctuations of streamwise velocity (denoted by  $\langle u \rangle_t$) at $(x-x_1)/\delta^\star_1=445$ near peak heating. The solid (dashed) lines indicate positive (negative) time-averaged fluctuations of the wall-normal velocity with contour spacing of 0.02 in dimensionless units [in panel~(a)], as well as positive (negative) time-averaged temperature fluctuations with contour spacing of 0.05 in dimensionless units [in panel~(b)].}}\label{perturb_U_445}
\end{center}
\end{figure}
\begin{figure}
\begin{center}
\includegraphics[width=1.35\textwidth, angle =90]{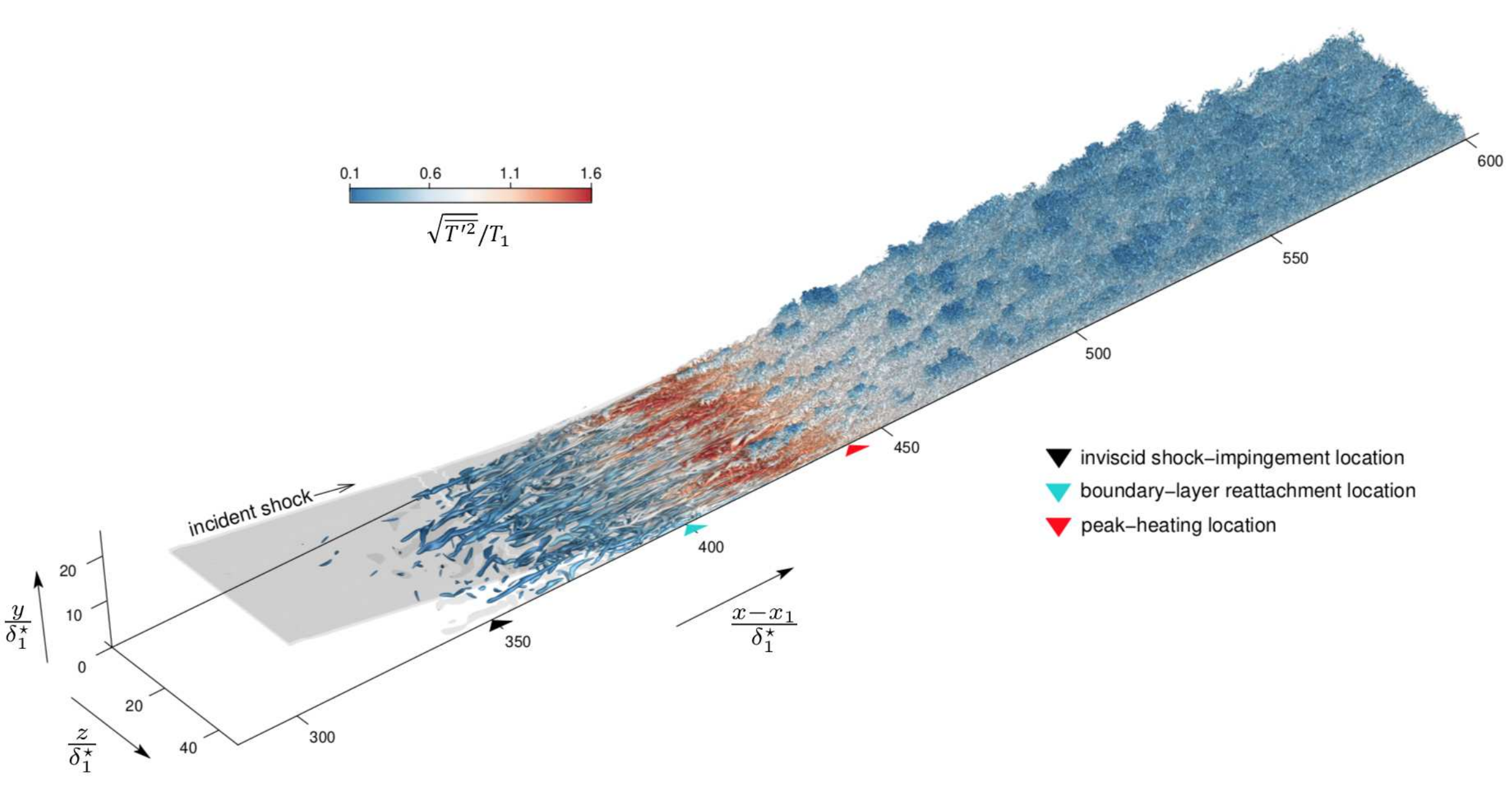}
\end{center}
\caption{{DNS instantaneous three-dimensional visualization of shock-induced transition in the case $\alpha=7^\circ$, including isosurfaces of the second invariant $Q$ of the velocity-gradient tensor colored by the magnitude of the dimensionless root-mean-square temperature. The value of $Q$ on the isosurfaces is set to  $Q=0.3 a_1^2/{{\delta^\star_1}^2}$, with $a_1$ being the speed of sound in the free stream at the inflow. The triangles mark key locations based on time and spanwise-averaged quantities.}}\label{DNS_777_Q}%
\end{figure}

{Similar considerations as those made above also apply downstream of the separation bubble}. In particular, the dashed lines in figure~\ref{UT_380_445} show the time- and spanwise-averaged velocity and temperature profiles for the downstream station $(x-x_1)/\delta^\star_1=445$ deep in the transitional zone where the Stanton number attains its maximum value. In figure~\ref{UT_380_445}(b), the temperature profile arrives at the wall with a positive slope because of a spike that is not visible in this vertical scale but is revealed later in \S\ref{Turbulence_statistics} by zoomed-up views near the wall. The corresponding time-averaged spatial fluctuations of the streamwise and wall-normal velocities on the cross-stream plane are shown in figure~\ref{perturb_U_445}(a). The signature of the four sets of streamwise streaks that were observed upstream in figure~\ref{perturb_U_380}(a) is still visible here. However, the structures are now distorted by harmonic interactions along the span (e.g., see $z/\delta^\star_1\approx 5$ and 40), and the magnitude of the fluctuations is much smaller compared with the upstream values shown in figure~\ref{perturb_U_380}(a). The time-averaged spatial fluctuations of the temperature and streamwise velocity in figure~\ref{perturb_U_445}(b) are almost anti-correlated along the entire cross-section except for a very thin region close to the wall below the temperature peak, where the positive temperature gradient induces a change in the sign of the correlation. As a result, the non-monotonicity of the mean temperature caused by the wall coldness has a fundamental effect that leads to the breakdown of Morkovin's hypothesis below the wall-normal location of the maximum temperature. This aspect is further analyzed in \S\ref{Turbulence_statistics}.

{The late stages of transition are visualized in figure~\ref{DNS_777_Q} using instantaneous iso-surfaces of the second invariant $Q$ of the velocity-gradient tensor for the case $\alpha=7^\circ$. The iso-surfaces are colored by the root-mean-square (rms) temperature and the incident shock is superimposed. The four large-scale spanwise structures discussed above are seen lingering between the reattachment and peak-heating location [$400\lesssim(x-x_1)/\delta^\star_1\lesssim 440$]. Downstream of the peak-heating location, a breakdown into much smaller scales is observed. For the case $\alpha=7^\circ$, the boundary layer ensuing from the shock-induced transition approaches the outflow in a turbulent state approximately at a post-interaction free-stream Mach number $Ma_2=4.2$ and a post-interaction momentum-based Reynolds number $Re_{2,\theta}=2,850$.}

\subsection{Comparisons between DNS and WMLES near the shock impingement} \label{WMLES1}

\begin{figure}
\begin{center}
\includegraphics[width=0.75\textwidth]{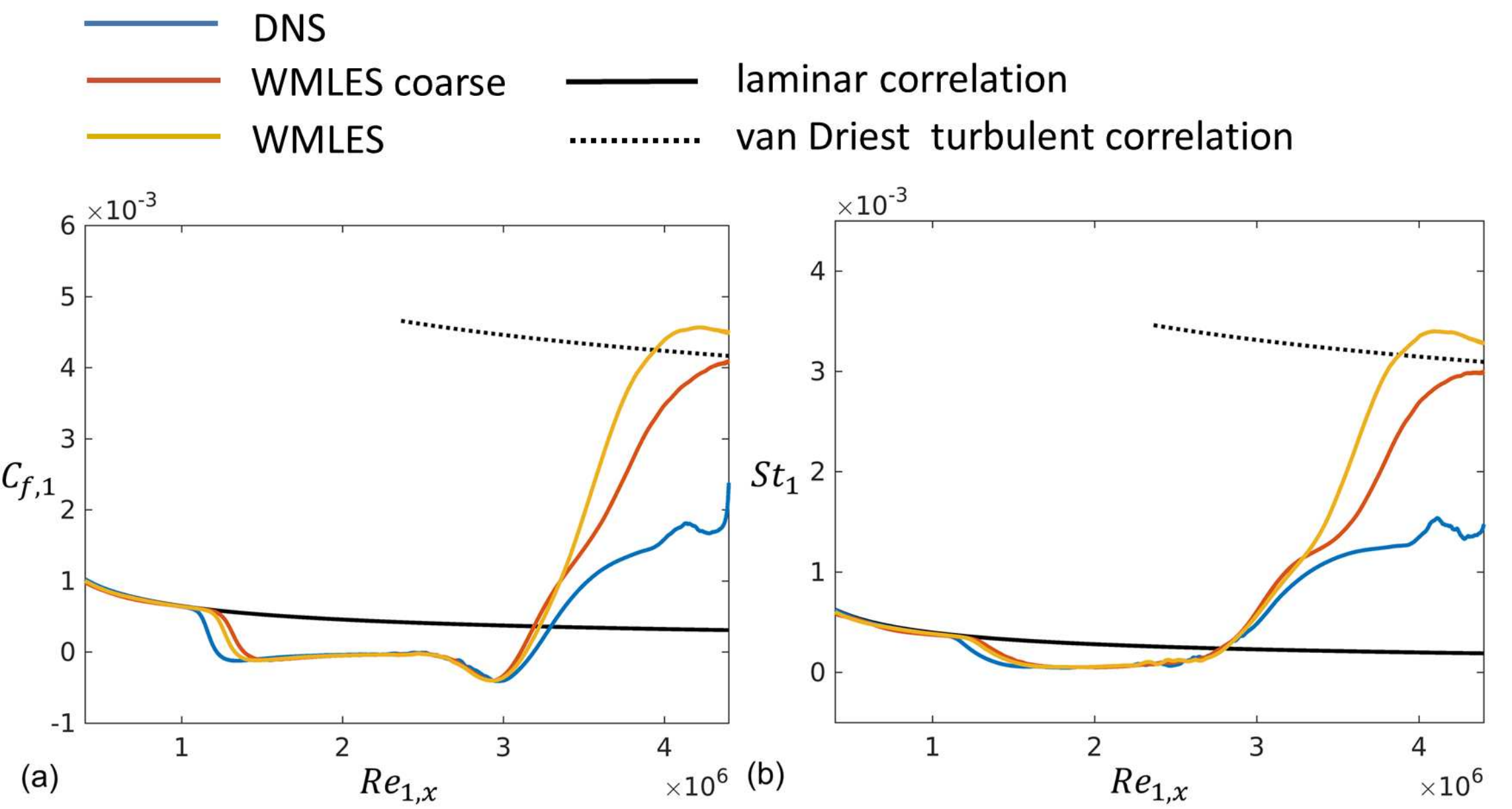}\\
\includegraphics[width=0.75\textwidth]{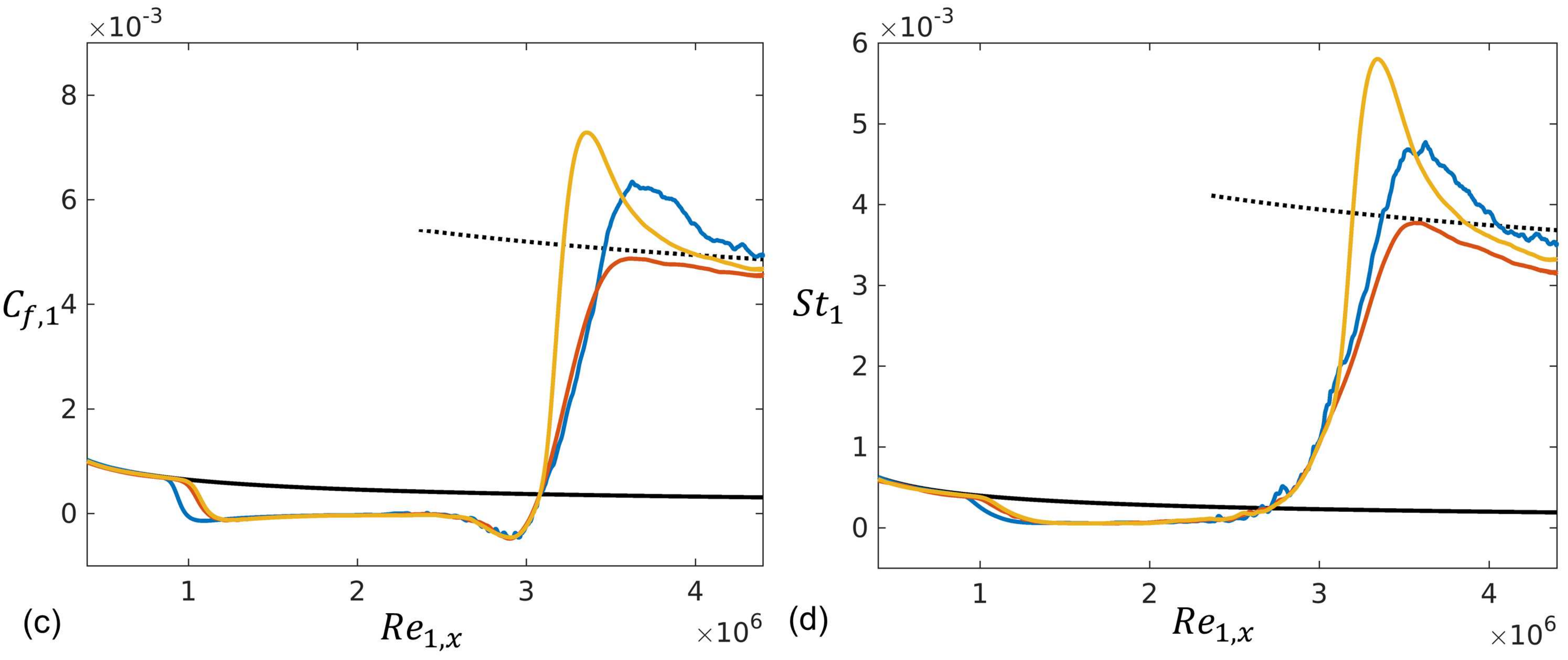}\\
\includegraphics[width=0.75\textwidth]{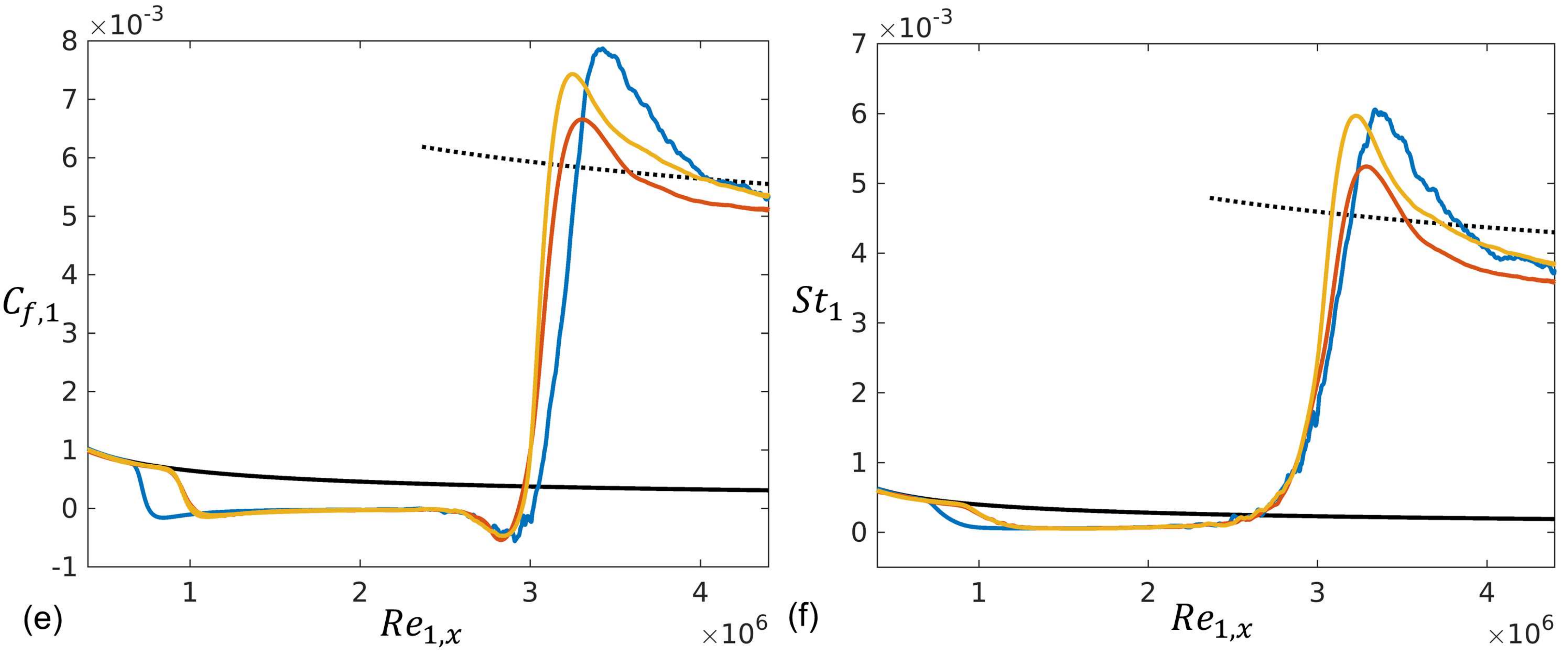}\\
\includegraphics[width=0.75\textwidth]{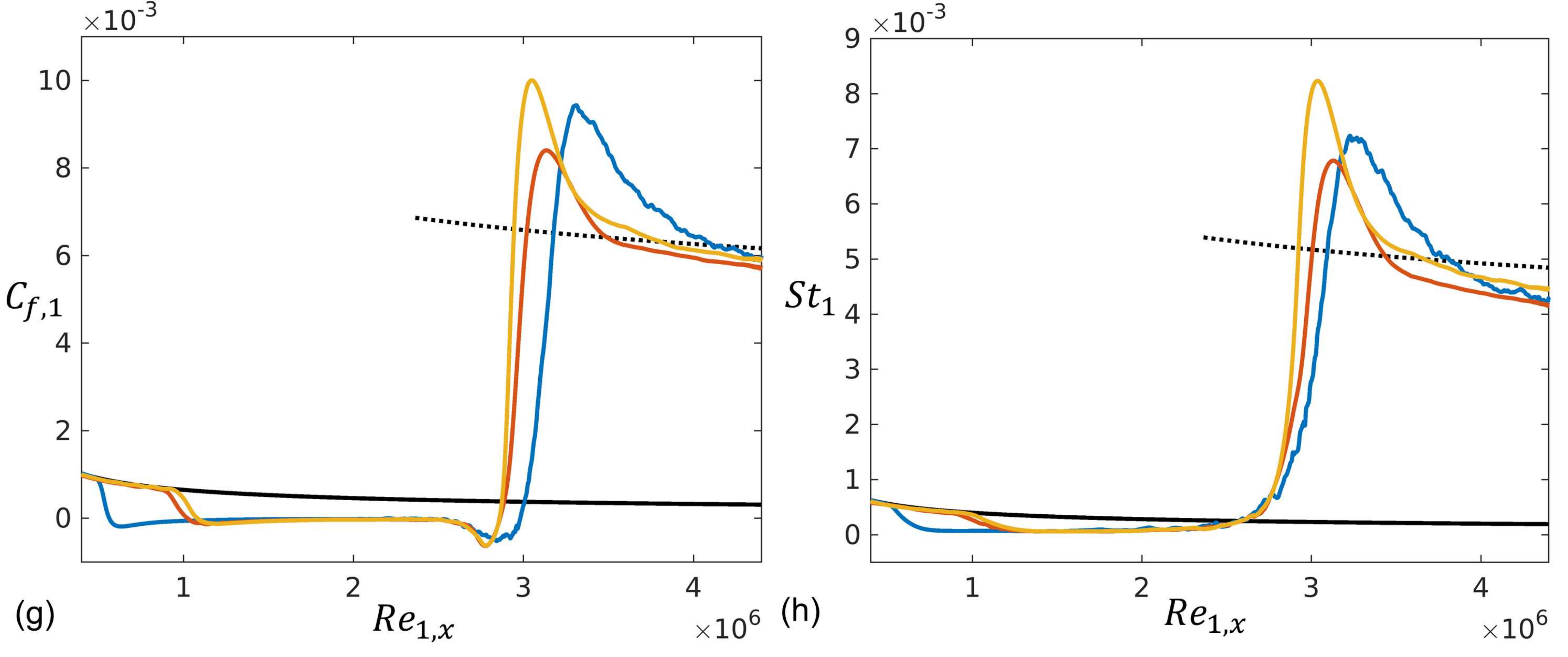}\\
\caption{{(a,c,e,g) Skin friction coefficient and (b,d,f,h) Stanton number as a function of the local Reynolds number $Re_{1,x}$ for wedge angles (a,b)~$\alpha = 5^\circ$, (c,d)~$\alpha = 6^\circ$, (e,f)~$\alpha = 7^\circ$, and (g,h)~$\alpha = 8^\circ$. Black solid and dotted lines denote, respectively, the laminar correlation and the van Driest turbulent correlation (see figure~\ref{Cf_St_all_angles} caption for details on the calculation of the turbulent correlation). Blue, yellow, and red lines denote, respectively, DNS, baseline WMLES, and coarse WMLES.}}
\label{DNS_WMLES_St_Cf_compare}
\end{center}
\end{figure}

{The DNS distributions of $C_{f,1}$ and $St_1$ as a function of the streamwise Reynolds number $Re_{1,x}$ are compared in figure~\ref{DNS_WMLES_St_Cf_compare} with those obtained using WMLES. In all cases, WMLES underpredicts the size of the separation bubble with respect to DNS. Additionally, WMLES delays separation (i.e., the streamwise coordinate at separation predicted by WMLES is always larger than that predicted by DNS). A good agreement is however observed between DNS and WMLES in the distributions of $C_{f,1}$ and $St_1$ upstream and within the separation zone.}

\begin{figure}
\vskip 0.2in
\begin{center}
\includegraphics[width=0.9\textwidth]{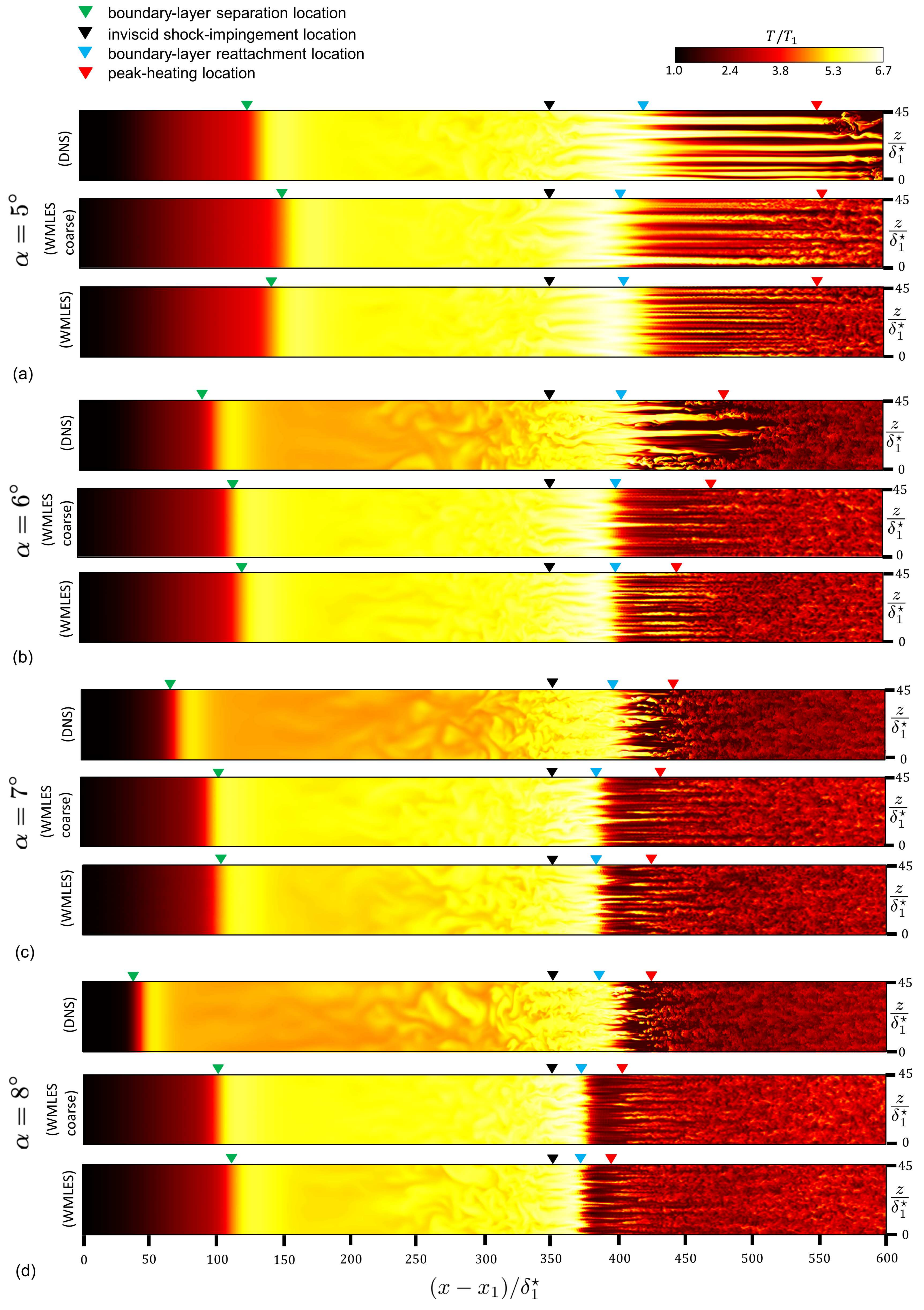}
\caption{{DNS and WMLES instantaneous temperature contours on a plane parallel to the wall at $y/{\delta^\star_1}= 1.65$ for (a)~$\alpha=5^\circ$, (b)~$\alpha=6^\circ$, (c)~$\alpha=7^\circ$, and (d)~$\alpha=8^\circ$. The streamwise locations indicated by triangles have been averaged in time and spanwise direction.}}
\label{DNS_WMLES_instant_all}
\end{center}
\end{figure}

{An additional aspect revealed by figure~\ref{DNS_WMLES_St_Cf_compare} is that WMLES prompts transition and peak heating with respect to DNS in all cases (i.e., transition starts always earlier along the streamwise coordinate in WMLES). The discrepancies between WMLES and DNS become clearly evident in the case $\alpha=5^\circ$. Specifically, the skin friction and Stanton number in WMLES rise to values significantly larger than the DNS. Closer examination of the solution in terms of the temperature contours in figure~\ref{DNS_WMLES_instant_all}(a) shows that the post-interaction boundary layer undergoes transition in both baseline and coarse WMLES, whereas no transition is observed in DNS within the present computational domain. At larger wedge angles, $\alpha=6^\circ$, $7^\circ$, and $8^\circ$, both DNS and WMLES predict transition to turbulence shortly downstream of the impingement by the shock, although WMLES always does it slightly in advance with respect to DNS, as shown in figure~\ref{DNS_WMLES_instant_all}(b-d).}

{Apparently, the incorrect transition predicted by WMLES at $\alpha=5^\circ$, when the shock has a relatively modest effect, is caused by  numerical errors, which spuriously influence the dynamics of the post-interaction boundary layer in absence of  competing disturbances of physical origin (see \S\ref{sect:introduction} for a discussion about the influences of inflow disturbances on transition when the shock angle is small). In contrast, at  higher incidence angles, $\alpha=6^\circ$, $7^\circ$, and $8^\circ$, when the shock has a stronger effect, the WMLES prediction of transition is reasonable, albeit spatially advanced with respect to DNS. That the transition predicted by WMLES at $\alpha=6^\circ$, $7^\circ$, and $8^\circ$ at both grid resolutions is not the result of a confabulation of numerical and modeling mishaps, is evidenced, for instance, by the correct spatial trend of the transition front moving upstream as $\alpha$ increases in figure~\ref{DNS_WMLES_instant_all}, or by the increase of the peak thermal load with $\alpha$ in figure~\ref{Cf_St_peaks}(b) in a manner that resembles the DNS.}

{The distribution of the mean pressure along the wall is predicted reasonably well by WMLES, as shown in figure~\ref{DNS_7_pressure}. Mismatches of 10\% are observed in the transitional region and near the separation point. Additionally, both WMLES and DNS agree well with the post-interaction pressure, velocity, and temperature anticipated by the inviscid solution, as observed in table~\ref{tab:jump_quantities}.}

\begin{figure}
\begin{center}
\vskip 0.1in
\includegraphics[width=0.85\textwidth]{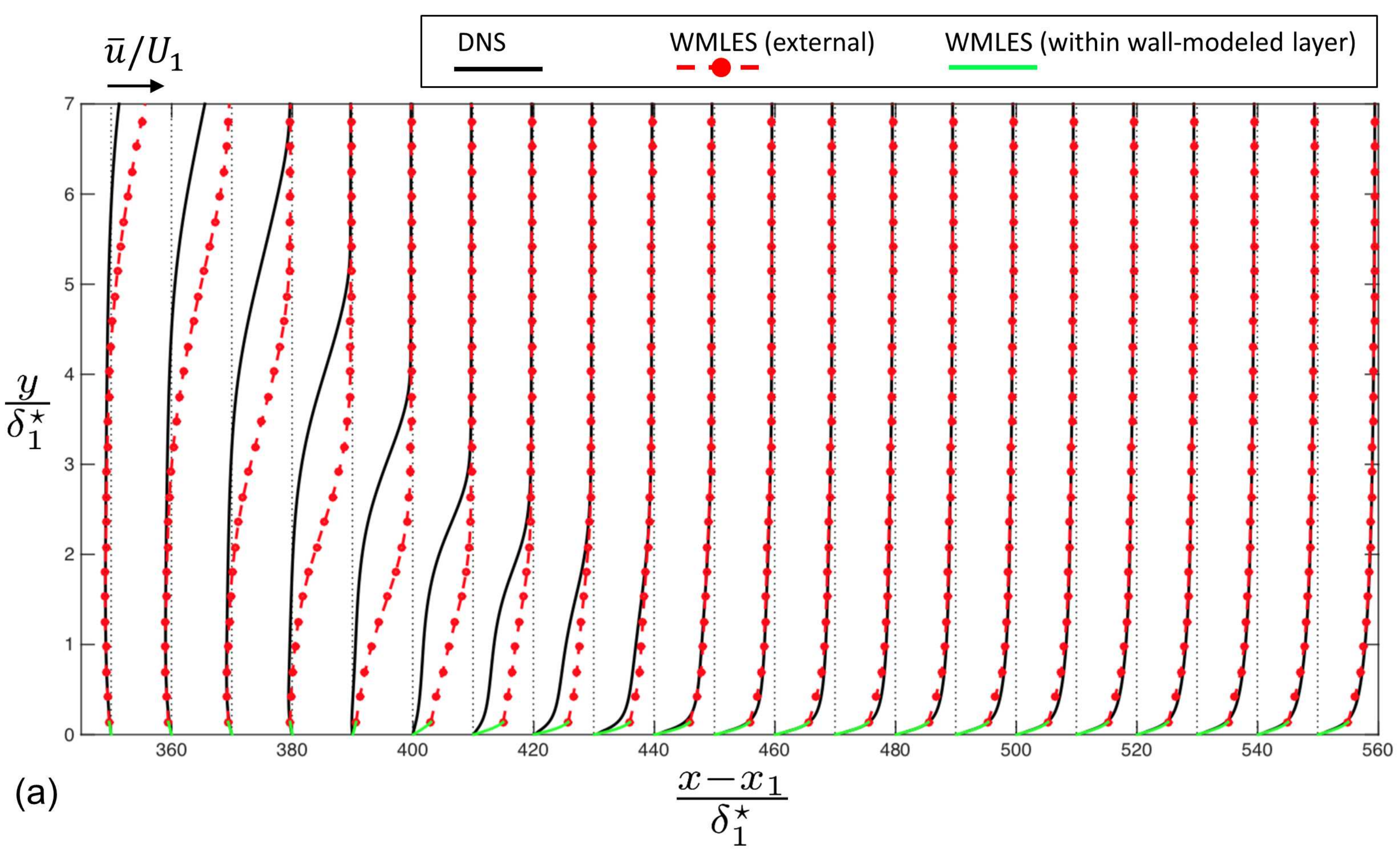}
\includegraphics[width=0.85\textwidth]{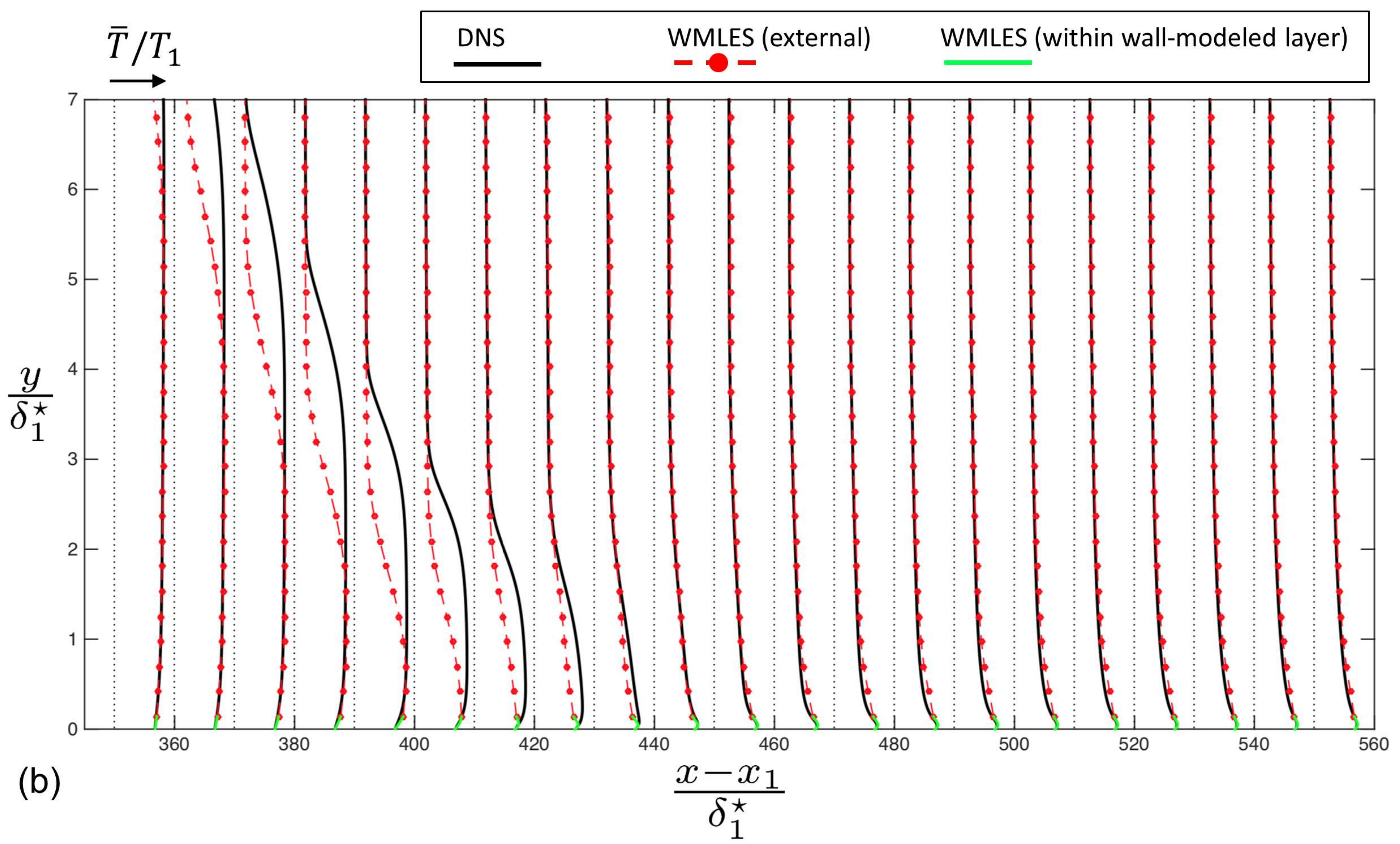}
\caption{{DNS (solid lines) and baseline WMLES (dot-dashed lines) of time- and spanwise-averaged profiles of (a)~streamwise velocity and (b)~temperature at several stations along the $x$ axis for $\alpha=7^\circ$. In panel~(a), the velocity is plotted as $10(\overline{u}/U_{1})+(x-x_1)/\delta_1^\star$, whereas in panel~(b) the temperature is plotted as $(3/2)(\overline{T} /T_{1})+(x-x_1)/\delta_1^\star$. Also included are the solutions of the equilibrium wall model (green lines) within the wall-modeled region $y\leq h_{wm}$.}}
\label{DNS_WMLES_mean_TU_compare}
\end{center}
\end{figure}
\begin{figure}
\begin{center}
\vskip 0.1in
\includegraphics[width=0.85\textwidth]{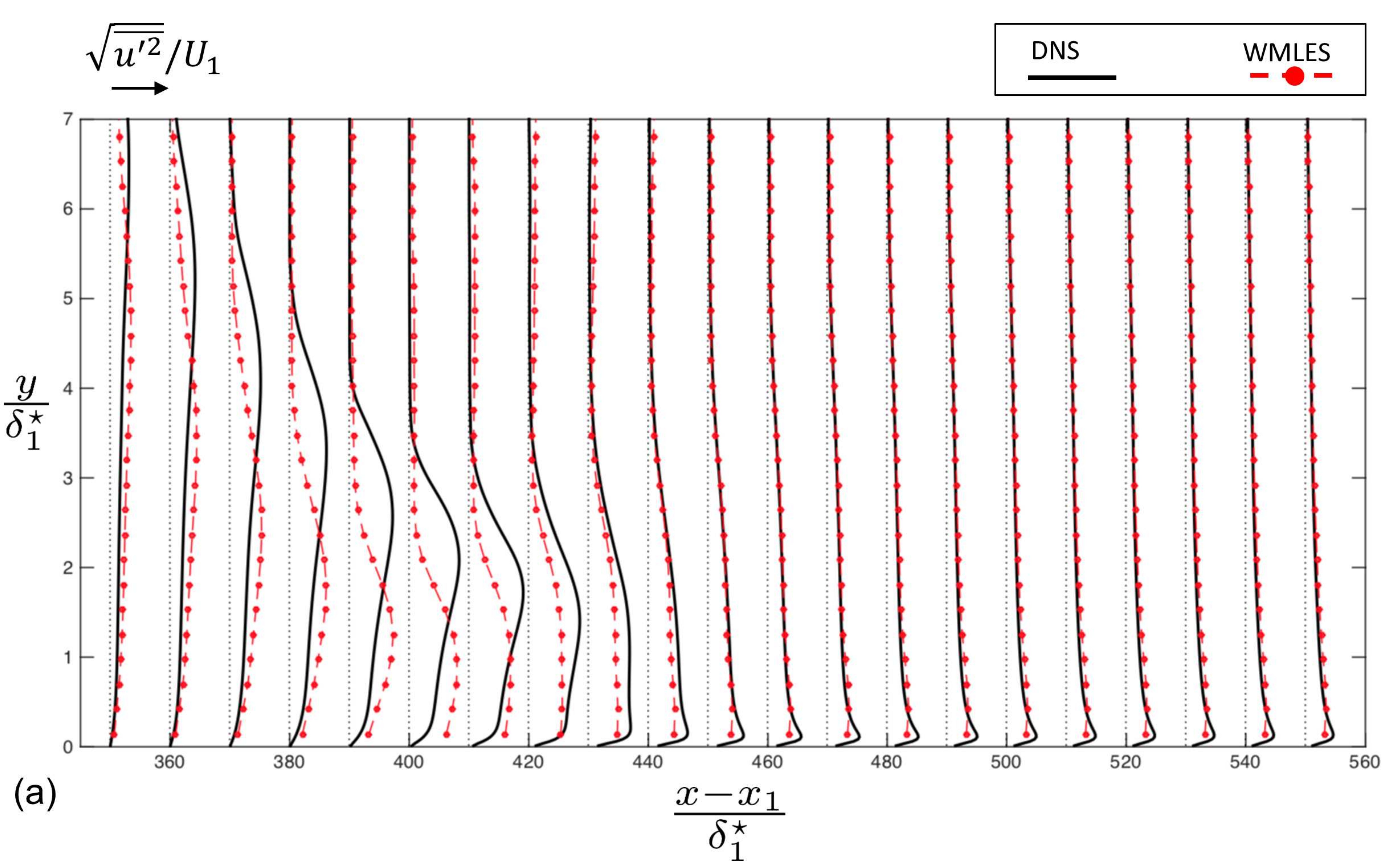}
\includegraphics[width=0.85\textwidth]{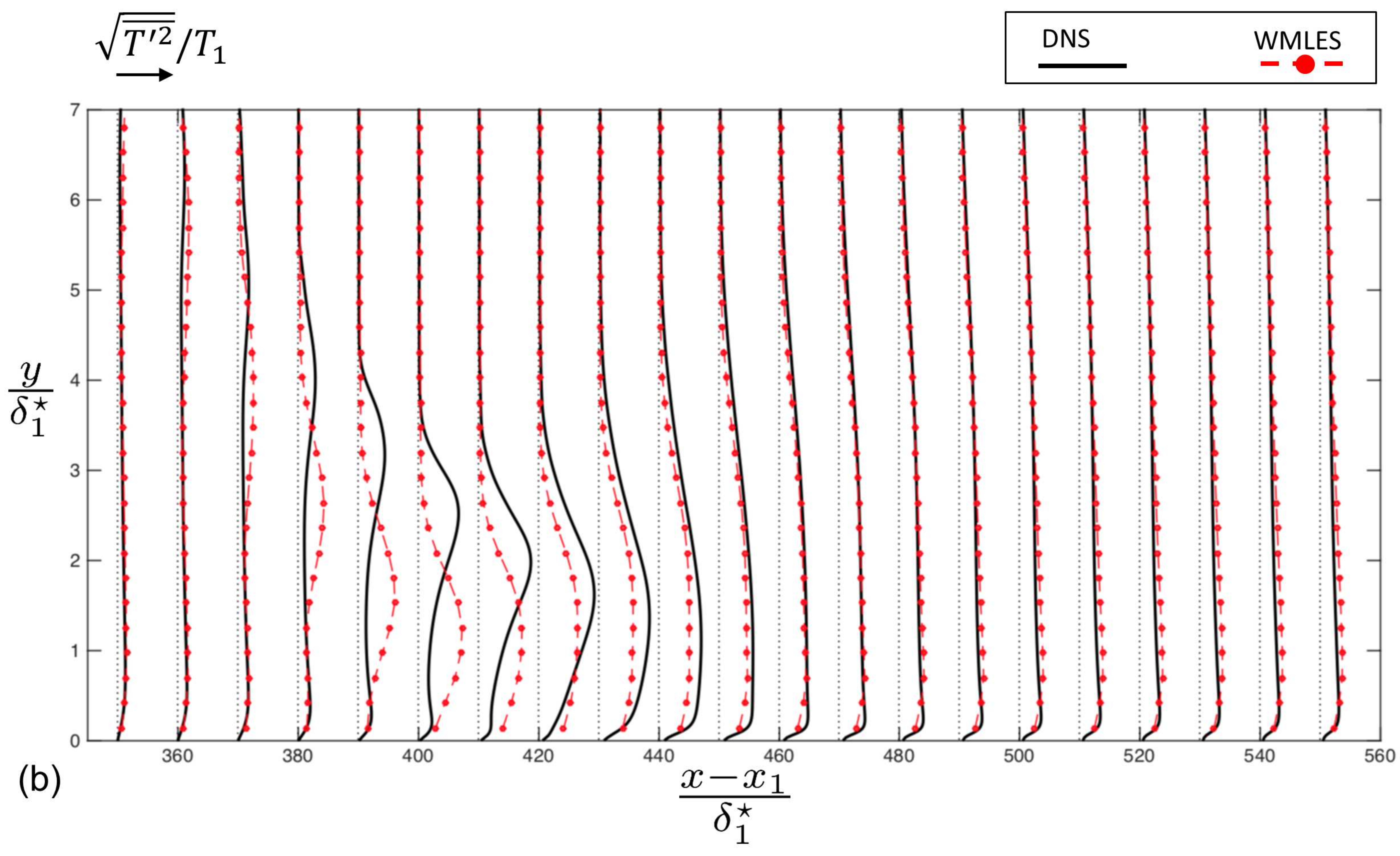}
\caption{{DNS (solid lines) and baseline WMLES (dot-dashed lines) of rms profiles of (a)~streamwise velocity and (b)~temperature at several stations along the $x$ axis for $\alpha=7^\circ$. In panel~(a), the velocity is plotted as $36\sqrt{\overline{u'^2}} /U_{1}+(x-x_1)/\delta_1^\star$, whereas in panel~(b) the temperature is plotted as $6\sqrt{\overline{T'^2}} /T_{1}+(x-x_1)/\delta_1^\star$.}}
\label{DNS_WMLES_RMS_TU_compare}
\end{center}
\end{figure}
\begin{figure}
\begin{center}
\includegraphics[width=0.8\textwidth]{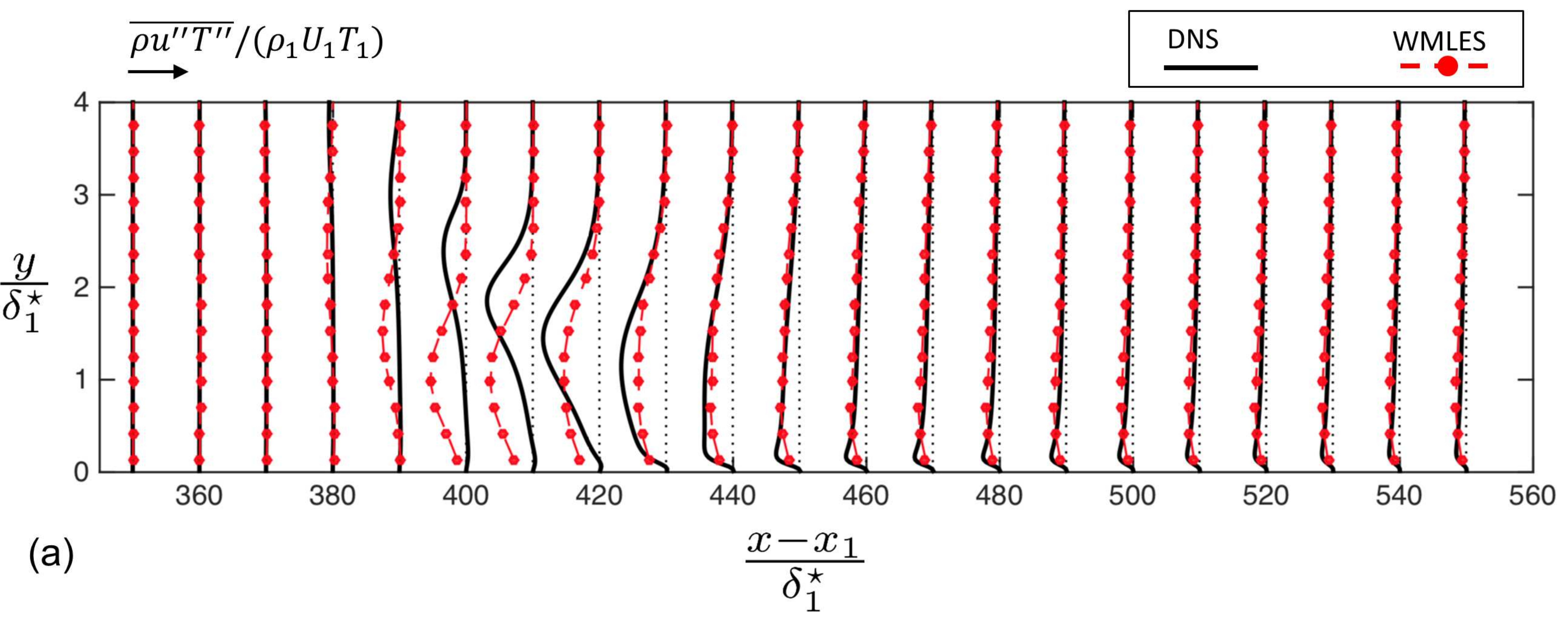}
\includegraphics[width=0.8\textwidth]{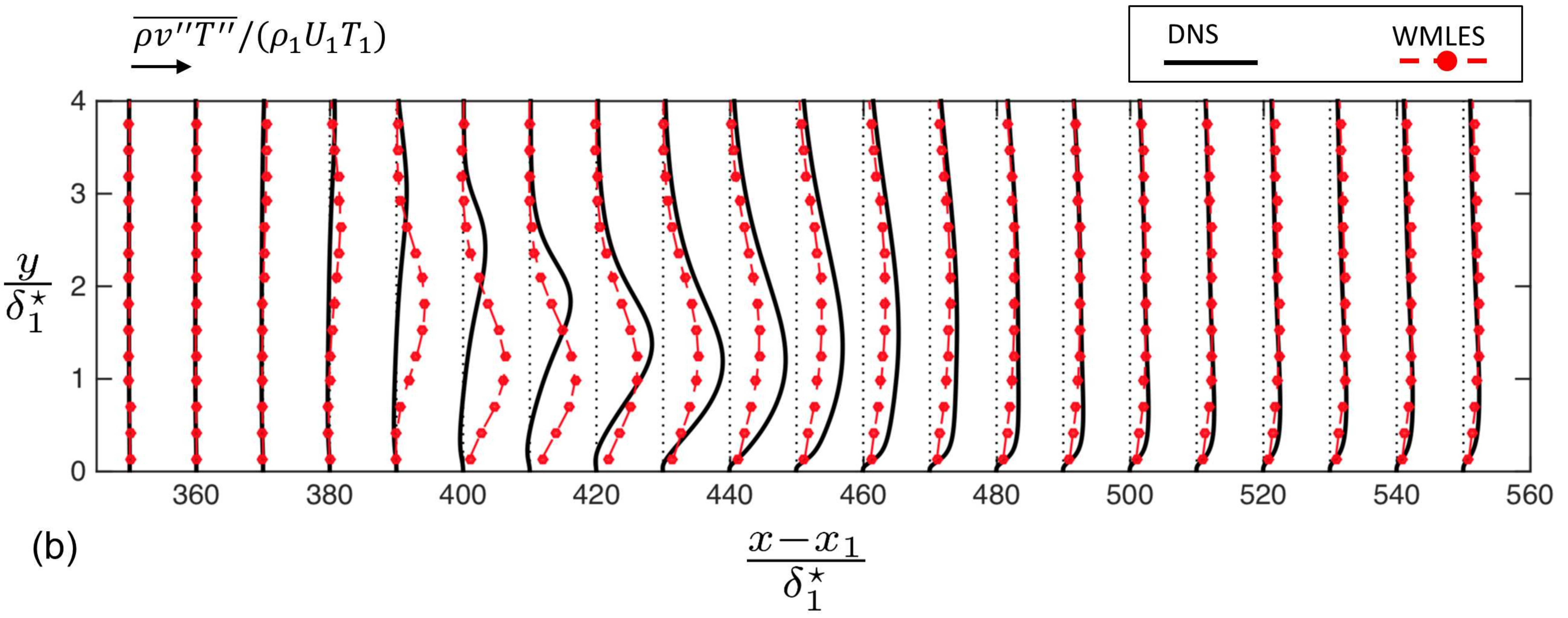}
\includegraphics[width=0.8\textwidth]{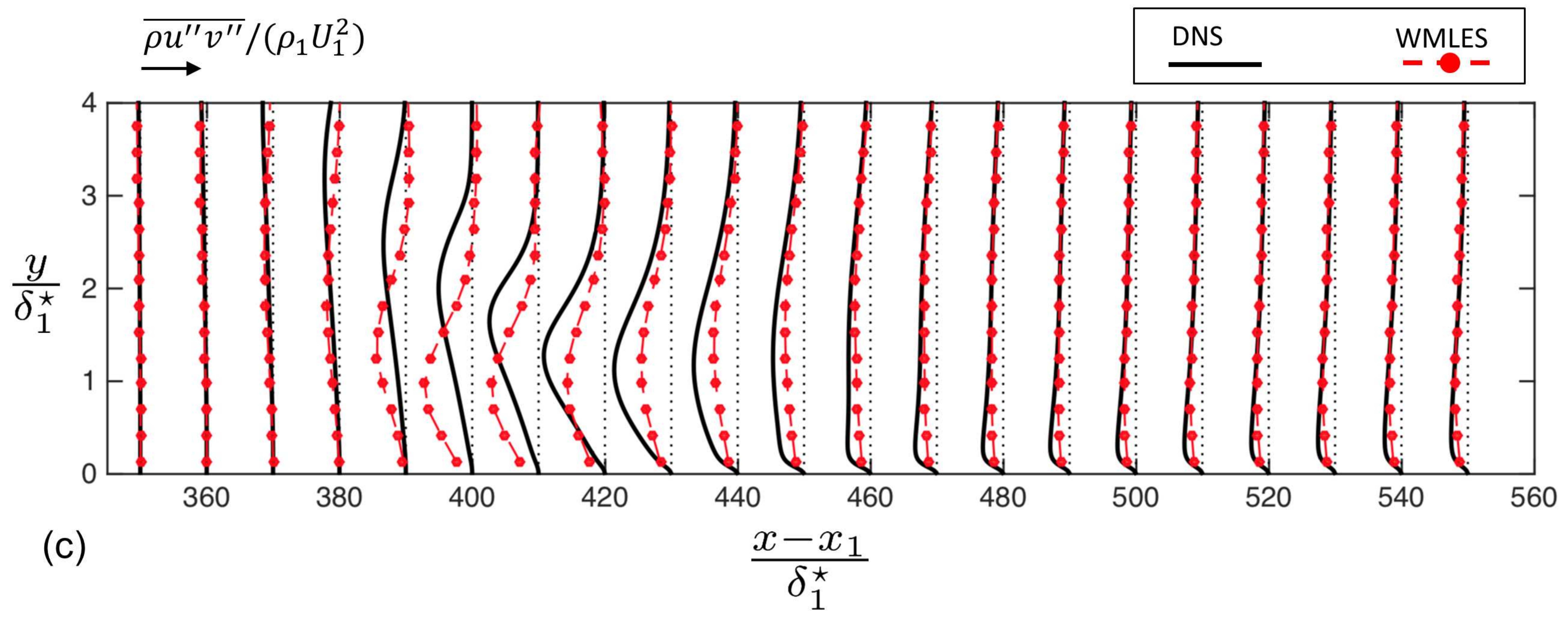}
\caption{{DNS (solid lines) and baseline WMLES (dot-dashed lines) of (a)~streamwise turbulent heat flux, (b)~wall-normal turbulent heat flux, and (c)~Reynolds stress at several stations along the $x$ axis for $\alpha=7^\circ$. The data is plotted as $20\overline{\rho u''T''} /(\rho_1U_1T_1)+(x-x_1)/\delta_1^\star$ in panel~(a), as  $150\overline{\rho v''T''} /(\rho_1U_1T_1)+(x-x_1)/\delta_1^\star$ in panel~(b), and as $1000\overline{\rho u''v''} /(\rho_1U_1^2)+(x-x_1)/\delta_1^\star$ in panel~(c).}}
\label{DNS_WMLES_stress_compare}
\end{center}
\end{figure}

The effect of the WMLES grid resolution is most significant in the transitional region. Specifically, a decrease in the WMLES grid resolution leads to shallower rises of wall pressure, $C_{f,1}$, and $St_1$ in the transitional region, along with smaller peak values of $C_{f,1}$ and $St_1$. Comparisons between DNS and WMLES peak values of $C_{f,1}$ and $St_1$ as a function of the wedge angle $\alpha$ in figure~\ref{Cf_St_peaks} indicate that improved agreement is obtained with the DNS results as the WMLES grid resolution is increased. Farther downstream of transition and peak heating, where the boundary layer becomes turbulent, only moderate differences are observed between the two resolutions, and the values of $C_{f,1}$ and $St_1$ predicted by WMLES nearly collapse on those of DNS. Overall, the baseline WMLES is closer to the DNS results in the transitional region than the coarse WMLES. {Additional results are provided in appendix~D, where the WMLES grid is coarsened isotropically in the three directions for $\alpha=7^\circ$.}



{A more detailed comparison is made between the DNS and WMLES flow fields in figure~\ref{DNS_WMLES_instant_all} by examining the instantaneous temperature contours on a plane parallel to the wall at $y/\delta^\star_1=1.65$. In the case $\alpha=5^\circ$ shown in figure~\ref{DNS_WMLES_instant_all}(a), the flow in the DNS contains organized streaks in the post-interaction region that persist downstream without undergoing breakdown. In contrast, the boundary layer in both WMLES cases involves unstable narrower streaks, and eventually transitions to turbulence. In the cases  $\alpha=6^\circ$, $7^\circ$, and $8^\circ$, the discrepancies are less significant, with both DNS and WMLES leading to transition. However, in those cases, narrower streaks are observed as the grid resolution of the WMLES is increased, although these structural discrepancies do not translate into severe mismatches neither in the location of breakdown nor in the location of peak heating. The latter, indicated by red triangles in figure~\ref{DNS_WMLES_instant_all} for each case, serves as a an accurate indicator of full breakdown to turbulence in both DNS and WMLES.}


%

{Further insights into the performance of the baseline WMLES are gained in figure~\ref{DNS_WMLES_mean_TU_compare} by comparing its time- and spanwise-averaged profiles of velocity and temperature near transition and peak heating with those from DNS for the case $\alpha=7^\circ$. There, the shear layer, which separates the recirculating flow and the downwards high-speed inviscid stream at the foot of the incident shock, is much closer to the wall in the WMLES, since the reattachment occurs noticeably more upstream in WMLES than in DNS [see figure~\ref{DNS_777}(b) for spatial localization of the shear layer]. The corresponding  profiles of the rms of the fluctuations of temperature and streamwise velocity  are presented in figure~\ref{DNS_WMLES_RMS_TU_compare}. The maximum values of those quantities occur in the shear layer in both WMLES and DNS. The turbulent heat fluxes in the streamwise and wall-normal directions, along with the Reynolds stress, are also maximized by the turbulent transport in the shear layer, as shown in figure~\ref{DNS_WMLES_stress_compare}. In all these profiles it is observed that WMLES provides an acceptable prediction of features such as shear-layer thickness, maximum shear, maximum values of rms fluctuations of temperature and streamwise velocity, and maximum values of turbulent heat fluxes and Reynolds stress. However, the WMLES profiles are clearly shifted in space near the shock impingement owing to different separation and reattachment locations.}

{In summary, mismatches between the WMLES and DNS statistics in the transitional region are mostly dominated by the erroneous spatial advance in the reattachment predicted by WMLES. However, the root cause of this discrepancy cannot be straightforwardly isolated. This can be understood by noticing that the spatial advance in the reattachment in WMLES is coupled with the spatial delay predicted in separation, the latter engendering a separation shock at an erroneous angle. In addition, as the boundary layer approaches separation, its velocity profile becomes more contorted and critically more under-resolved by the WMLES grid. The separation shock in WMLES then  intersects the main shock, which is refracted towards the boundary layer at an erroneous angle. The resulting accumulation of errors is germane to the present configuration that involves widely different, interacting flow structures communicated by an adverse pressure gradient, and is not as severely observed in WMLES predictions of less complex configurations such as shock-free flat-plate turbulent boundary layers or turbulent channel flows. These considerations highlight the closely coupled contributions of all these phenomena in setting the overall performance of WMLES in the present problem.}

\begin{figure}
\begin{center}
\includegraphics[width=1\textwidth]{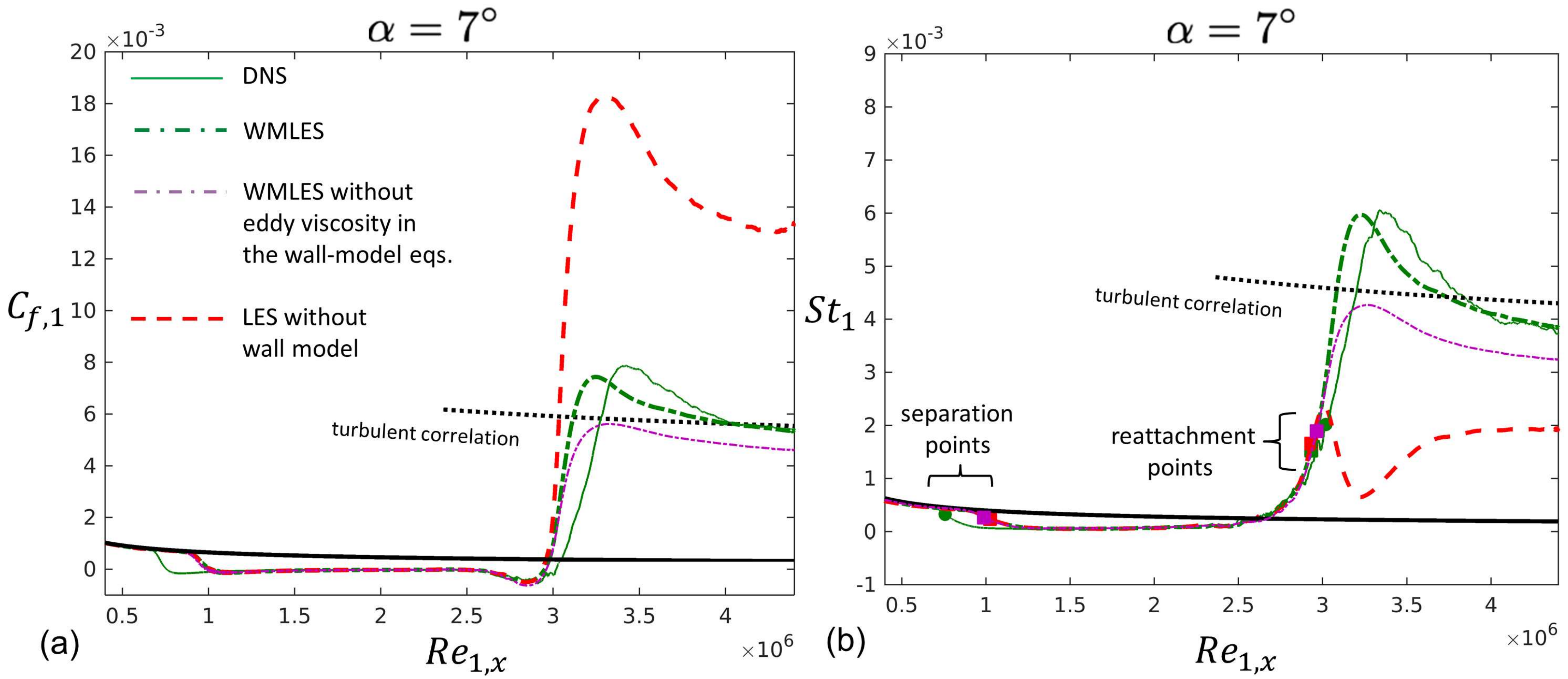}
\caption{{(a)~Skin friction coefficient and (b)~Stanton number as a function of the local Reynolds number $Re_{1,x}$ in DNS (solid lines), baseline WMLES (thick dot-dashed lines), baseline WMLES with eddy viscosity set to zero in the wall-model equations (thin dot-dashed lines), and no-slip LES performed on the baseline WMLES grid but without wall model (dashed lines). All results in this figure correspond to the case $\alpha=7^\circ$.} }
\label{DNS_WMLES_7_no_model}
\vskip 5mm
\end{center}
\end{figure}

{In order to isolate the role of the equilibrium wall model and its eddy viscosity in predicting the spatial distributions of the skin friction coefficient and Stanton number, two tests are performed in figure~\ref{DNS_WMLES_7_no_model} for the case $\alpha=7^\circ$ using the baseline WMLES grid. In the first test, the equilibrium wall model is replaced by a non-slip boundary condition. The resulting curves are denoted by the tag ``LES without wall model" in figure~\ref{DNS_WMLES_7_no_model}. Negligible changes are observed in the laminar portion upstream of the shock-impingement zone, which indicates that the wall model does not have any significant effect there. In contrast, in the post-interaction boundary layer, the  skin-friction coefficient is significantly over-predicted while the Stanton number is under-predicted by a factor of two. In conclusion, without the wall model, the predictions of the transitional and turbulent portions of the boundary layer are largely degraded due to deficient physical modeling and to numerical errors enabled by the coarse LES mesh.}

{In the second test, the equilibrium wall model is activated but its eddy viscosity $\mu_{t,wm}$, defined in equation \eqref{eq:mixinglength} in appendix~B, is turned off. The resulting curves are denoted by the tag ``WMLES without eddy viscosity in the wall-model equations" in figure~\ref{DNS_WMLES_7_no_model}. By turning off the eddy viscosity, the wall model provides only the viscous continuations of the velocity and temperature profiles within the wall-modeled region. In this case, the performance of the equilibrium wall model also deteriorates significantly. In particular, despite the fact that the average matching location $h_{wm}^+$ is within the damped spatial range of the wall-normal coordinate (i.e., see figure~\ref{yplusWMLES}), the results in figure~\ref{DNS_WMLES_7_no_model} suggest that $\mu_{t,wm}$ plays an important role in the prediction of $C_f$ and $St$ not only in turbulent boundary layer ensuing downstream of the interaction, as expected, but also in the transitional zone near the shock impingement. Note that these considerations do not imply neither that the eddy-viscosity model \eqref{eq:mixinglength} is the correct one to use in the transitional zone, nor that the eddy-viscosity hypothesis is appropriate for transitional flows, but that the eddy viscosity model \eqref{eq:mixinglength} appears to have a beneficial effect on the solution compared to setting $\mu_{t,wm}=0$. This effect is particularly relevant in transitional spots of large skin friction, where the local instantaneous values of $h_{wm}^+$ can be as high as four times the van Driest damping constant, as shown in figure~\ref{yplusWMLES}.}

\begin{table}
\centering
\vskip 0.1in
\begin{tabular}{cc|ccc|ccc|ccc|ccc|ccc}
wedge angle $\alpha$ [deg] & & & $Re_{2,\theta}$ & & & $Re_{2,\tau}$ & & & $Ma_2$  & & &  {$T_w/T_2$}\\
 \hline
 6 & & & $ 2,796 $ & & & $595$ & & & 4.5 & & &  {2.7} \\
  7 & & & $ 2,846$ & & & $682$ & & & 4.2 & & &  {2.5} \\
  8 & & & $  2,912 $ & & & $788$  & & & 4.0  & & &  {2.3}\\
  \hline
\end{tabular}
\caption{{Conditions corresponding to the turbulent boundary layer at $(x-x_1)/\delta_1^\star=580$ far away downstream of the recompression shock. The table includes the post-interaction Reynolds number based on the momentum thickness $Re_{2,\theta}$, the Reynolds number based on the wall-friction velocity $Re_{2,\tau}$, the post-interaction Mach number $Ma_2$, and the ratio of the wall temperature and the post-interaction free-stream temperature $T_w/T_2$.}} \label{tableturbulent}
\end{table}

\subsection{The turbulent boundary layer far downstream of the impingement by the shock} \label{Turbulence_statistics}

{While \S\ref{structures} and \S\ref{WMLES1} were focused on the transitional region, this section analyzes the turbulent boundary layer downstream of the interaction region for the cases $\alpha=6^\circ$, $7^\circ$, and $8^\circ$.  Table~\ref{tableturbulent} summarizes relevant parameters characterizing the state of the turbulent boundary layer at a representative streamwise location $(x-x_1)/\delta_1^\star=580$, on which most of the statistical analysis outlined below is based.}

{The physical characteristics of the ensuing turbulent boundary layer are determined by the post-interaction values of the free-stream Mach number $Ma_2$, the momentum-based Reynolds number $Re_{2,\theta}$ (or the friction Reynolds number $Re_{2,\tau}$), and the ratio of the wall temperature to the free-stream temperature $T_w/T_2$. In the three cases shown in table~\ref{tableturbulent}, $Ma_2$ is smaller than the inflow free-stream value $Ma_1=6$ because of the net deceleration of the free stream as it crosses the incident shock and the train of shocks and expansion fans induced by the impingement. Similarly, the net heating of the free stream across the interaction zone leads to temperature ratios $T_w/T_2$ smaller than the corresponding inflow value $T_w/T_1=4.5$. In the turbulent boundary layer, the overall consequences of increasing the wedge angle are a decrease in $Ma_2$, an increase in $Re_{2,\theta}$ (or $Re_{2,\tau}$), and a decrease in $T_w/T_2$.}


{As indicated by the third plateau of the wall pressure shown in figure~\ref{DNS_7_pressure}, the turbulent boundary layer far downstream of the impingement by the shock is one under negligible mean streamwise gradient of static pressure. The wall-normal gradient of the static pressure is similarly weak, since the wall pressure in that third plateau is well described by the free-stream static pressure calculated from an inviscid interaction, as expected from the moderate values of $Ma_2$. The remainder of this section is dedicated to assessments of analogies and hypotheses traditionally developed for zero-pressure-gradient compressible boundary layers, such as Reynolds analogies, mean-velocity transformations, and Morkovin's hypothesis.}

\begin{figure}
\begin{center}
\includegraphics[width=1\textwidth]{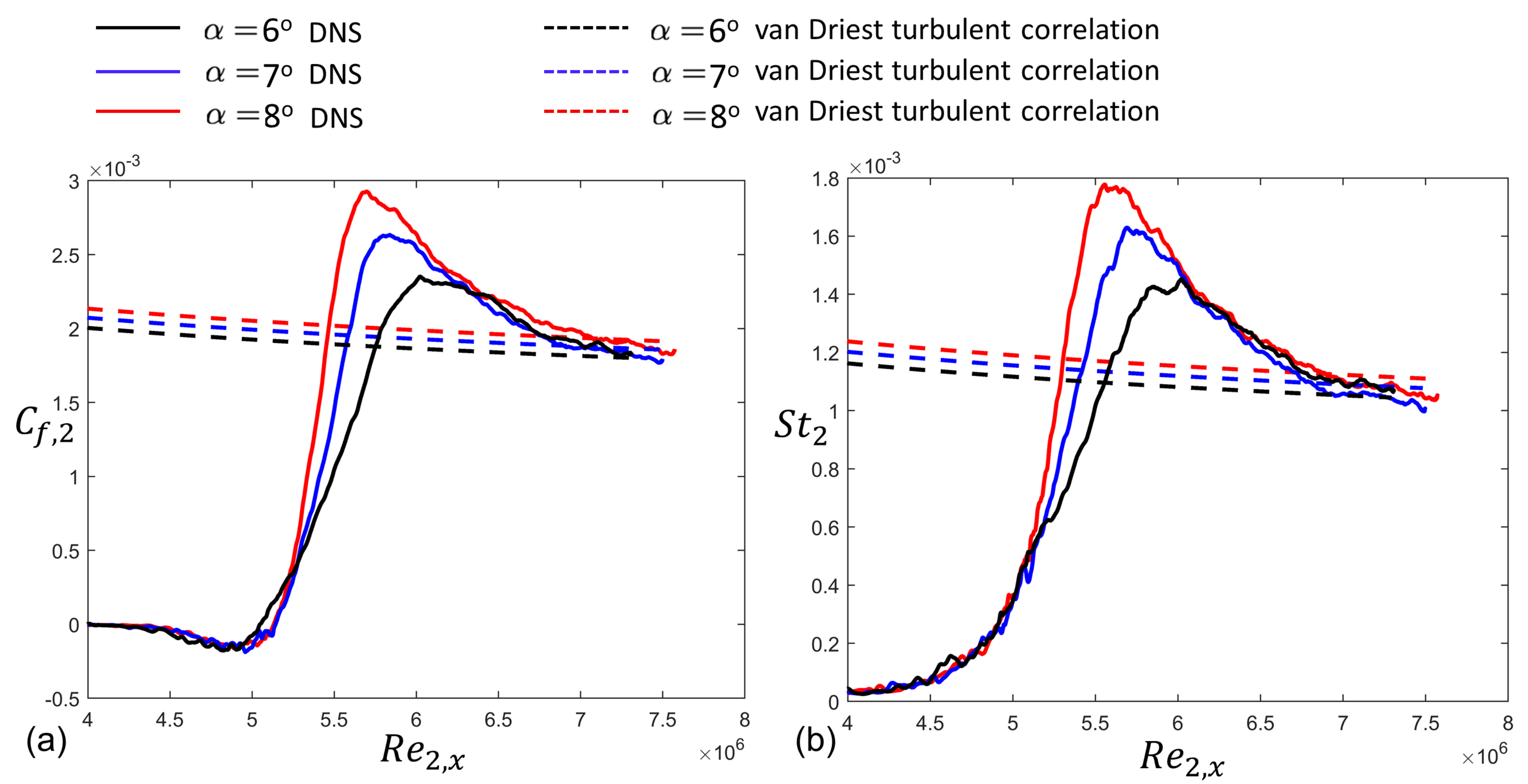}
 \caption{{(a) Skin friction coefficient and (b) Stanton number as a function of the post-interaction local Reynolds number $Re_{2,x} = U_2 x/\nu_2$ for $\alpha=6^\circ$ (black lines), $\alpha=7^\circ$ (blue lines), and $\alpha=8^\circ$ (red lines), including DNS (solid lines) and the van Driest turbulent correlations (dashed lines). In this figure, the van Driest turbulent correlation for the skin friction coefficient $C_{f,2}$ is calculated based on post-interaction free-stream conditions, with a virtual origin equated to the leading edge of the plate. Similarly, the van Driest turbulent correlation for the Stanton number $St_2$ is calculated from $C_{f,2}$ using the Reynolds analogy factor  $2St_2/C_{f,2}=1.16$ from \cite{chi1966influence}.}}
\label{DNS_678_converge}
\end{center}
\end{figure}

{The three different wedge angles considered in table~\ref{tableturbulent} unfold in dimensionless space as three different sets of values of $Re_{2,\theta}$, $Ma_2$, and $T_w/T_2$. Although the sensitivity of the solutions to the particular value of $Re_{2,\theta}$ is expected to be small at the relatively large values of $Re_{2,\theta}$ considered here, some of the statistics of the different metrics described below do not collapse among the three turbulent boundary layers because of additional dependencies of the solution on $Ma_2$ and $T_w/T_2$. Notable exceptions that remain relatively robust to changes of the wedge angle within the range tested here are the skin friction coefficient $C_{f,2}$ and the Stanton number $St_2$ downstream of the recompression shock when scaled with post-interaction free-stream values, as defined in equations \eqref{Cf2} and \eqref{St2}. This is elicited by the improved collapse of the turbulent portion of the profiles observed in figure~\ref{DNS_678_converge}, as opposed to the significant dispersion in figure~\ref{Cf_St_peaks} when pre-interaction free-stream conditions are employed instead. These considerations suggest that the effects of the variations of $\alpha$ on the wall shear stress and wall heat flux in the turbulent portion of the boundary layer can be approximately scaled out despite the different values of $Re_{2,\theta}$, $Ma_2$, and $T_w/T_2$ in each case.}




{In figure~\ref{DNS_678_converge}(b), the agreement between the van Driest turbulent correlation for $St_2$ and the DNS solution is greatly enhanced by using the Reynolds analogy factor $2St_2/C_{f,2}=1.16$ proposed by \cite{chi1966influence} based on correlation of experimental data for turbulent boundary layers with Mach numbers less than 5 and near-adiabatic wall boundary conditions. This is in contrast to the traditional Reynolds analogy factor $2St_2/C_{f,2}=Pr^{-2/3}=1.24$ utilized in figure~\ref{Cf_St_all_angles}(b) for boundary layers with non-unity Prandtl numbers, which leads to significant mismatch between the van Driest turbulent correlation for $St_2$ and the DNS solution. That the Reynolds analogy factor $2St_2/C_{f,2}=1.16$ proposed by \cite{chi1966influence} is a more accurate model of the DNS results presented here can be seen in table~\ref{tab:weak_reynolds_analogy}.}

\begin{table}
\centering
\vskip 0.1in
\begin{tabular}{cc|ccc|ccc|ccc}
  Wedge angle $\alpha$ [deg] & & & DNS & & & WMLES & & & WMLES coarse\\
 \hline
  6 & & &  1.157 & & &  1.170 & & &  1.146\\
  7 & & &  1.127 & & &  1.153 & & &  1.129\\
  8 & & &  1.125 & & &  1.192 & & &  1.152\\
 \hline
\end{tabular}
\caption{{The Reynolds analogy factor, $2St_2/C_{f,2}$, for DNS, baseline WMLES and coarse WMLES, calculated at station $(x-x_1)/\delta^\star_1= 580$ for various wedge angles.}}
\label{tab:weak_reynolds_analogy}
\end{table}

\begin{figure}
\begin{center}
\vskip 0.1in
\includegraphics[width=0.7\textwidth]{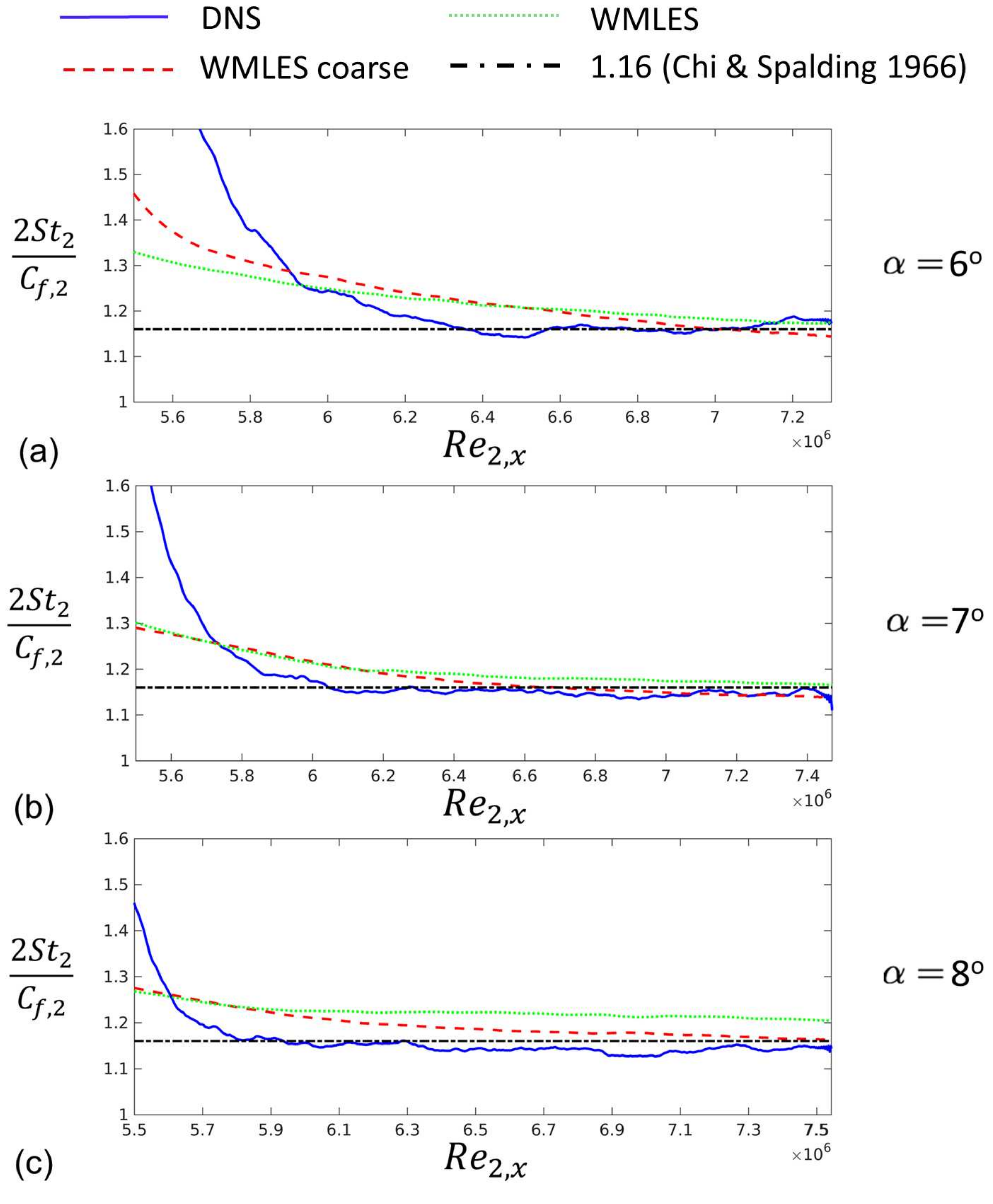}
\caption{{DNS (blue solid lines), baseline WMLES (green dotted lines), and coarse WMLES (red dashed lines) distributions of the Reynolds analogy factor as a function of the post-interaction local Reynolds number $Re_{2,x} = U_2 x/\nu_2$ for (a)~$\alpha=6^\circ$, (b)~$\alpha=7^\circ$, and (c)~$\alpha=8^\circ$, along with the reference value 1.16 (black dot-dashed line) proposed by \cite{chi1966influence}.}}
\label{Reynolds_analogy}
\end{center}
\end{figure}

{Both DNS and WMLES results settle increasingly earlier on a value of the Reynolds analogy factor as the wedge angle increases, as observed in figure~\ref{Reynolds_analogy}, because the boundary layer transitions correspondingly earlier along the streamwise coordinate. The differences between the Reynolds analogy factors predicted by WMLES and DNS, and between them and the value 1.16 experimentally correlated by \cite{chi1966influence}, are small and remain within a 5\% error for all the conditions tested here. However, as the wedge angle increases, the DNS results predict a slight decrease in the mean value of the Reynolds analogy factor, whereas the trend of the WMLES results is less clear. As observed in previous experimental studies collected by \cite{Cary} and discussed in \cite{Bradshaw}, increasing the wall cooling leads to a slight decrease in the Reynolds analogy factor below that proposed by \cite{chi1966influence}. Although the behavior of the DNS results observed in table~\ref{tab:weak_reynolds_analogy} as the wedge angle is increased is reminiscent of an increase in wall cooling, it should be mentioned that the ratio $T_w/T_{0}$, with $T_{0}$ being the stagnation temperature, is mostly the same in both DNS and WMLES within a 0.1\% error, and is independent of the wedge angle, since the wall temperature is fixed and the stagnation temperature of the free stream is constant across the interaction region. As a result, the decrease in the Reynolds analogy factor $2St_2/C_{f,2}$ observed as the wedge angle increases in the DNS cannot be easily reconciliated with the observations made by \cite{Cary}, and may instead  be attributed to the intrinsic dependency of the solution on the parameters $Re_{2,\theta}$, $Ma_{2}$, and $T_w/T_2$ listed in table~\ref{tableturbulent}, which differ slightly among the three cases.}

\begin{figure}
\begin{center}
\vskip 0.1in
\includegraphics[width=0.99\textwidth]{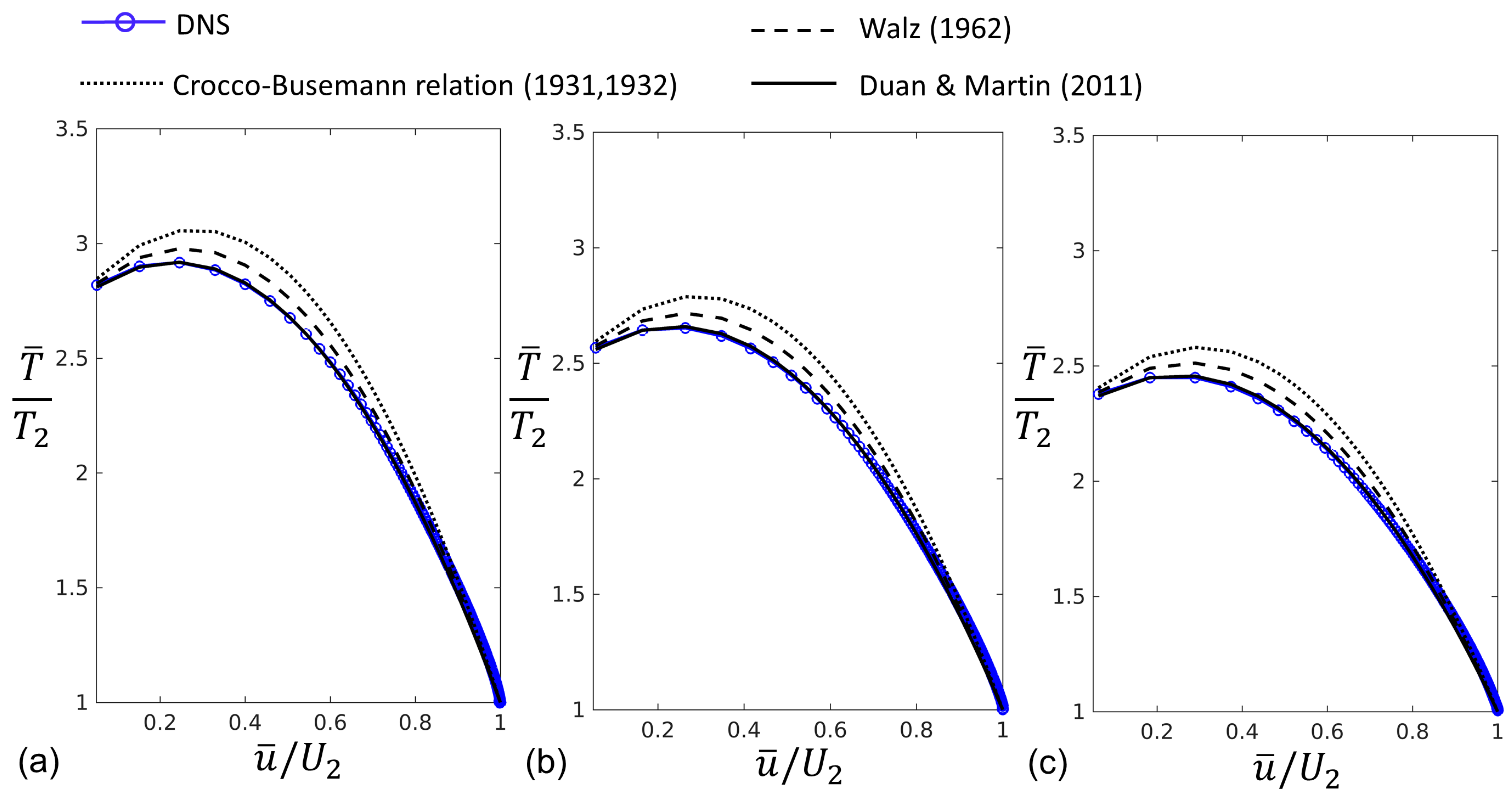}
\caption{{DNS relation between the time- and spanwise-averaged profiles of temperature and velocity at $(x-x_1)/\delta^\star_1=580$ for wedge angle of (a)~$\alpha=6^\circ$, (b)~$\alpha=7^\circ$, and (c)~$\alpha=8^\circ$. Also included are the model relations from \cite{Walz1962}, \cite{Busemann1931}, \cite{crocco1932sulla}, and \cite{duan2011direct}. The values used for normalization are the edge velocity $U_2$ and edge temperature $T_2$ for the corresponding angle.}}
\label{Mean_TU_relation_DNS}
\end{center}
\end{figure}

%
%

{For all three cases, figure~\ref{Mean_TU_relation_DNS} indicates that the present DNS results best match with the temperature-velocity relation proposed by \cite{duan2011direct}, which is nonetheless based on correlation of DNS data of a different configuration involving temporally-evolving turbulent boundary layers. In contrast, the Crocco-Busemann formula for $Pr=1$ \citep{Busemann1931,crocco1932sulla}, and the Walz relation that accounts for $Pr\neq 1$ \citep{Walz1962,Walz1966}, depart from the DNS data by amounts of order 10\% and 5\%, similarly to previous observations by \cite{zhang2014generalized} and \cite{duan2010direct}.}

{In figure~\ref{Mean_TU_relation_DNS}, the model for the temperature-velocity relation proposed by \cite{duan2011direct} requires a calibration parameter $\theta=0.8259$, which was connected later through analysis by \cite{zhang2014generalized} with the Reynolds analogy factor multiplied by the Prandtl number, namely $\theta=2St_2Pr/C_{f,2}$. Evaluation of the latter using the DNS results in table~\ref{tab:weak_reynolds_analogy} indicates that $2St_2Pr/C_{f,2}$ differs from the model parameter $\theta=0.8259$ by small amounts of order $0.9\%$ (for $\alpha=6^\circ$), $1.7\%$ (for $\alpha=7^\circ$), and $1.9\%$ (for $\alpha=8^\circ$), thereby corroborating the analysis made by \cite{zhang2014generalized}.}

\begin{figure}
\begin{center}
\includegraphics[width=0.6\textwidth]{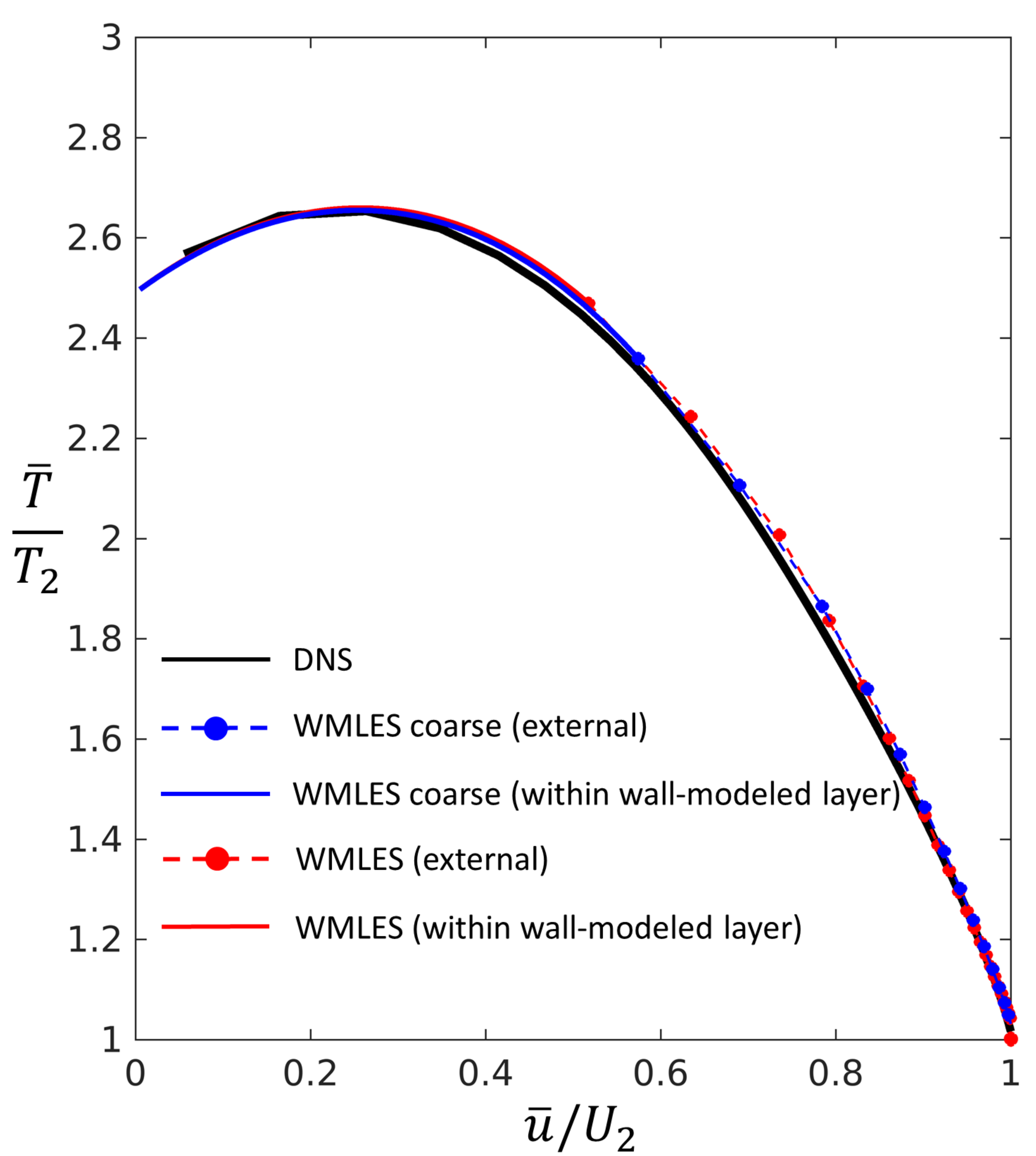}
\caption{{DNS (black solid line), baseline WMLES (red dot-dashed line), and coarse WMLES (blue dot-dashed line) relations between the time- and spanwise-averaged profiles of temperature and velocity at  $(x-x_1)/\delta^\star_1=580$ for $\alpha=7^\circ$, including the solution predicted within the wall-modeled region $y\leq h_{wm}$ (blue and red solid lines).}}
\label{Mean_TU_relation}
\end{center}
\end{figure}

{The comparisons between the mean temperature-velocity relations from DNS and WMLES presented in figure~\ref{Mean_TU_relation} for the case $\alpha=7^\circ$ show an encouraging agreement over the entire range of velocities. This is despite the fact that a significant portion of the momentum of the turbulent boundary layer is unresolved by the LES grid. However, the equilibrium wall model correctly captures the DNS mean temperature-velocity relations within the wall-modeled region even in the coarser WMLES case.}

%
%

\begin{figure}
\begin{center}
\includegraphics[width=1\textwidth]{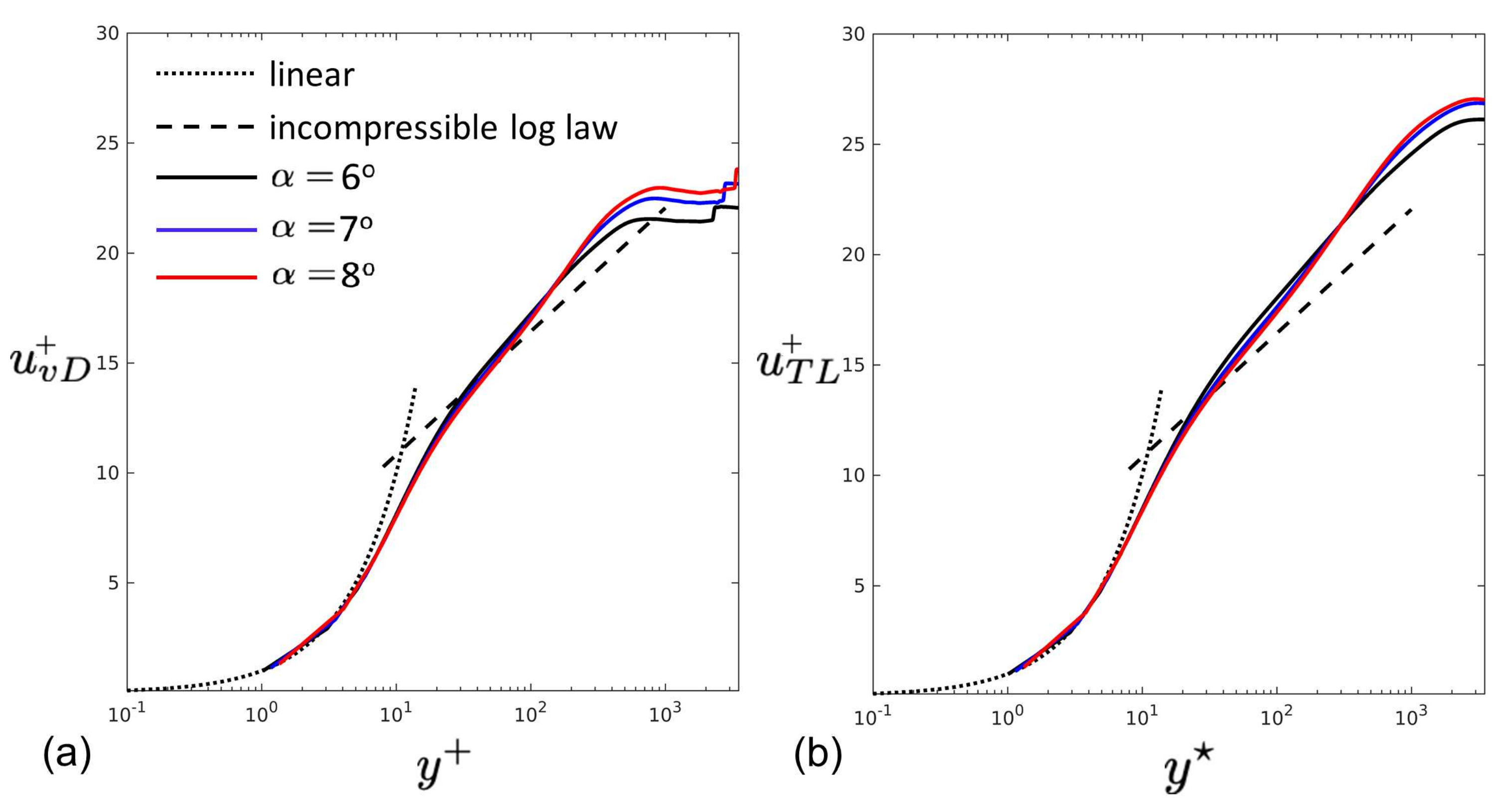}
 \caption{{Transformed mean streamwise velocity as a function of the wall-normal coordinate for $\alpha=6^\circ$, $7^\circ$, and $8^\circ$ at $(x-x_1)/\delta^\star_1=580$ using the transforms by (a)~\cite{vandriest} and (b)~\cite{trettel2016mean}. The figure includes DNS results for $\alpha=6^\circ$ (black solid lines), $7^\circ$ (blue solid lines), and $8^\circ$ (red solid lines), along with the incompressible profiles in the viscous sublayer (black dotted line) and log layer (dashed lines). In panel~(b), $y^\star  = \overline {\rho} (\overline {\tau_w}/\overline {\rho})^{1/2}y/\overline \mu$ represents a semi-locally scaled wall-normal coordinate.}}
\label{DNS_velocity_transformation_all_angles}
\end{center}
\end{figure}

{
Some understanding of the structure of the mean streamwise velocity profile can be gained by transforming it in such a way as to resemble as much as possible the mean velocity profile of an incompressible turbulent boundary layer. This is the objective of the velocity transformations shown in figure~\ref{DNS_velocity_transformation_all_angles}, which includes those proposed by \cite{vandriest} and \cite{trettel2016mean}, the latter being a revision of the former to account for both viscosity and density variations in boundary layers over non-adiabatic walls. Both transformations reveal the presence of viscous- and log-like layers in the transformed velocity profiles. A lack of collapse among the transformed mean velocity profiles corresponding to the three different wedge angles in figure  \ref{DNS_velocity_transformation_all_angles} is clearly noticeable in the outer layer, where the sensitivity of the wake parameter to changes in the post-interaction Mach numbers and heating rates appears to be significant.}

{Neither one of the two transformations employed in figure  \ref{DNS_velocity_transformation_all_angles} lead to collapse of the log-layer mean velocity profile on the incompressible log law. Specifically, figure~\ref{DNS_velocity_transformation_all_angles} indicates that, for the three angles tested here, the effective K\'{a}rm\'{a}n constant of the transformed mean velocity profile is smaller than the nominal K\'{a}rm\'{a}n constant $0.42$ of the incompressible log law.}


%
%

\begin{figure}
\begin{center}
\includegraphics[width=0.99\textwidth]{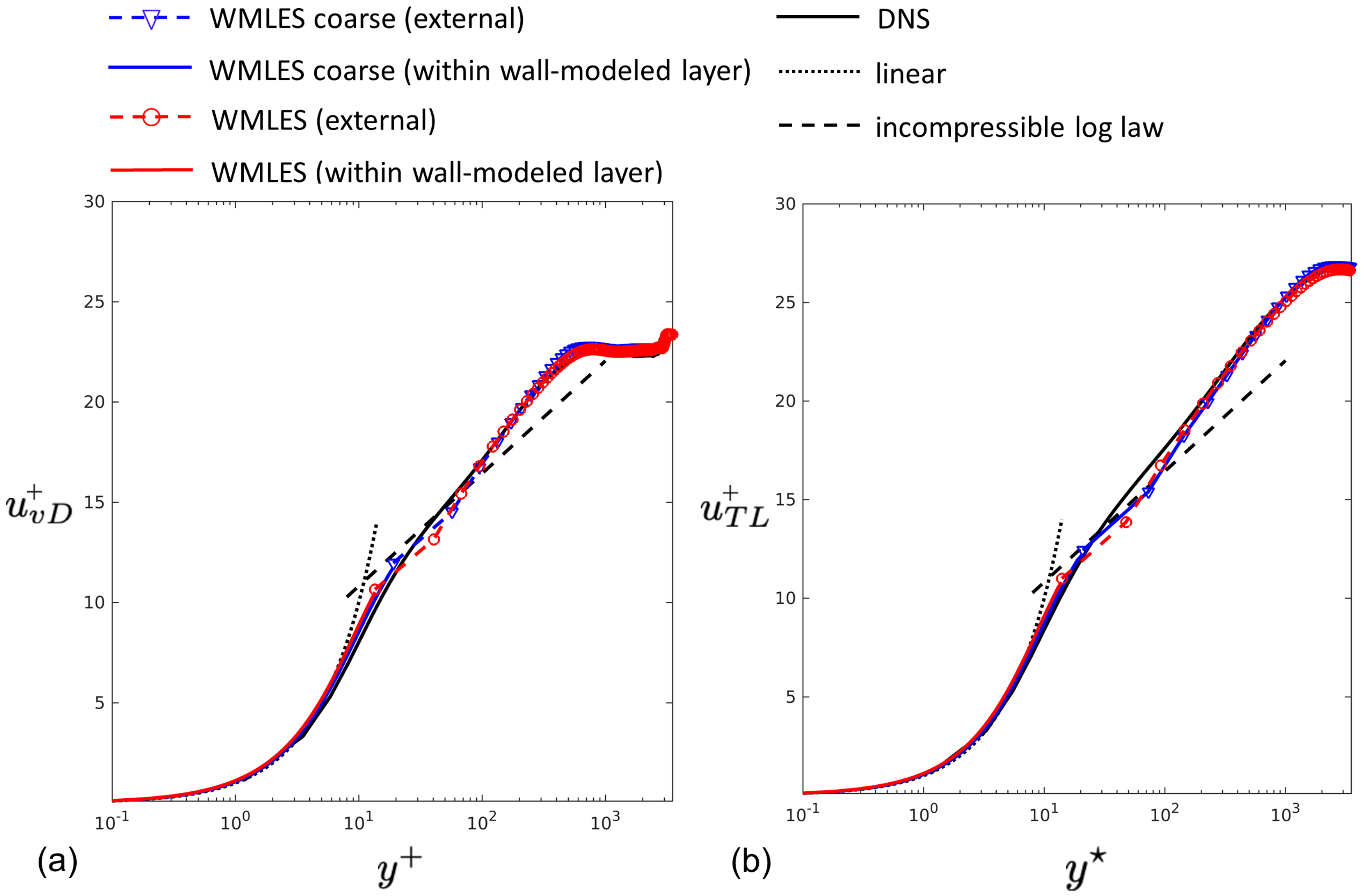}
 \caption{{Transformed mean streamwise velocity as a function of the wall-normal coordinate for $\alpha=7^\circ$ at $(x-x_1)/\delta^\star_1=580$ using the transforms by (a)~\cite{vandriest} and (b)~\cite{trettel2016mean}. The figure includes DNS (black solid lines), baseline WMLES (red dashed and solid lines), coarse WMLES (blue dashed and solid lines), along with the incompressible profiles in the viscous sublayer (black dotted line) and log layer (dashed lines). In panel~(b), $y^\star  = \overline {\rho} (\overline {\tau_w}/\overline {\rho})^{1/2}y/\overline \mu$ represents a semi-locally scaled wall-normal coordinate.}}
\label{WMLES_DNS_velocity_transformation}
\end{center}
\end{figure}

{The transformed mean velocity profiles obtained by the WMLES agree well with those of the DNS for the most part, as suggested by the comparisons provided in figure~\ref{WMLES_DNS_velocity_transformation} for $\alpha=7^\circ$. However, discrepancies are observed in  the first and second grid points of the LES grid, where the WMLES are expected to be influenced by numerical errors. The two other angles $\alpha=6^\circ$ and $8^\circ$ lead to similar conclusions and are not included here for brevity. The differences caused by coarsening the resolution of WMLES grid in the wall-normal direction are small and do not degrade the agreement between DNS and WMLES in any signficant way.}


%

\begin{figure}
\begin{center}
\includegraphics[width=0.65\textwidth]{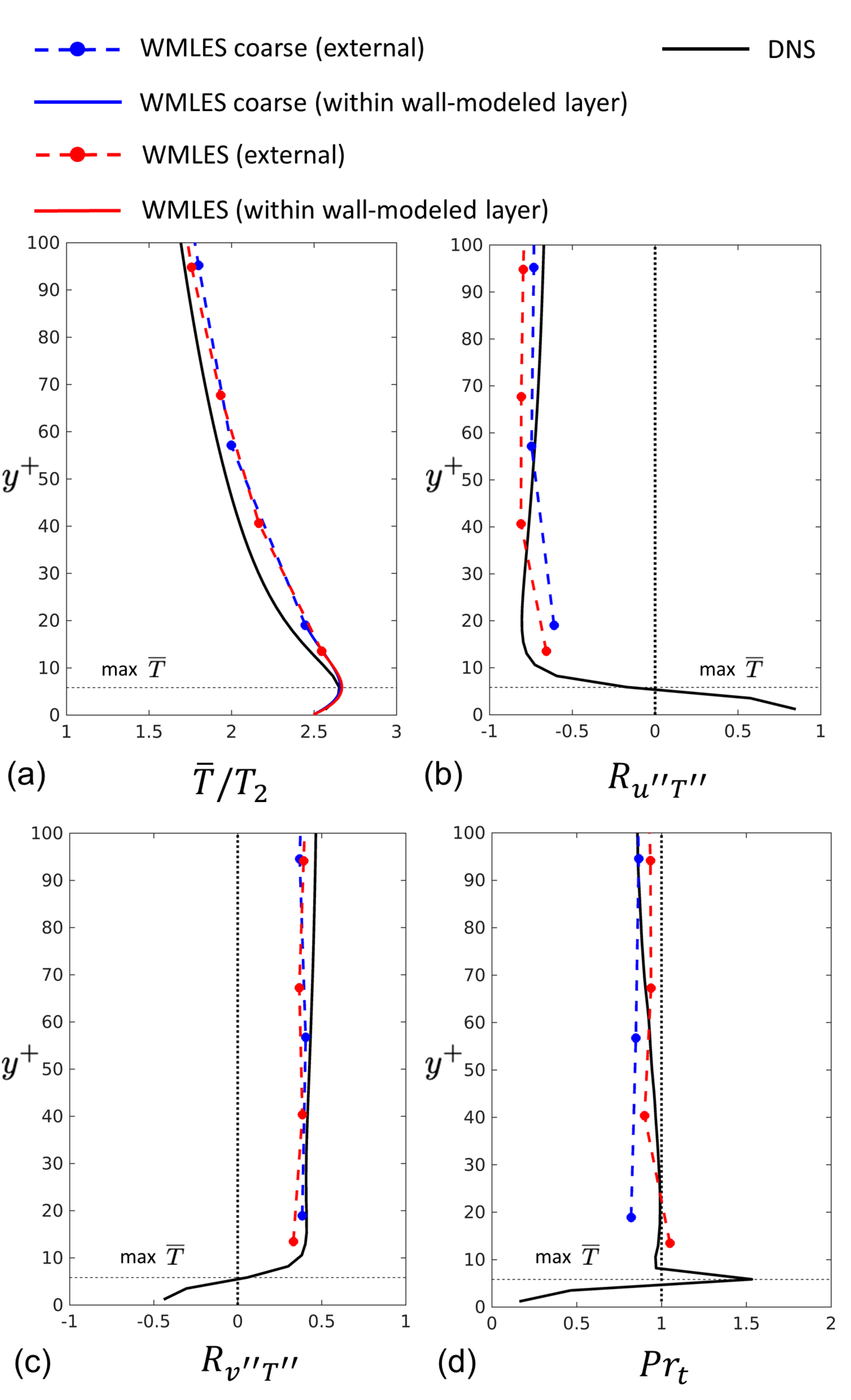}
\caption{{DNS (black solid lines), baseline WMLES (red dashed lines), and coarse WMLES (blue dashed lines)   wall-normal profiles of (a) time- and spanwise-averaged temperature, (b)~temperature/streamwise-velocity correlation coefficient, (c)~temperature/wall-normal-velocity correlation coefficient, and (d)~turbulent Prandtl number, all profiles being obtained for $\alpha=7^\circ$ at $(x-x_1)/\delta_1^\star=580$. In panel~(a), the mean temperature predicted by the equilibrium wall model within the wall-modeled region $y\leq h_{wm}$ is indicated by the blue and red solid lines. In all panels, the horizontal dashed lines indicate the wall-normal location of the maximum mean temperature.}}
\label{Prandtl_number_correlation}
\end{center}
\end{figure}

{The good agreement shown in figure~\ref{DNS_WMLES_St_Cf_compare} between the Stanton numbers predicted by DNS and WMLES in the turbulent boundary layer downstream of the recompression shock for the transitioning cases $\alpha=6^\circ$, $7^\circ$, and $8^\circ$ must rely on the correct WMLES prediction of the mean temperature profile near the wall. This is corroborated by the comparison between the mean temperature profiles obtained from DNS and WMLES provided in figure~\ref{Prandtl_number_correlation}(a) for $7^\circ$. Although discrepancies of order 10\% are observed between the DNS and WMLES mean temperature profiles at wall-normal distances $y^+$ corresponding to the log and outer layers of the transformed mean velocity profile, the wall model captures  correctly the mean temperature profile in the buffer zone and in the viscous sublayer. There, the temperature reaches its maximum value because of the heat generated by friction. This maximum value is not directly resolved by the LES grid but modeled successfully by the equilibrium wall model, thereby yielding a correct approximation of the magnitude and sign of the wall heat flux. The comparisons of the mean temperatures pertaining to the two other angles $\alpha=6^\circ$ and $8^\circ$ lead to similar conclusions and are not included here for brevity.

The Morkovin hypothesis appears to provide unsatisfactory results in the present configuration. First, whereas the Morkovin hypothesis establishes perfect anticorrelation between $T''$ and $u''$ \citep{Morkovin1962}, both DNS and WMLES unisonally indicate that $T''$ and $u''$ in the present configuration are not fully anticorrelated, as shown in figure~\ref{Prandtl_number_correlation}(b) for $\alpha=7^\circ$. Away from the wall, this non-perfect anticorrelation is explained by the approximately $10\%$ fluctuations observed in the stagnation temperature across the boundary layer. Additionally, as anticipated in figures~\ref{perturb_U_380} and \ref{perturb_U_445}, the sign of the temperature/streamwise velocity correlation changes near the wall at the wall-normal location where the maximum of the mean temperature is attained. A similar change of sign in the temperature/wall-normal velocity correlation is also observed in figure~\ref{Prandtl_number_correlation}(c) for $\alpha=7^\circ$ at the same location. The WMLES results provide excellent predictions for the correlations of the temperature with the streamwise and wall-normal velocities across the entirety of the resolved portion of the turbulent boundary layer.

The turbulent Prandtl number $Pr_t$ shown in figure~\ref{Prandtl_number_correlation}(d) for $\alpha=7^\circ$ varies between 0.7 and 1.0 across the boundary layer in both DNS and WMLES, whereas a peak value of 1.5 appears to be attained at the maximum temperature location. Although not shown here for brevity, similar conclusions about the temperature/velocity correlations and the turbulent Prandtl number also hold for the other two wedge angles $\alpha=6^{\circ}$ and $8^\circ$.}

\begin{figure}
\begin{center}
\includegraphics[width=0.8\textwidth]{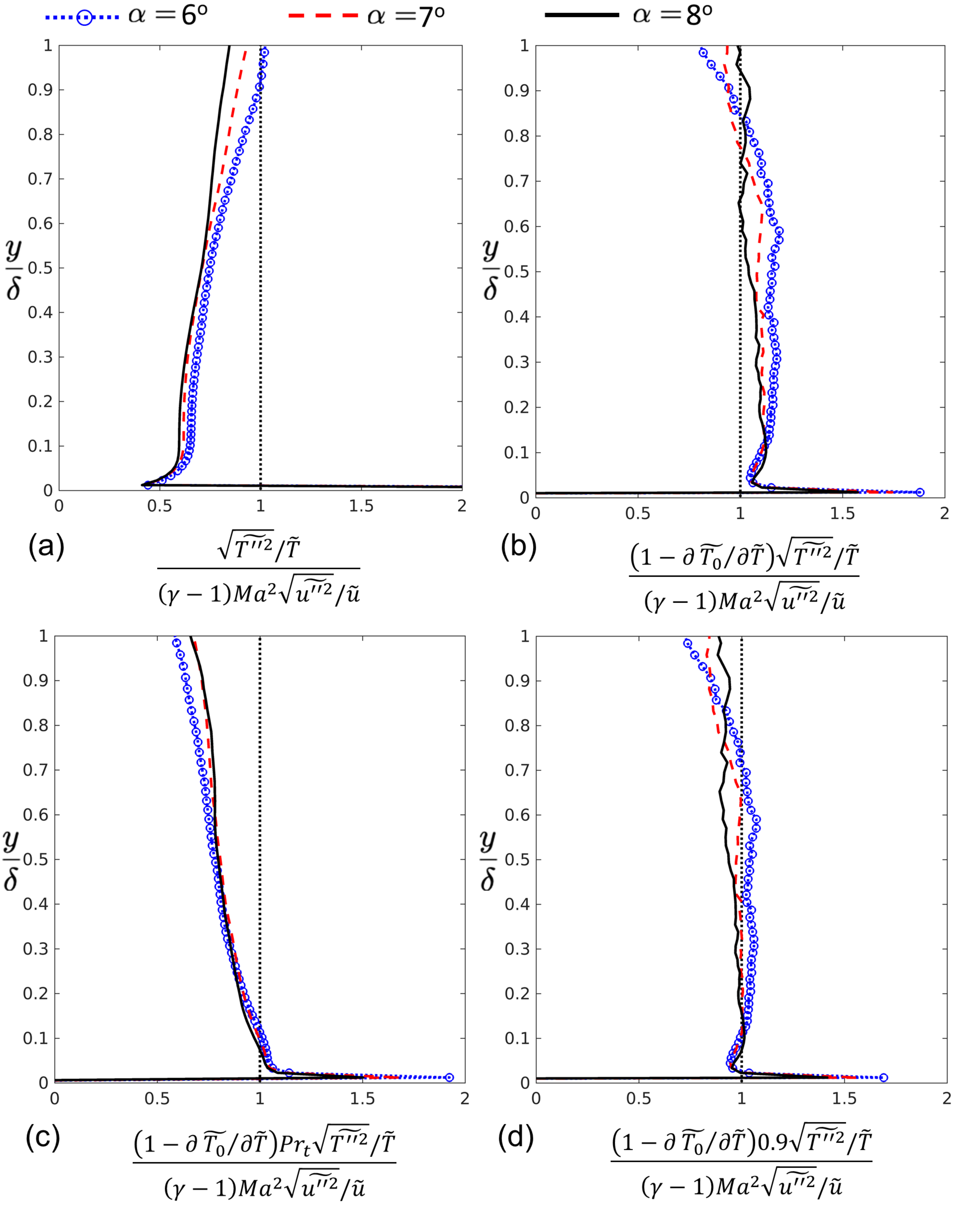}\\
\caption{{DNS wall-normal distributions of the (a)~strong Reynolds analogy by \cite{Morkovin1962}, (b)~modified strong Reynolds analogy GSRA by \cite{gaviglio1987reynolds}, (c)~modified strong Reynolds analogy HSRA by \cite{huang1995compressible} based on the local turbulent Prandtl number, and (d)~modified strong Reynolds analogy HSRA by \cite{huang1995compressible} based on a constant turbulent Prandtl number equal to $0.9$. The blue, red, and black lines denote, respectively, the results for $\alpha=6^\circ$, $7^\circ$, and $8^\circ$ at $(x-x_1)/\delta^\star_1=580$. The vertical dotted lines indicate unity ratios, and correspondingly, total validity of the proposed analogy.In the labels, $Ma = [\widetilde{\rho} (\widetilde{u}^2+\widetilde{v}^2)/(\gamma \widetilde{P})]^{1/2}$ denotes the local Mach number, and $\widetilde{T_{0}}$ indicates the Favre average of the local stagnation temperature $T_{0} = T[1+(\gamma-1)(\rho|\mathbf{v}|^2/(2\gamma P)]$ based on the modulus of the streamwise velocity vector $|\mathbf{v}|$. Additionally, $\delta$ is the local boundary layer thickness at $(x-x_1)/\delta^\star_1=580$ defined as the height where $\overline{u}=0.99U_2$.}}
\label{DNS_strong_Reynolds_analogy}
\end{center}
\end{figure}

\begin{figure}
\begin{center}
\includegraphics[width=\textwidth]{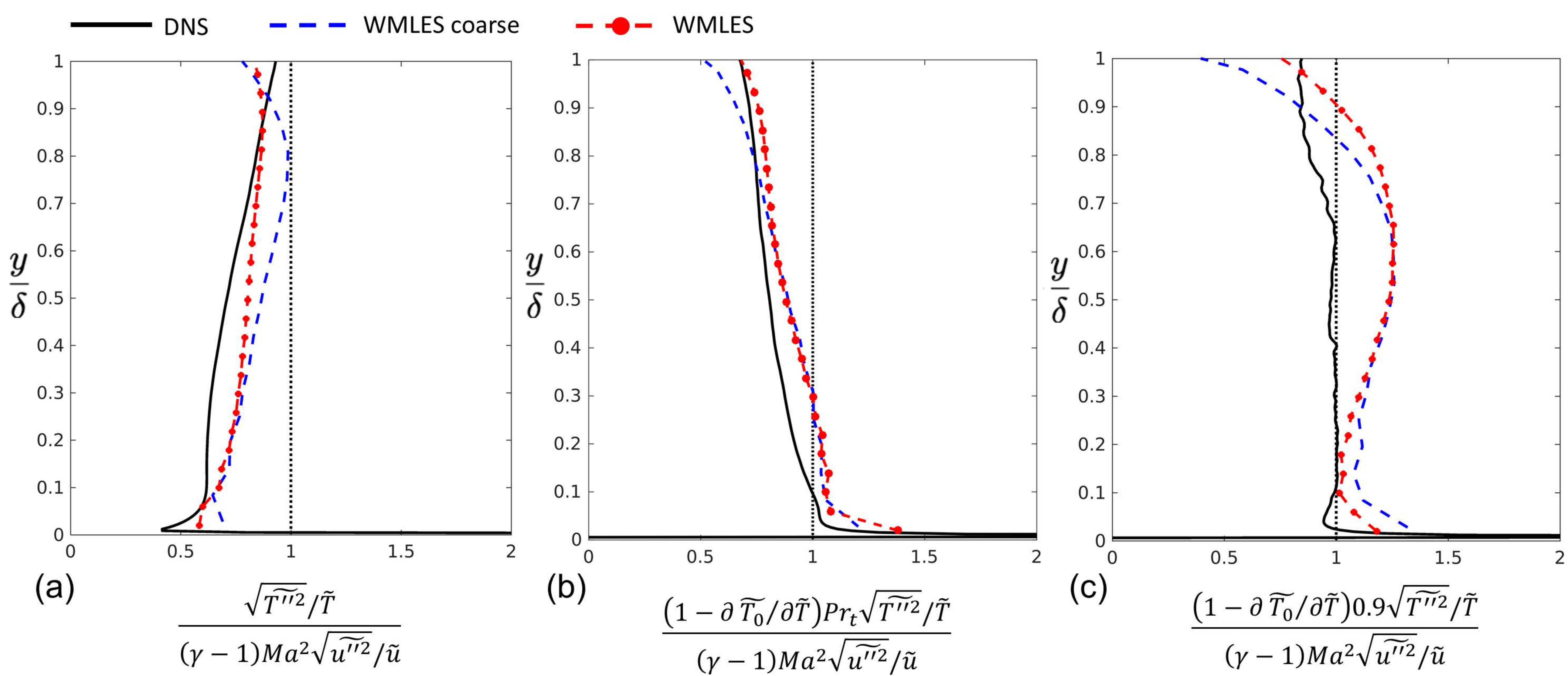}
\caption{{DNS (black solid lines), baseline WMLES (red dot-dashed lines), and coarse WMLES (blue dashed lines) wall-normal distributions of the (a)~strong Reynolds analogy by \cite{Morkovin1962}, (b)~modified strong Reynolds analogy HSRA by \cite{huang1995compressible} based on the local turbulent Prandtl number, and (c)~modified strong Reynolds analogy HSRA by \cite{huang1995compressible} based on a constant turbulent Prandtl number equal to $0.9$. All profiles correspond to the case $\alpha=7^\circ$ at $(x-x_1)/\delta^\star_1=580$. The vertical dotted lines indicate unity ratios, and correspondingly, total validity of the analogy. In the labels,  $Ma = [\widetilde{\rho} (\widetilde{u}^2+\widetilde{v}^2)/(\gamma \widetilde{P})]^{1/2}$ denotes the local Mach number, and $\widetilde{T_{0}}$ indicates the Favre average of the local stagnation temperature $T_{0} = T[1+(\gamma-1)(\rho|\mathbf{v}|^2/(2\gamma P)]$ based on the modulus of the streamwise velocity vector $|\mathbf{v}|$. Additionally, $\delta$ is the local boundary layer thickness at $(x-x_1)/\delta^\star_1=580$ defined as the height where $\overline{u}=0.99U_2$.}}
\label{strong_Reynolds_analogy}
\end{center}
\end{figure}

{That the strong Reynolds analogy (SRA), proposed by \cite{Morkovin1962} to relate in a directly proportional way the rms values of the streamwise velocity fluctuations and the temperature fluctuations, is not a good approximation in the present configuration is shown in figure~\ref{DNS_strong_Reynolds_analogy}(a), where departures of about 50\% from SRA behavior are observed.  Over the years, the SRA has been improved in different studies that account for wall heat transfer and stagnation-temperature fluctuations. For instance, \cite{gaviglio1987reynolds} proposed a revised SRA (referred to as GSRA below) by assuming that the characteristic length scales of the fluctuations of temperature and velocity are similar. In a different approach, \cite{huang1995compressible} proposed another revised SRA (referred to as HSRA below) by including the local turbulent Prandtl number on the basis of a mixing-length model. Using the DNS flow fields for $\alpha=6^\circ$, $7^\circ$, and $8^\circ$, figure~\ref{DNS_strong_Reynolds_analogy}(b,c) provides an evaluation of the GSRA and HSRA expressions found in \cite{gaviglio1987reynolds} and  \cite{huang1995compressible}, respectively, in such a way that the total validity of the corresponding relation would imply a unity value on the vertical axis across the entire boundary layer. While the classical SRA in figure~\ref{DNS_strong_Reynolds_analogy}(a) fails to reproduce the DNS data, as also observed previously by \cite{duan2010direct} and \cite{zhang2014generalized}, the discrepancies are greatly reduced for all wedge angles by using the GSRA. However, the performance of the HSRA can be greatly enhanced by setting the Prandtl number in the HSRA relation to $Pr_t=0.9$, as shown in figure~\ref{DNS_strong_Reynolds_analogy}(d).}

{The predictive capabilities of the WMLES to recreate the SRA relations is assessed in figure~\ref{strong_Reynolds_analogy} for the representative wedge angle of $7^\circ$. Good agreement between DNS and WMLES is observed for the SRA and HSRA in figure~\ref{strong_Reynolds_analogy}(a,b). However, the WMLES deviates significantly from the DNS in the outer portion of the boundary layer when the HSRA is used with $Pr_t=0.9$, as shown in figure~\ref{strong_Reynolds_analogy}(c). These errors are commensurate with the errors incurred by the WMLES in predicting the turbulent Prandtl number calculated a-posteriori from the DNS results.}

\begin{figure}
\begin{center}
\includegraphics[width=\textwidth]{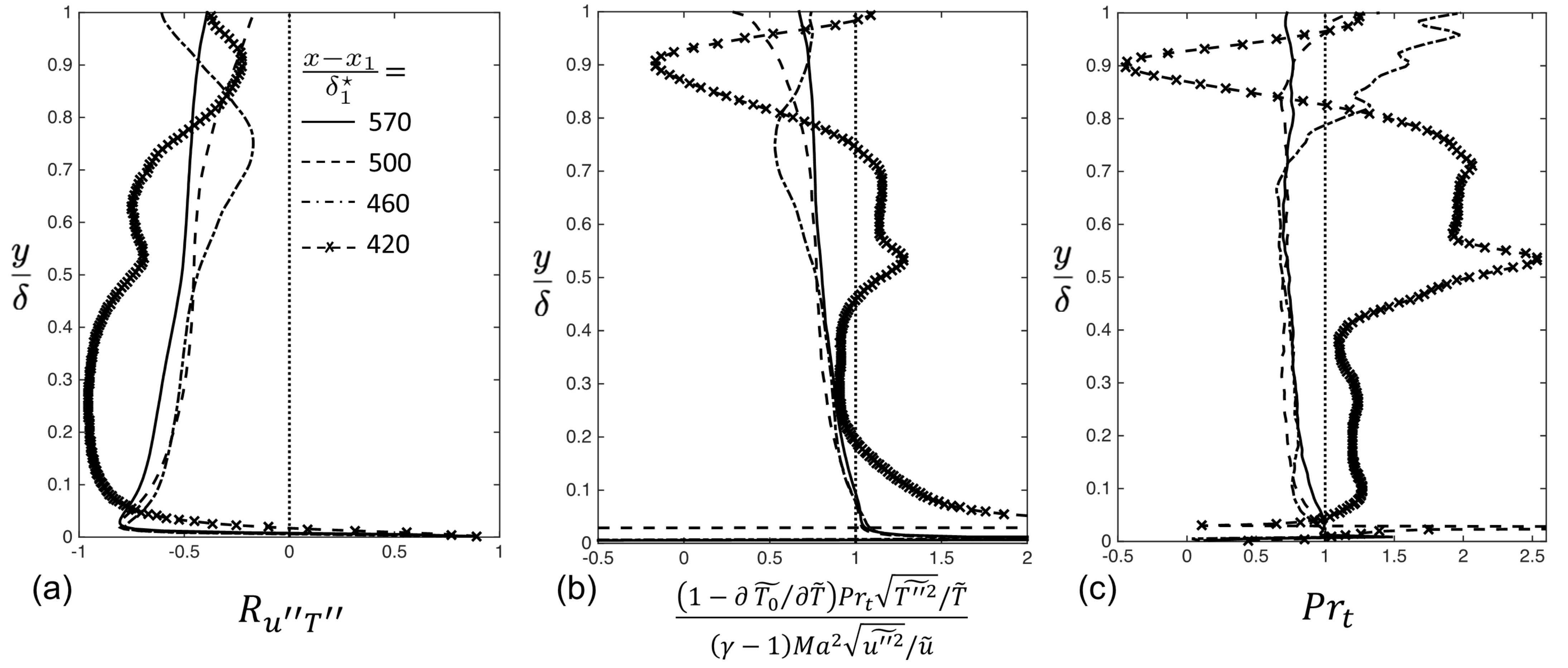}
\caption{{DNS wall-normal profiles at several streamwise stations for $\alpha=7^\circ$, including (a)~temperature/streamwise-velocity correlation coefficient, (b)~modified strong Reynolds analogy HSRA by \cite{huang1995compressible} based on the local turbulent Prandtl number, and (c)~turbulent Prandtl number. In the labels,  $Ma = [\widetilde{\rho} (\widetilde{u}^2+\widetilde{v}^2)/(\gamma \widetilde{P})]^{1/2}$ denotes the local Mach number, and $\widetilde{T_{0}}$ indicates the Favre average of the local stagnation temperature $T_{0} = T[1+(\gamma-1)(\rho|\mathbf{v}|^2/(2\gamma P)]$ based on the modulus of the streamwise velocity vector $|\mathbf{v}|$. Additionally, $\delta$ is the boundary layer thickness defined at $(x-x_1)/\delta^\star_1=580$ as the height where $\overline{u}=0.99U_2$.}}
\label{Prandtl_correlation_upstreams}
\end{center}
\end{figure}
\begin{figure}
\begin{center}
\includegraphics[width=0.6\textwidth]{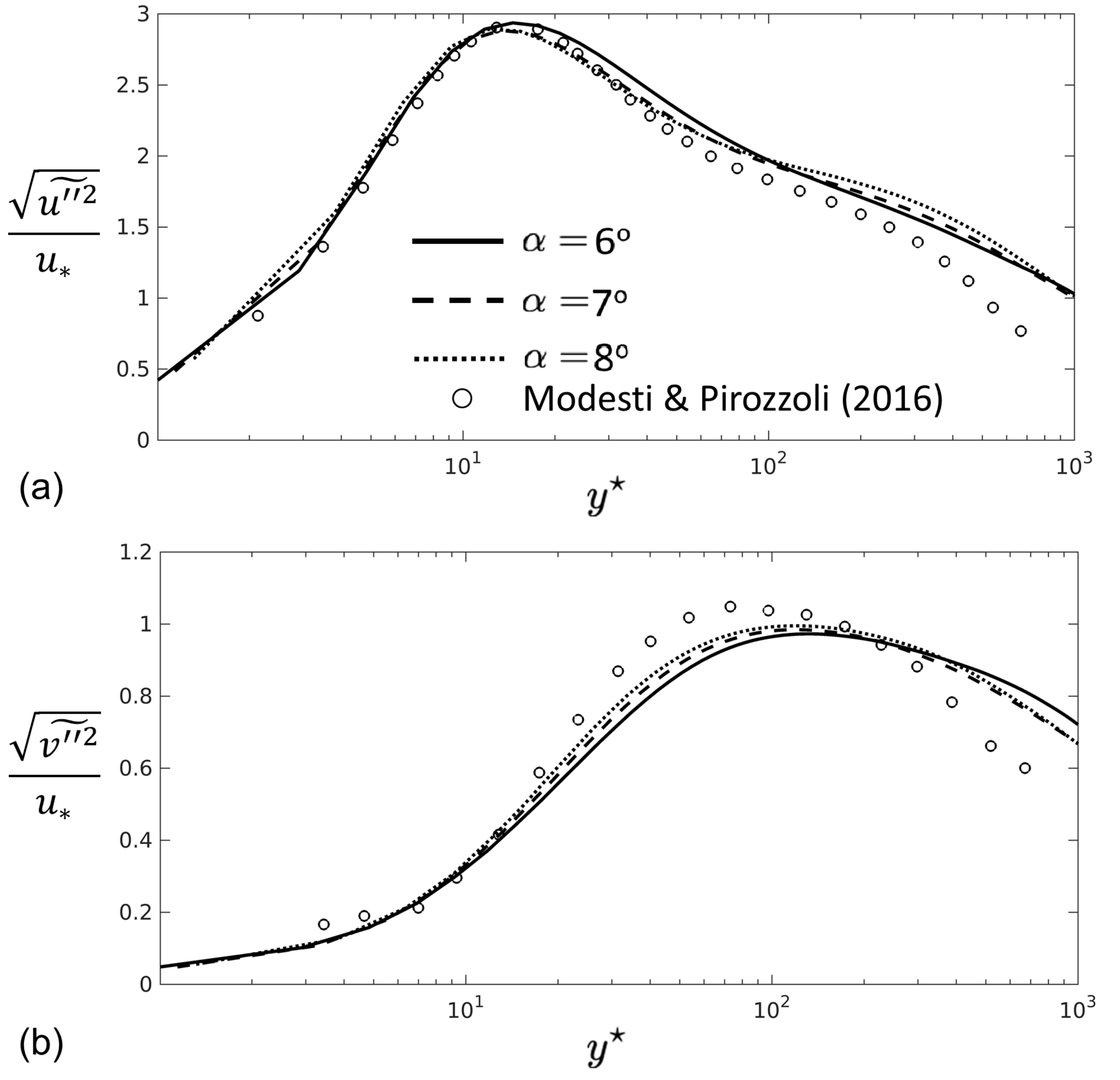}
\caption{{(a) Streamwise and (b) Wall-normal turbulence intensities at the streamwise location $(x-x_1)/{\delta_1^\star}= 580$ for $\alpha= 6^\circ$, $7^\circ$, and $8^\circ$, defined in terms of the density-weighted velocity scale $u_*=\sqrt{\overline{\tau_w}/\overline{\rho}}$, and plotted as a function of the semi-locally scaled wall-normal coordinate $y^\star = \overline {\rho} (\overline {\tau_w}/\overline {\rho})^{1/2}y/\overline \mu$. Also included is the reference data of channel case CH15C from \cite{modesti2016reynolds} corresponding to $Re_{2,\tau} = 1,015$ and $Ma = 1.5$.}}
\label{DNS_turbulence_intensities}
\end{center}
\end{figure}

{A comparison between profiles from DNS at several streamwise stations, $(x-x_1)/\delta^\star_1=420$, $460$, and $500$ in the transitional region, and $570$ in the fully turbulent region, is presented in figure~\ref{Prandtl_correlation_upstreams} for the representative wedge angle of $7^\circ$. The correlation between the streamwise velocity fluctuation $u''$ and the temperature fluctuation $T''$ is presented in figure~\ref{Prandtl_correlation_upstreams}(a), the HSRA based on the local turbulent Prandtl number is shown in figure~\ref{Prandtl_correlation_upstreams}(b), and the turbulent Prandtl number is given in figure~\ref{Prandtl_correlation_upstreams}(c). While significant variations of these metrics are observed upstream deep in the transitional region, as expected for HSRA and $Pr_t$ because of their lack of clear physical meaning there, the variations among different streamwise stations tend to decrease as the turbulent portion of the boundary layer is approached, or equivalently, as the local Reynolds number $Re_{2,x}$ increases.}


%

%

{The streamwise and wall-normal rms velocities, normalized with semi-local inner scalings, are presented in figure~\ref{DNS_turbulence_intensities} for $\alpha=6^{\circ}$, $7^{\circ}$, and $8^\circ$. Similarly to the transformed mean velocities in figure~\ref{DNS_velocity_transformation_all_angles},  a collapse of the curves corresponding to the three wedge angles is observed except in the outer layer, where the changes in $Ma_{2}$, $Re_{2,\theta}$, and $T_w/T_2$ across the three cases may have an appreciable effect. Additional results from supersonic channel flow simulations by \cite{modesti2016reynolds} at lower Mach numbers overlaid on figure~\ref{DNS_velocity_transformation_all_angles} corroborate the common observation that the streamwise and wall-normal rms velocities close to the wall do not depend significantly on the Mach number when scaled with appropriate inner units.}

\begin{figure}
\begin{center}
\vskip 0.1in
\includegraphics[width=\textwidth]{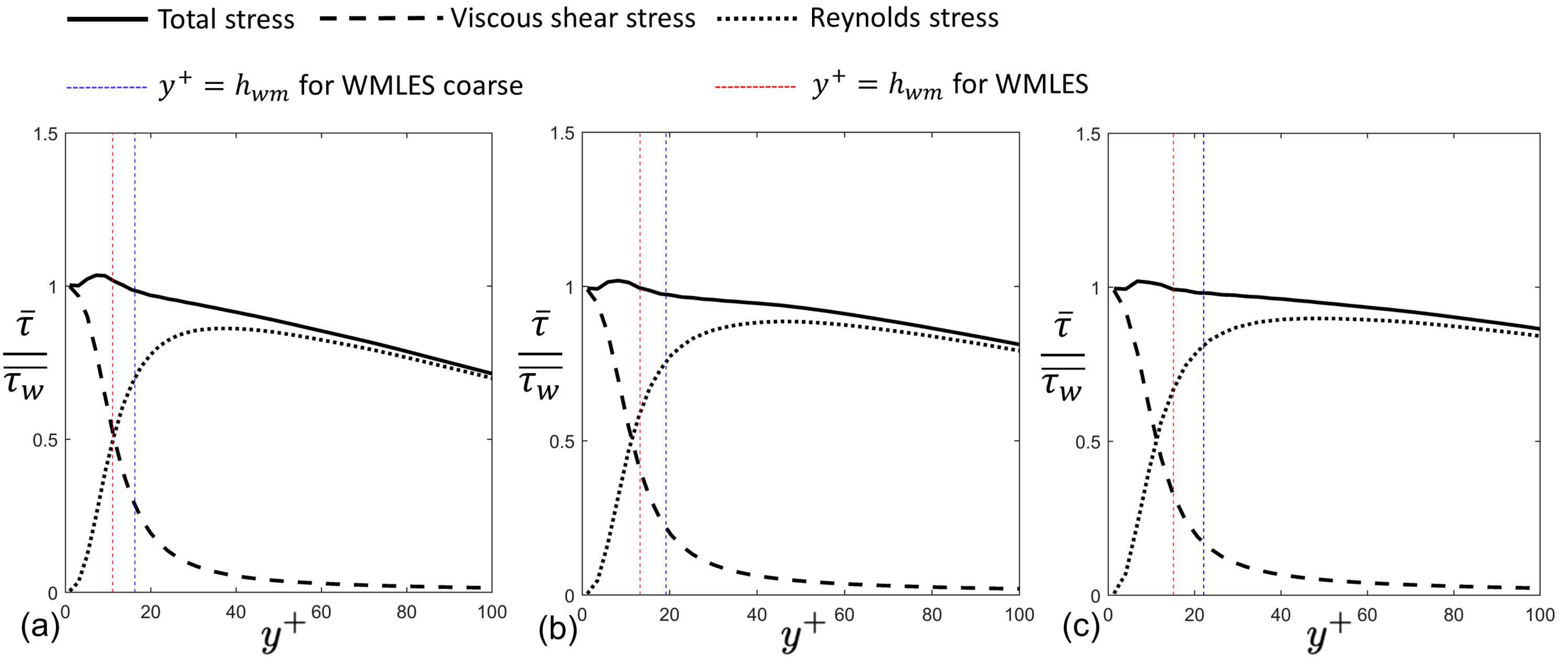}
\caption{{DNS wall-normal profiles of time- and spanwise-averaged shear stress $\overline{\tau}$ normalized with its value on the wall $\overline{\tau_w}$ for wedge angles $\alpha=6^\circ$, $7^\circ$, and $8^\circ$ at  $(x-x_1)/\delta^\star_1=580$ for wedge angles $\alpha=6^\circ$, $7^\circ$ and $8^\circ$. The dashed, dotted and solid lines denote the distributions of the viscous shear stress $\mu (\widetilde{T}) (\partial \widetilde{u}/\partial y)$, the turbulent shear stress $-\overline{\rho u''v''}$, and the total stress $-\overline{\rho u''v''} + \mu (\widetilde{T}) (\partial \widetilde{u}/\partial y)$. The vertical lines denote the matching location in the baseline WMLES (red lines) and in the coarse WMLES (blue lines).}}
\label{DNS_stress}
\end{center}
\end{figure}

{Despite the reasonable agreements between WMLES and DNS outlined throughout this section, figure~\ref{DNS_stress} shows that the core assumption of constant-stress-layer, represented by the momentum equation \eqref{mom} of the equilibrium wall model, is not strictly satisfied within the wall-modeled layer for any of the three angles $\alpha=6^{\circ}$, $7^{\circ}$, and $8^{\circ}$. Specifically, the evaluation of the mean shear stress $\overline{\tau}$ provided in figure~\ref{DNS_stress} using the DNS shows that the total stress $\overline{\tau}$ varies by amounts of order 10\% across the wall-modeled layer. These variations increase as the wedge angle decreases, or equivalently, as the friction Reynolds number decreases. In addition, as shown in figure~\ref{DNS_stress}, the ratio of the total and wall shear stresses, $\overline{\tau}/\overline{\tau_w}$, is not bounded by unity when the stresses are defined consistently with the Favre-averaged streamwise momentum equation. Instead, it features a maximum within the wall-modeled layer that was also observed in early computational work at lower Mach numbers by \cite{gatski2002numerical}.}


%
\section{Conclusions}\label{conc}
In this study, DNS and WMLES are employed to investigate the problem of an oblique shock wave impinging on a Mach-6 undisturbed laminar boundary layer over a cold wall that has a temperature of 55\% of the free-stream stagnation temperature. The incident shock leads to boundary-layer separation far upstream of the shock-impingement region. If the angle $\alpha$ of the wedge used to generate the incident shock is sufficiently large, and more particularly, if $\alpha\geq 6^{\circ}$ in the present DNS, the incident shock causes boundary-layer transition via breakdown of near-wall streaks shortly downstream of the impingement zone even in the absence of inflow free-stream disturbances. The transition causes a localized significant increase in the Stanton number and skin-friction coefficient. Increasing incidence angles lead to earlier transition, longer separation bubble, and higher peak values of wall heat transfer and wall shear stress. Specifically, the peak thermomechanical loads increase approximately linearly with the wedge angle.

In the DNS, transition and peak heating occur downstream of the shock on the leeward side of the separation bubble, where stationary streaks are visible in the Stanton number contours that give rise to broadband turbulence downstream upon reattachment of the overriding shear layer to the wall. The turbulent boundary layer ensuing downstream from the interaction has a Mach number within the range $4.0$ to $4.5$ depending on the wedge angle. Conventional transformations fail to collapse the mean velocity profiles on the incompressible log law. The Reynolds analogy factor of the turbulent boundary layer is close to the value 1.16 proposed by \cite{chi1966influence}. The Morkovin's hypothesis of perfect anticorrelation between velocity and temperature breaks down profusely near the wall in the viscous sublayer, below the wall-normal coordinate $y^{+}\sim 4-5$ corresponding to the maximum temperature, where the correlation becomes positive. A modified strong Reynolds analogy based on that proposed by \cite{huang1995compressible}, but with a calibrated turbulent Prandtl number of 0.9, becomes the most appropriate relation between the rms fluctuations of velocity and temperature.

The DNS data is used as benchmark to test predictions from WMLES. In particular, an equilibrium wall model is employed along the entire plate (including the laminar zone) to partially model the effects of near-wall turbulence. For all considered wedge angles, WMLES prompts transition and peak heating, delays separation, and advances reattachment, thereby shortening the separation bubble. The WMLES results depart strongly from DNS at the lowest wedge angle tested here, {which is below the threshold $\alpha\geq 6^{\circ}$ mentioned above.} In this case, DNS does not show transition, whereas WMLES predicts a spurious transition driven by numerical errors that remain unchallenged because of the absence of competing physical disturbances, since no inflow perturbations are employed in any of the cases analyzed in this study. In contrast, at higher angles, the effects of the shock on the boundary layer, including the absolute instability that is triggered  in the separation bubble, are sufficiently strong to override the numerical errors, and WMLES predicts transition in reasonable agreement with DNS. Specifically, WMLES correctly captures the advancement of the transition front along with the increase of the peak thermal load as the wedge angle increases. In the transitioning cases, WMLES provides predictions of peak skin friction and Stanton number within $\pm 10\%$ error with respect to DNS, but at a significantly reduced computational cost by a factor of approximately 150.

In the turbulent boundary layer ensuing downstream of the shock impingement, WMLES reproduces a number of key DNS statistics, including the Reynolds analogy factor, the outer portion of the temperature-velocity correlation profile, the mean velocity-temperature relation, and the profiles of mean velocity and temperature. Furthermore, the WMLES results reproduce the value and location of the maximum temperature resulting from viscous heating, which is concealed in the wall-modeled layer. These considerations remain mostly unaltered after coarsening the WMLES grid by a factor of 1.4 in the wall-normal direction.

Although it is traditionally asserted that WMLES is inadequate for transitional flows, numerical experiments performed in this work show that turning off the wall model everywhere leads to a severe degradation of the WMLES predictions with regards to transition and peak thermomechanical loads in the interaction region. Similarly, turning off the eddy viscosity model in the momentum and energy conservation equations of the wall model has a significant negative impact not only in the turbulent boundary layer, as expected, but also in the transitional zone where spots rendering large skin friction develop, thereby suggesting that the eddy viscosity in the wall model has a beneficial effect on the predictions of transition. It should be stressed that the transitional aspects of the flow considered in this study depart considerably from shock-free, unmolested boundary layers on flat plates that take long distances for eigenmodes to grow from inflow disturbances and trigger transition. In those, the WMLES grid, and the wall model itself, cannot faithfully support the spatiotemporal dynamics associated with the long growth of the disturbances. In contrast, in the present configuration, transition occurs rather compactly in space due to the sudden flow distortion caused by the shock, and does not necessitate any long spatiotemporal development of disturbances along the laminar portion of the boundary layer. As a result, the WMLES grid only needs to warrant a reasonable resolution of the steady two-dimensional laminar boundary layer upstream of the interaction with 4 to 5 grid points across the wall-normal dimension. These considerations suggest that WMLES may perform comparatively better in this type of problems than in shock-free transitional boundary layers. {Additionally, high Mach number flows necessarily entail hot boundary layers. Correspondingly, the matching location in WMLES can be easily set near the buffer zone or within the viscous sublayer in the turbulent boundary layer ensuing downstream of the shock, while still leading to a drastic reduction in computational cost relative to DNS.}

\section*{Acknowledgements}

This work was funded by the US Air Force Office of Scientific Research (AFOSR), Grant \#~FA9550-16-1-0319. Supercomputing resources were provided by the US Department of Energy through the INCITE Program.  The authors are grateful to Dr.~Jeffrey O'Brien and Dr.~Christopher Ivey for useful technical discussions on this subject.

\section*{Declaration of Interests}
The authors report no conflict of interest.

\appendix

\begin{figure}
\begin{center}
\includegraphics[width=0.95\textwidth]{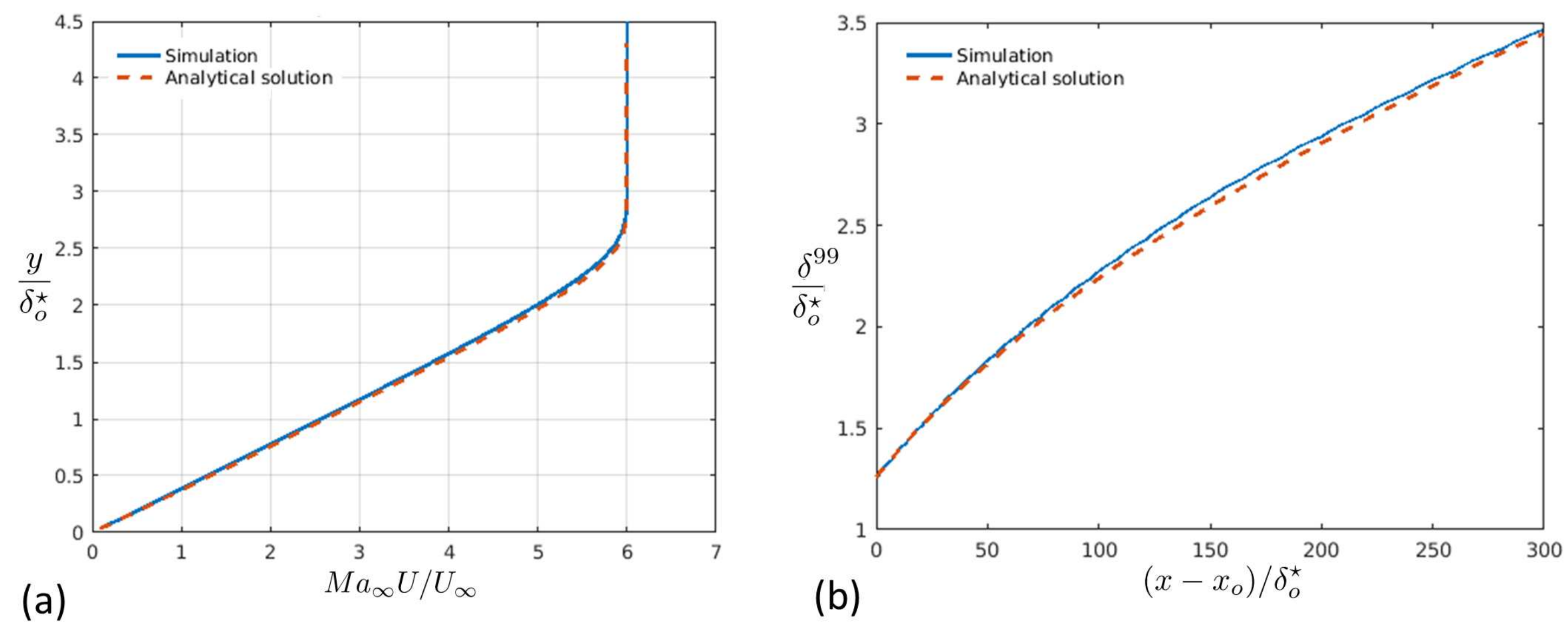}
\caption{{Comparisons between 2D numerical simulations of a Mach-6 hypersonic laminar boundary layer and the similarity solution for (a) the streamwise velocity evaluated at $(x-x_o)/{\delta^ \star_o}=150$, and (b) the streamwise evolution of the $99\%$ boundary-layer thickness.}}
\vskip -0.1in
\label{Laminar_boundary_layer}
\end{center}
\end{figure}

{
\section*{Appendix A. Code validation and verification}\label{sect:code_validation}

This appendix presents examples employed to verify and validate the charLES code in the context of hypersonic flows. The results shown below pertain to hypersonic laminar boundary layers and channels, along with the hypersonic flow around the BOLT subscale vehicle geometry.

\subsection*{A1. Mach-6 laminar boundary layer}\label{sect:code_validation_laminar_boundary_layer}

Figure~\ref{Laminar_boundary_layer} shows comparisons between the similarity solution for a compressible boundary layer at a free-stream Mach number $Ma_{\infty}=6$ and inflow Reynolds number $Re_{\delta_o^\star}=6830$ and results obtained using the present code in two-dimensional (2D) numerical simulations. The computational domain is $300\delta_{o}^{\star} \times 25\delta_{o}^{\star}$ in the streamwise and wall-normal directions respectively, which corresponds to $1500 \times 150$ cells. The results show that the code reproduces reasonably well the similarity solution for the streamwise velocity component $U$ and the $99\%$ boundary-layer thickness $\delta^{99}$.}


\begin{figure}
\begin{center}
\vskip 0.1in
\includegraphics[width=0.9\textwidth]{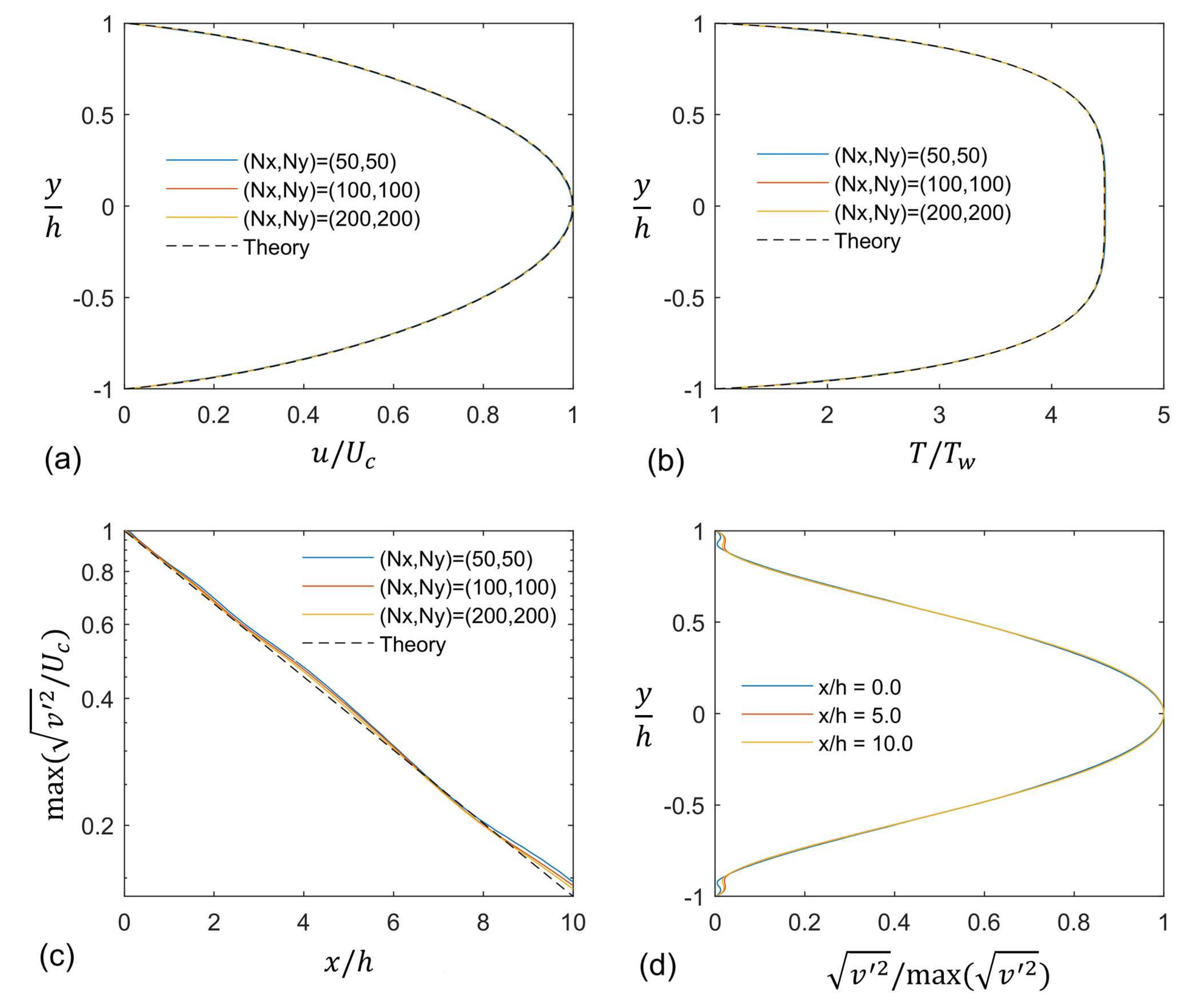}
\caption{{Results of the solver verification for hypersonic laminar channels: (a) base flow streamwise velocity at $x/h=10$; (b) base flow temperature at $x/h=10$; (c) maximum magnitude of the vertical velocity at $y/h=0$ normalized with its inflow value; and (d) vertical velocity profiles normalized by their maximum values at each streamwise station.}}
\label{channel_stability}
\end{center}
\end{figure}

{
\subsection*{A2. Mach-6 laminar channel flow stability}\label{sect:code_validation_laminar_channel_flows}

A verification exercise is performed in this section using a fully developed laminar channel with isothermal walls, for which an analytical solution exists. All quantities are normalized with the following reference scales: the half-channel height $h$, the centerline streamwise velocity $U_c$, and the wall temperature $T_w$. The centerline Mach number is $Ma_{\infty} = 6$, whereas the Reynolds number is $Re_h = 1000$. The verification is conducted by injecting an eigenfunction, obtained from a spatial stability analysis, at the inlet of the computational domain and comparing the resulting spatial decay rate with the prediction of linear stability theory. The disturbance frequency is $\omega h/U_c = 0.5$, for which the spatial wavenumber of the eigenfunction is $\alpha h = 0.9877 + 0.1998i$. The domain size is $(L_x/h, L_y/h) = (10, 2)$. Three grid resolutions are studied: $(N_x, N_y) = (50, 50)$, $(N_x, N_y) = (100, 100)$ and $(N_x, N_y) = (200, 200)$. The chosen disturbance amplitude is small enough to ensure that the nonlinear terms remain inactive.

The base-flow profiles for velocity and temperature at the station $x/h = 10$ are shown in figure~\ref{channel_stability}(a,b). Colors indicate the 2D numerical solution obtained on different grids, whereas the dashed lines correspond to the analytical solution. The streamwise evolution of the maximum value of the magnitude of the vertical component of the perturbation velocity,  normalized by its value at the inflow, is shown in figure~\ref{channel_stability}(c), showing good agreement with the prediction from linear stability theory. The profiles of the magnitude of the vertical component of the the perturbation velocity at streamwise stations, $x/h = 0$, $5$ and $10$, are shown in figure~\ref{channel_stability}(d) confirming the invariance of the eigenfunction shape with downstream distance consistent with linear stability theory. }

{

\subsection*{A3. Mach-6 hypersonic flow over BOLT}\label{sect:code_validation_BOLT}

Comparisons between experiments and the 3D numerical solution provided by charLES for the Mach-6 hypersonic flow over the BOLT subscale vehicle geometry are outlined in this section. This case is thoroughly described by \cite{wheaton2018boundary} and \cite{thome2019boundary}, and therefore the details are omitted here. Briefly, the temperature, velocity, density, and Mach number in the free stream are $T_{\infty}=52$~K, $U_{\infty}=864$ m/s, $\rho_{\infty}=3.8\cdot 10^{-2}$ kg/m$^3$, and $Ma_\infty = 6$, respectively, whereas the wall temperature is $T_w = 300$ K and the unit Reynolds number is $Re_\infty =9.9 \times 10^6$~m$^{-1}$.



A $1/3$-scale model of the BOLT vehicle considered here is meshed with an unstructured grid consisting of $518$M Voronoi elements. The grid is stretched with a stretching ratio of 40 near the wall and it becomes gradually isotropic away from the wall. In the vicinity of the nose, the ratio of the nose radius to the minimum grid spacing in the wall tangent direction is 32, indicating sufficient resolution to resolve the locally large curvature of the vehicle edges.  The simulations are compared with experiments performed in the Boeing-AFOSR Mach-6 Quiet Tunnel (BAM6QT) at Purdue University, which is known to have a very low level of free-stream disturbances \citep{schneider2008development,berridge2018hypersonic}. For this reason, the simulations employ an undisturbed laminar inflow.}



\begin{figure}
    \centering
    \includegraphics[width=1\textwidth]{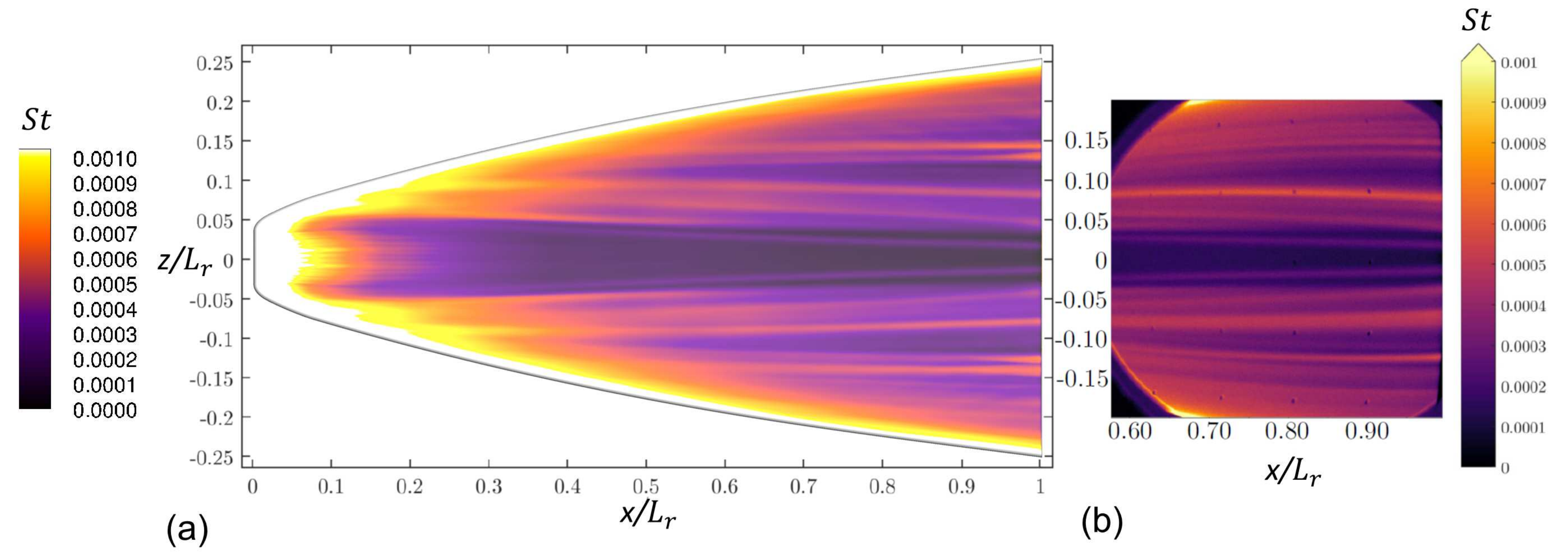}
    \caption{{Spatial distribution of the Stanton number over the surface of the BOLT subscale vehicle; (a) simulations and (b) experiments \citep{berridge2018hypersonic,thome2019boundary}. In the notation, $L_r$ represents the streawmise length of the vehicle.}}
    \label{fig:BOLT_heat_flux_contourlines}
\end{figure}
\begin{figure}
    \centering
    \includegraphics[width=\textwidth]{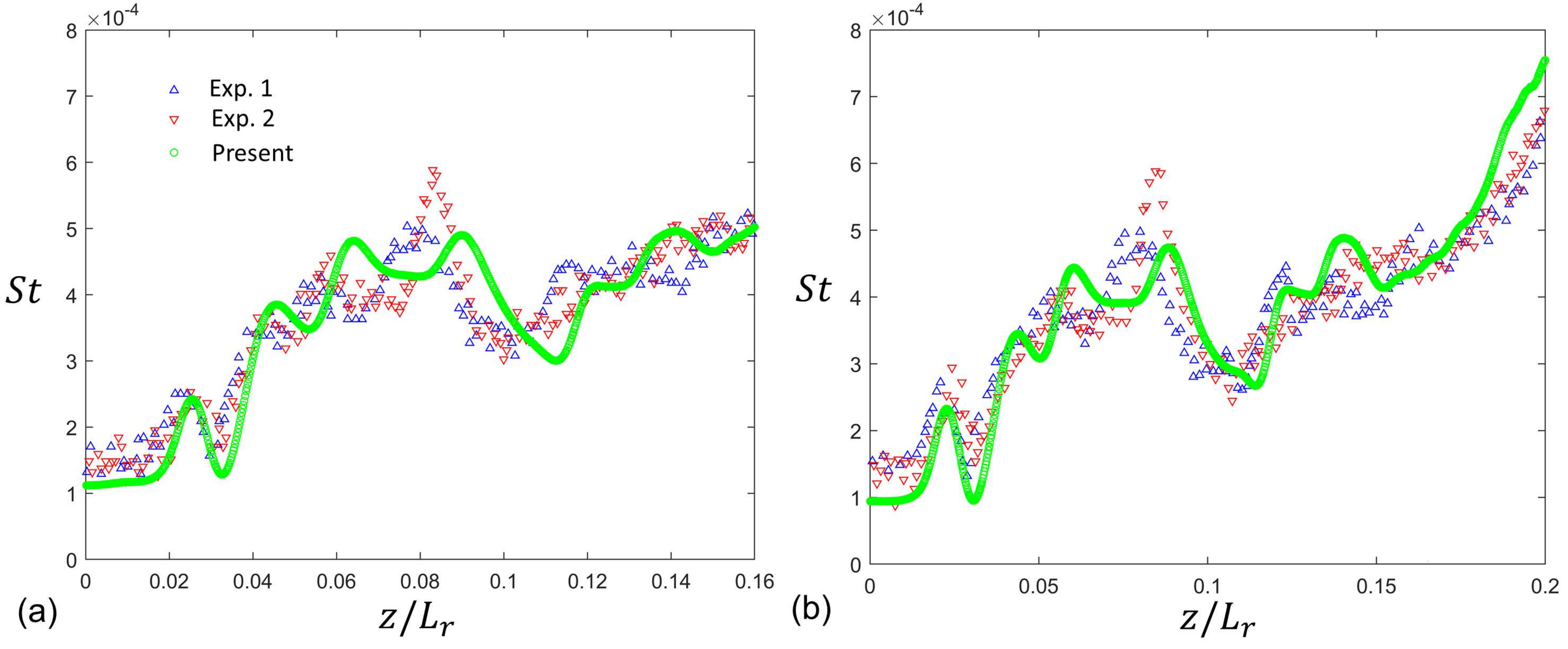}
    \includegraphics[width=\textwidth]{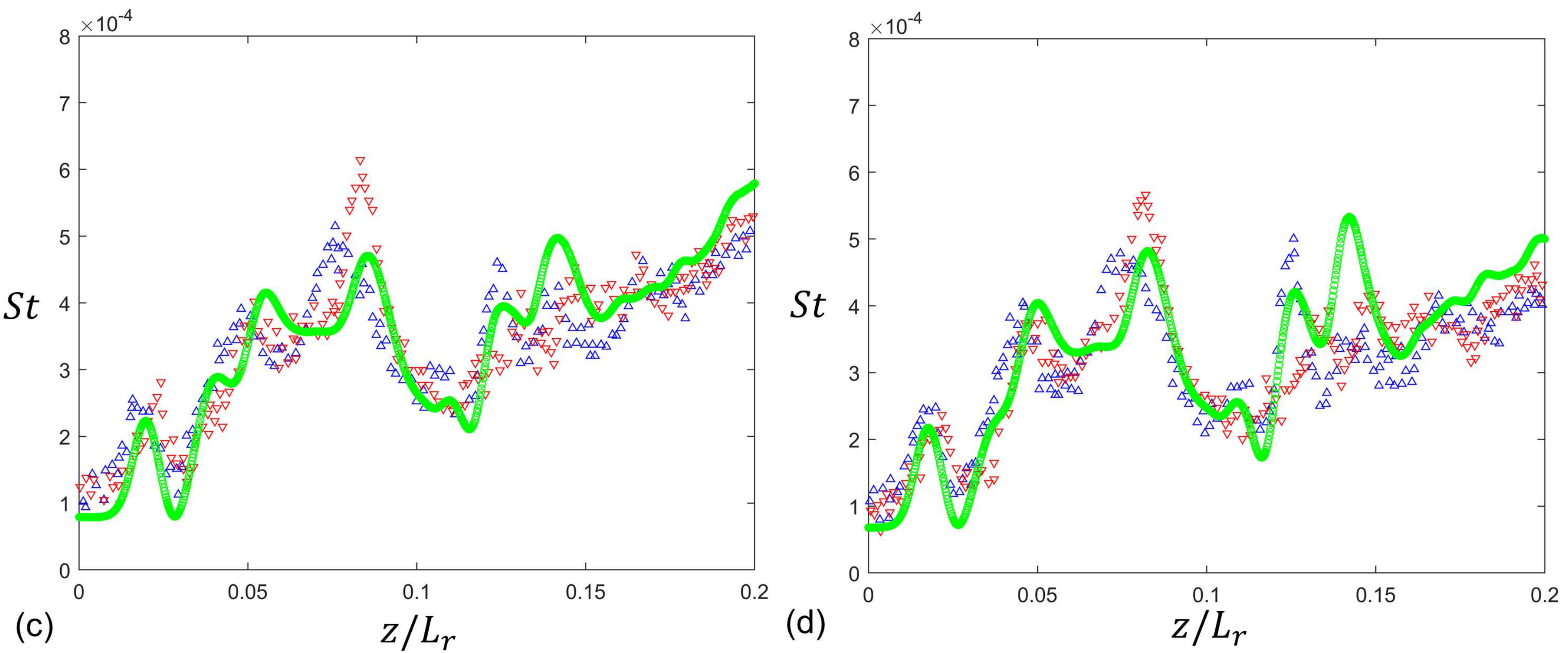}
    \caption{{Spanwise profiles of the Stanton number obtained from the  simulations (green circles) and experiments (red and blue triangles; \cite{berridge2018hypersonic,thome2019boundary}) at four streamwise stations corresponding to (a) $x/L_r =$ $0.64$, (b) $0.73$, (c) $0.82$, and (d) $0.91$. The blue and red triangles denote experimental data extracted on each side of the surface around the centerline of the BOLT vehicle $z/L_r = 0$. In the notation, $L_r$ represents the streawmise length of the vehicle.}}
\label{fig:BOLT_heat_flux}
\end{figure}

{Figure~\ref{fig:BOLT_heat_flux_contourlines} shows good agreement between the Stanton number distribution obtained from the simulations using charLES and from the experiments reported in \cite{berridge2018hypersonic} and \cite{thome2019boundary}. The streaky structures in the Stanton number distribution, caused by cross-flow instabilities, are predicted by the simulations, particularly near the centerline, where the boundary layer is lifted by the stationary streamwise vortices with mushroom-like structures. {Further quantitative comparisons are provided in figure~\ref{fig:BOLT_heat_flux} by the spanwise profiles of the Stanton number at four streamwise stations. Two sets of experimental data points are provided that correspond to each side of the surface around the vehicle centerline. Although the overall agreement between simulations and experiments is satisfactory, it is noted by \cite{thome2019boundary} and \cite{wheaton2018boundary} that the experimental results are influenced by uncertainties associated with surface roughness, thermal inertial of the vehicle model, and imperfect alignment with the free stream.}}



%
%


\section*{Appendix B. The equilibrium wall model}\label{sect:wm}

The equilibrium wall model integrates the momentum and total-energy conservation equations
\begin{equation}\tag{A.1}
\frac{d}{dy}\left[\left(\mu+\mu_{t,wm}\right)\frac{du_{||}}{dy}\right]=0,
\label{mom}\\
\end{equation}
\begin{equation}\tag{A.2}
\frac{d}{dy}\left[\left(\mu+\mu_{t,wm}\right)u_{||}\frac{du_{||}}{dy}+c_p\left(\frac{\mu}{Pr}+\frac{\mu_{t,wm}}{Pr_{t,wm}}\right)\frac{dT}{dy}\right]=0,
\label{ener}
\end{equation}
within a layer spanning from the wall to a matching location, where appropriate boundary conditions are applied, as indicated below.
In this formulation, $y$ is the wall-normal coordinate, $u_{||}$ is the total wall-parallel velocity including both the streamwise and spanwise components, $T$ is the static temperature, $c_p$ is the specific heat at constant pressure, $Pr=0.72$ is the molecular Prandtl number, $\mu$ is the molecular dynamic viscosity, and the subscript ``$wm$'' indicates variables in the wall model. The molecular viscosity $\mu$ is a function of the temperature, with the exact dependence being provided in \S\ref{setup}. Additionally, the eddy viscosity $\mu_{t,wm}$ is specified according to the mixing-length model
\begin{equation}\tag{A.3}
\mu_{t,wm}=\kappa \rho y \sqrt{\frac{\tau_w}{\rho}}D,
\label{eq:mixinglength}
\end{equation}
where $\kappa=0.4$ is the K\'{a}rm\'{a}n constant, $\rho$ is the density and $\tau_w$ is the local wall shear stress.
The damping function $D$ is given by
\begin{equation}\tag{A.4}
D=\left[1-\exp\left(-\frac{y^+}{A^+}\right)\right]^2,
\label{eq:damping}
\end{equation}
where the superscript ``+'' indicates lengths in wall units and the constant $A^+=17$.
The density and the temperature are related by the equation of state
\begin{equation}\tag{A.5}
P=\rho R_g T, \label{state}
\end{equation}
where $R_g$ is the gas constant and $P$ is the static pressure, the latter of which is modeled as a constant across the wall-modeled region and matches with the LES outside.
Lastly, $Pr_{t,wm}=0.9$ is the eddy Prandtl number and is the same for all WMLES cases in this work. Note that the model does not include the wall-normal velocity component, streamwise pressure gradient, nor time variations of momentum and energy, and does not account for energy transfer by pressure work.

Equations \eqref{mom} and \eqref{ener}, along with \eqref{eq:mixinglength}-\eqref{state} are numerically integrated on a one-dimensional grid between $0\leq y\leq h_{wm}$ bounded by the wall at $y=0$ and by a LES/wall-model matching location at $y=h_{wm}$.
Specifically, the wall-model solution matches with the LES solution at $y=h_{wm}$ corresponding to the first LES grid point from the wall.
The boundary conditions for the wall model at the wall $y=0$ are
\begin{equation}\tag{A.6}
u_{||}=0, \quad \quad T=T_w,  \label{bc1}
\end{equation}
where $T_w$ is the wall temperature.
The corresponding boundary conditions at the matching location $y=h_{wm}$ are
\begin{equation}\tag{A.7}
u_{||}=\widetilde{U}_{||}, \quad T=\widetilde{T} , \quad P= \overline{P},\label{bc2}
\end{equation}
where $\widetilde{U}_{||}$, $\widetilde{T}$ and $\overline{P}$ are the resolved LES values of wall-parallel velocity, static temperature and static pressure.
{The time-filtering approach proposed by \cite{yang2017log} is employed for calculating the boundary conditions (\ref{bc2}) at the matching location.}

\begin{figure}
\begin{center}
\includegraphics[width=1\textwidth]{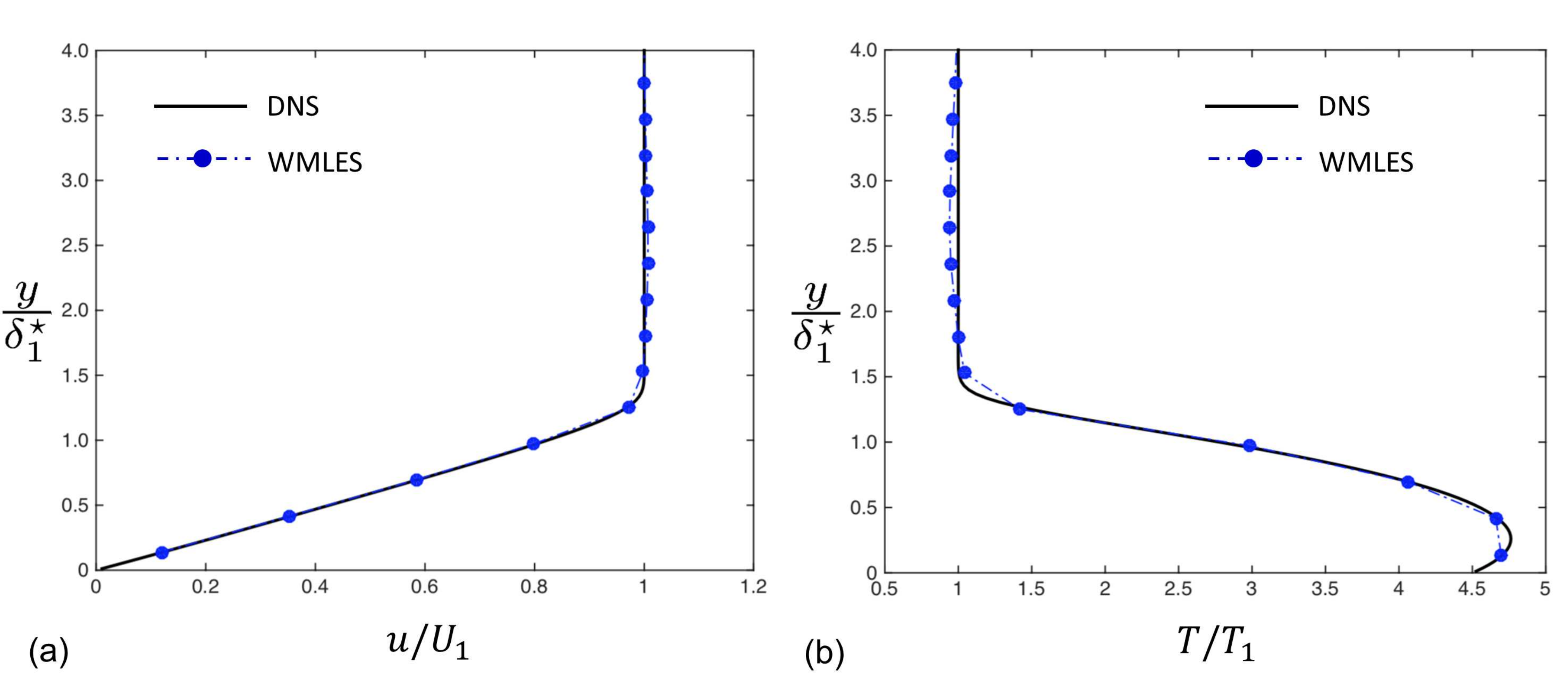}
\caption{DNS (solid lines) and WMLES (dot-dashed line) of (a)~streamwise velocity and (b)~temperature profiles at $(x-x_1)/\delta^\star_1=6$ in the laminar region of the boundary layer upstream of the separation location.}
\label{inlet_laminar}
\end{center}
\end{figure}

\section*{Appendix C. WMLES performance in the laminar region}
The performance of WMLES in predicting the laminar portion of the boundary layer upstream of the separation bubble is illustrated in figure~\ref{inlet_laminar}, where profiles of streamwise velocity and temperature from DNS and WMLES are compared at a representative location close to the inlet, i.e. $(x-x_1)/\delta_1^\star=6$. Approximately five points across the boundary layer at this station prove to be sufficient resolution for the WMLES to capture the steady laminar profiles there.

While the equilibrium wall model formulation is fundamentally different from the conservation equations of the laminar boundary layer, good agreement is obtained due to the fact that the turbulent eddy viscosity is negligible in the boundary layer at this early station, including within the wall-modeled region, because $h^{+}_{wm}\ll A^{+}$ close to the inlet, as shown in figure~\ref{yplusWMLES}. As a result, the role of wall model in the laminar portion of the boundary layer is limited to providing viscous approximations of the velocity and temperature profiles very close to the wall.

\section*{Appendix D. Grid-resolution study of WMLES }\label{sect:WMLES_resolution_Study}

\begin{figure}
\begin{center}
\vskip 0.1in
\includegraphics[width=\textwidth]{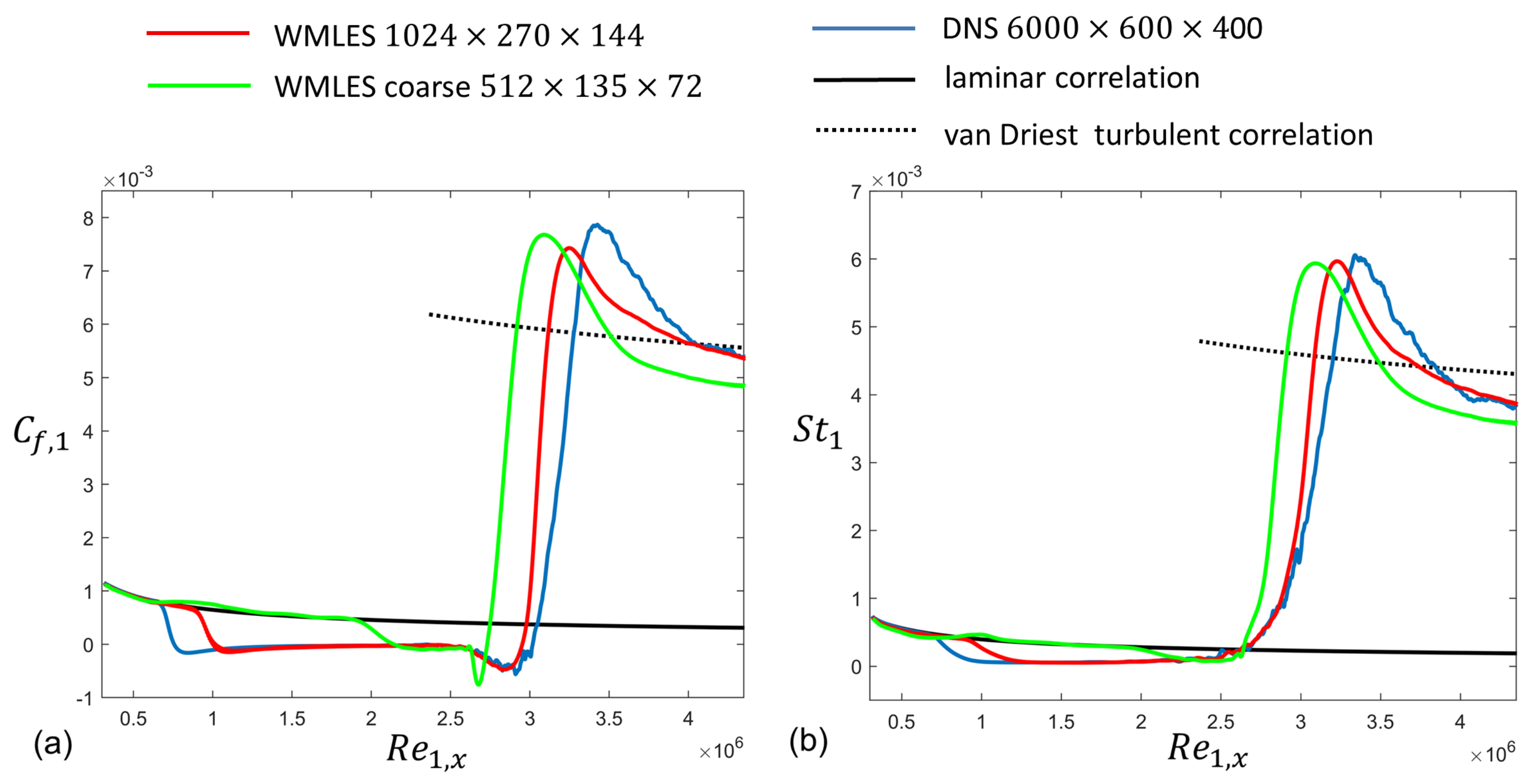}
\caption{{Grid-resolution study for the case $\alpha=7^\circ$ including the WMLES baseline grid (red line), an isotropically coarsened WMLES grid (green line), along with the DNS grid (blue line), and the laminar (black line) and turbulent (dotted line) correlations for (a) skin friction coefficient and (b) Stanton number as a function of the streamwise distance Reynolds number.}}
\label{WMLES_convergence}
\end{center}
\end{figure}

Results for WMLES on a grid coarsened by factors of 2 in every direction (isotropic grid coarsening) relative to the baseline grid are shown in figure~\ref{WMLES_convergence} for the case $\alpha=7^\circ$. The comparisons suggest a clear trend of convergence toward DNS. Specifically, as the WMLES is increasingly coarsened, the size of the separation bubble is increasingly underpredicted, the separation is increasingly delayed, and the transition occurs increasingly farther upstream.


\bibliographystyle{jfm}
\bibliography{SBLI}

\begin{thebibliography}{73}
\expandafter\ifx\csname natexlab\endcsname\relax\def\natexlab#1{#1}\fi
\def\au#1{#1} \def\ed#1{#1} \def\yr#1{#1}\def\at#1{#1}\def\jt#1{\textit{#1}}
  \def\bt#1{#1}\def\bvol#1{\textbf{#1}} \def\vol#1{#1} \def\pg#1{#1}
  \def\publ#1{#1}\def\arxiv#1{#1}\def\org#1{#1}\def\st#1{\textit{#1}}

\bibitem[Adams(2000)]{adams2000direct}
{\sc \au{Adams, N.~A.}} \yr{2000}  \at{{Direct simulation of the turbulent
  boundary layer along a compression ramp at $M$=3 and $Re_\theta$= 1685}}.
  \jt{J. Fluid Mech.}  \bvol{420},  \pg{47--83}.

\bibitem[Adler \& Gaitonde(2018)]{adler2018dynamic}
{\sc \au{Adler, M.~C.} \& \au{Gaitonde, D.~V.}} \yr{2018}  \at{Dynamic linear
  response of a shock/turbulent-boundary-layer interaction using constrained
  perturbations}.  \jt{J. Fluid Mech.}  \bvol{840},  \pg{291--341}.

\bibitem[Bermejo-Moreno {\em et~al.\/}(2014)Bermejo-Moreno, Campo, Larsson,
  Bodart, Helmer \& Eaton]{bermejo2014confinement}
{\sc \au{Bermejo-Moreno, I.}, \au{Campo, L.}, \au{Larsson, J.}, \au{Bodart,
  J.}, \au{Helmer, D.} \& \au{Eaton, J.~K.}} \yr{2014}  \at{Confinement effects
  in shock wave/turbulent boundary layer interactions through wall-modelled
  large-eddy simulations}.  \jt{J. Fluid Mech.}  \bvol{758},  \pg{5--62}.

\bibitem[Berridge {\em et~al.\/}(2018)Berridge, McKiernan, Wadhams, Holden,
  Wheaton, Wolf \& Schneider]{berridge2018hypersonic}
{\sc \au{Berridge, Dennis~C}, \au{McKiernan, Gregory}, \au{Wadhams, Tim~P},
  \au{Holden, Michael}, \au{Wheaton, Bradley~M}, \au{Wolf, Thomas~D} \&
  \au{Schneider, Steven~P}} \yr{2018} Hypersonic ground tests in support of the
  boundary layer transition (bolt) flight experiment.  \bt{In {\em 2018 Fluid
  Dynamics Conference\/}},  \pg{p. 2893}.

\bibitem[Bose \& Park(2018)]{bose2018wall}
{\sc \au{Bose, S.~T.} \& \au{Park, G.~I.}} \yr{2018}  \at{Wall-modeled
  large-eddy simulation for complex turbulent flows}.  \jt{Annu. Rev. Fluid
  Mech.}  \bvol{50},  \pg{535--561}.

\bibitem[Bradshaw(1977)]{Bradshaw}
{\sc \au{Bradshaw, P.}} \yr{1977}  \at{Compressible turbulent shear layers}.
  \jt{Annu. Rev. Fluid Mech.}  \bvol{9},  \pg{33--54}.

\bibitem[Bres {\em et~al.\/}(2018)Bres, Bose, Emory, Ham, Schmidt, Rigas \&
  Colonius]{bres2018large}
{\sc \au{Bres, G.~A.}, \au{Bose, S.~T.}, \au{Emory, M.}, \au{Ham, F.~E.},
  \au{Schmidt, O.~T.}, \au{Rigas, G.} \& \au{Colonius, T.}} \yr{2018}
  Large-eddy simulations of co-annular turbulent jet using a {V}oronoi-based
  mesh generation framework.  \bt{In {\em 2018 AIAA/CEAS Aeroacoustics
  Conference\/}},  \pg{p. 3302}.

\bibitem[Busemann(1931)]{Busemann1931}
{\sc \au{Busemann, A.}} \yr{1931}  \bt{Handbuch der experimentalphysik}. ,
  \vol{vol.~4}.  \publ{Geest und Port}.

\bibitem[Candler(2019)]{Candler2019}
{\sc \au{Candler, G.~V.}} \yr{2019}  \at{Rate effects in hypersonic flows}.
  \jt{Annu. Rev. Fluid Mech.}  \bvol{51}~(1),  \pg{379--402}.

\bibitem[Cary(1970)]{Cary}
{\sc \au{Cary, A.~M.}} \yr{1970}  \bt{Summary of available information on
  {R}eynolds analogy for zero pressure gradient, compressible, turbulent
  boundary-layer flow}. {\em Tech. Rep.\/}.  \org{NASA TN D-5560}.

\bibitem[Chandrashekar(2013)]{chandrashekar2013}
{\sc \au{Chandrashekar, P.}} \yr{2013}  \at{Kinetic energy preserving and
  entropy stable finite volume schemes for compressible {E}uler and
  {N}avier-{S}tokes equations}.  \jt{Commun. Comput. Phys.}  \bvol{14}~(5),
  \pg{1252--1286}.

\bibitem[Chi \& Spalding(1966)]{chi1966influence}
{\sc \au{Chi, S.~W.} \& \au{Spalding, D.~B.}} \yr{1966} Influence of
  temperature ratio on heat transfer to a flat plate through a turbulent
  boundary layer in air.  \bt{In {\em Int. Heat Trans. Conf., Chicago IL\/}},
  \pg{pp. 41--49}.

\bibitem[Crocco(1932)]{crocco1932sulla}
{\sc \au{Crocco, L.}} \yr{1932}  \at{Sulla trasmissione del calore da una
  lamina piana a un fluido scorrente ad alta velocita}.  \jt{L'Aerotecnica}
  \bvol{12},  \pg{181--197}.

\bibitem[Currao {\em et~al.\/}(2020)Currao, Choudhury, Gai, Neely \&
  Buttsworth]{currao2020hypersonic}
{\sc \au{Currao, G. M.~D.}, \au{Choudhury, R.}, \au{Gai, S.~L.}, \au{Neely,
  A.~J.} \& \au{Buttsworth, D.~R.}} \yr{2020}  \at{Hypersonic transitional
  shock-wave--boundary-layer interaction on a flat plate}.  \jt{AIAA J.}
  \bvol{58}~(2),  \pg{814--829}.

\bibitem[Davidson \& Babinsky(2015)]{Babinsky}
{\sc \au{Davidson, T.~S.~C.} \& \au{Babinsky, H.}} \yr{2015}  \at{{Transition
  location effects on normal shock wave-boundary layer interactions}}.
  \jt{AIAA Paper AIAA 2015-1975} .

\bibitem[Di~Renzo \& Urzay(2019)]{Renzo}
{\sc \au{Di~Renzo, M.} \& \au{Urzay, J.}} \yr{2019}  \at{{An a priori study of
  the accuracy of an equilibrium wall model for dissociating air in supersonic
  channel flows}}.  \jt{Annual Research Briefs, Center for Turbulence Research}
   \pg{pp. 29--40}.

\bibitem[van Driest(1956)]{vandriest}
{\sc \au{van Driest, E.}} \yr{1956}  \at{The problem of aerodynamic heating}.
  \jt{Aeronaut. Eng. Rev.}  \pg{pp. 26--41}.

\bibitem[Duan {\em et~al.\/}(2010)Duan, Beekman \& Martin]{duan2010direct}
{\sc \au{Duan, L.}, \au{Beekman, I.} \& \au{Martin, M.~P.}} \yr{2010}
  \at{Direct numerical simulation of hypersonic turbulent boundary layers.
  {P}art 2. {E}ffect of wall temperature}.  \jt{J. Fluid Mech.}  \bvol{655},
  \pg{419--445}.

\bibitem[Duan \& Martin(2011)]{duan2011direct}
{\sc \au{Duan, L.} \& \au{Martin, M.~P.}} \yr{2011}  \at{Direct numerical
  simulation of hypersonic turbulent boundary layers. {P}art 4. {E}ffect of
  high enthalpy}.  \jt{J. Fluid Mech.}  \bvol{684},  \pg{25--59}.

\bibitem[Dupont {\em et~al.\/}(2005)Dupont, Haddad, Ardissone \&
  Debi{\`e}ve]{dupont2005space}
{\sc \au{Dupont, P.}, \au{Haddad, C.}, \au{Ardissone, J.~P.} \&
  \au{Debi{\`e}ve, J.~F.}} \yr{2005}  \at{Space and time organisation of a
  shock wave/turbulent boundary layer interaction}.  \jt{Aerosp. Sci. Technol.}
   \bvol{9}~(7),  \pg{561--572}.

\bibitem[Dupont {\em et~al.\/}(2006)Dupont, Haddad \&
  Debi{\`e}ve]{dupont2006space}
{\sc \au{Dupont, P.}, \au{Haddad, C.} \& \au{Debi{\`e}ve, J.~F.}} \yr{2006}
  \at{{Space and time organization in a shock-induced separated boundary
  layer}}.  \jt{J. Fluid Mech.}  \bvol{559},  \pg{255--277}.

\bibitem[Dupont {\em et~al.\/}(2008)Dupont, Piponniau, Sidorenko \&
  Debi{\`e}ve]{dupont2008investigation}
{\sc \au{Dupont, P.}, \au{Piponniau, S.}, \au{Sidorenko, A.} \&
  \au{Debi{\`e}ve, J-F.}} \yr{2008}  \at{{Investigation by particle image
  velocimetry measurements of oblique shock reflection with separation}}.
  \jt{AIAA J.}  \bvol{46}~(6),  \pg{1365--1370}.

\bibitem[Dussauge {\em et~al.\/}(2006)Dussauge, Dupont \&
  Debi{\`e}ve]{dussauge2006unsteadiness}
{\sc \au{Dussauge, J-P.}, \au{Dupont, P.} \& \au{Debi{\`e}ve, J-F.}} \yr{2006}
  \at{{Unsteadiness in shock wave boundary layer interactions with
  separation}}.  \jt{Aerosp. Sci. Technol.}  \bvol{10}~(2),  \pg{85--91}.

\bibitem[Fernholz \& Finley(1980)]{Fernholz1980}
{\sc \au{Fernholz, H.~H.} \& \au{Finley, P.~J.}} \yr{1980}  \bt{A critical
  commentary on mean flow data for two-dimensional compressible turbulent
  boundary layers}. {\em Tech. Rep.\/}.  \org{AGARD-AG-253}.

\bibitem[Gaitonde(2013)]{gaitonde2013progress}
{\sc \au{Gaitonde, D.~V.}} \yr{2013}  \at{{Progress in shock wave/boundary
  layer interactions}}.  \jt{AIAA Paper 2013-2607} .

\bibitem[Gatski \& Erlebacher(2002)]{gatski2002numerical}
{\sc \au{Gatski, T.~B.} \& \au{Erlebacher, G.}} \yr{2002}  \at{{Numerical
  simulation of a spatially evolving supersonic turbulent boundary layer}}.
  \jt{NASA/TM-2002-211934} .

\bibitem[Gaviglio(1987)]{gaviglio1987reynolds}
{\sc \au{Gaviglio, J.}} \yr{1987}  \at{Reynolds analogies and experimental
  study of heat transfer in the supersonic boundary layer}.  \jt{Int. J. Heat
  Mass Tran.}  \bvol{30}~(5),  \pg{911--926}.

\bibitem[Gottlieb {\em et~al.\/}(2001)Gottlieb, Shu \& Tadmor]{gottlieb2001}
{\sc \au{Gottlieb, S.}, \au{Shu, C.-W.} \& \au{Tadmor, E.}} \yr{2001}
  \at{Strong stability-preserving high-order time discretization methods}.
  \jt{SIAM review}  \bvol{43}~(1),  \pg{89--112}.

\bibitem[Guarini {\em et~al.\/}(2000)Guarini, Moser, Shariff \&
  Wray]{guarini2000direct}
{\sc \au{Guarini, S.~E.}, \au{Moser, R.~D.}, \au{Shariff, K.} \& \au{Wray, A.}}
  \yr{2000}  \at{Direct numerical simulation of a supersonic turbulent boundary
  layer at {M}ach 2.5}.  \jt{J. Fluid Mech.}  \bvol{414},  \pg{1--33}.

\bibitem[Hildebrand {\em et~al.\/}(2018)Hildebrand, Dwivedi, Nichols,
  Jovanovi{\'c} \& Candler]{hildebrand2018simulation}
{\sc \au{Hildebrand, N.}, \au{Dwivedi, A.}, \au{Nichols, J.~W.},
  \au{Jovanovi{\'c}, M.~R.} \& \au{Candler, G.~V.}} \yr{2018}  \at{{Simulation
  and stability analysis of oblique shock-wave/boundary-layer interactions at
  Mach 5.92}}.  \jt{Phys. Rev. Fluids}  \bvol{3}~(1),  \pg{013906}.

\bibitem[Huang {\em et~al.\/}(1995)Huang, Coleman \&
  Bradshaw]{huang1995compressible}
{\sc \au{Huang, P.~G.}, \au{Coleman, G.~N.} \& \au{Bradshaw, P.}} \yr{1995}
  \at{Compressible turbulent channel flows: {DNS} results and modelling}.
  \jt{J. Fluid Mech.}  \bvol{305},  \pg{185--218}.

\bibitem[Iyer \& Malik(2019)]{Malik}
{\sc \au{Iyer, P.~S.} \& \au{Malik, M.~R.}} \yr{2019}  \at{{Analysis of the
  equilibrium wall model for high-speed turbulent flows}}.  \jt{Phys. Rev.
  Fluids}  \bvol{25},  \pg{074604}.

\bibitem[Kawai \& Larsson(2012)]{kawai2012wall}
{\sc \au{Kawai, S.} \& \au{Larsson, J.}} \yr{2012}  \at{{Wall-modeling in large
  eddy simulation: Length scales, grid resolution, and accuracy}}.  \jt{Phys.
  Fluids}  \bvol{24}~(1),  \pg{015105}.

\bibitem[Knight \& Mortazavi(2017)]{Knight}
{\sc \au{Knight, D.} \& \au{Mortazavi, M.}} \yr{2017}  \at{{Hypersonic shock
  wave transitional boundary layer interactions - A review}}.  \jt{AIAA Paper
  2017-3124} .

\bibitem[Lakebrink {\em et~al.\/}(2019)Lakebrink, Mani, Rolfe, Spyropoulos,
  Philips, Bose \& Mace]{lakebrink2019}
{\sc \au{Lakebrink, M.~T.}, \au{Mani, M.}, \au{Rolfe, E.~N.}, \au{Spyropoulos,
  J.~T.}, \au{Philips, D.~A.}, \au{Bose, S.~T.} \& \au{Mace, J.~L.}} \yr{2019}
  \at{Toward improved turbulence-modeling techniques for internal-flow
  applications}.  \jt{AIAA Paper 2019-3703} .

\bibitem[Larsson {\em et~al.\/}(2015)Larsson, Laurence, Bermejo-Moreno, Bodart,
  Karl \& Vicquelin]{Johan}
{\sc \au{Larsson, J.}, \au{Laurence, S.}, \au{Bermejo-Moreno, I.}, \au{Bodart,
  J.}, \au{Karl, S.} \& \au{Vicquelin, R.}} \yr{2015}  \at{{Incipient thermal
  choking and stable shock-train formation in the heat-release region of a
  scramjet combustor. {P}art II: {L}arge eddy simulations}}.  \jt{Combust.
  Flame}  \bvol{162},  \pg{907--920}.

\bibitem[Lash {\em et~al.\/}(2016)Lash, Combs, Kreth, Beckman \&
  Schmisseur]{Lash}
{\sc \au{Lash, E.}, \au{Combs, C.}, \au{Kreth, P.}, \au{Beckman, E.} \&
  \au{Schmisseur, J.}} \yr{2016}  \at{{Image-based analysis of the dynamics of
  transitional shock wave-boundary layer interactions}}.  \jt{AIAA Paper
  2016-4320} .

\bibitem[Lehmkuhl {\em et~al.\/}(2018)Lehmkuhl, Park, Bose \&
  Moin]{lehmkuhl2018}
{\sc \au{Lehmkuhl, O.}, \au{Park, G.~I.}, \au{Bose, S.~T.} \& \au{Moin, P.}}
  \yr{2018}  \at{Large-eddy simulation of practical aeronautical flows at stall
  conditions}.  \jt{Proceedings of the 2018 Summer Program, Center for
  Turbulence Research, Stanford University}  \pg{pp. 87--96}.

\bibitem[Leyva(2017)]{Leyva2017}
{\sc \au{Leyva, I.~A.}} \yr{2017}  \at{{The relentless pursuit of hypersonic
  flight}}.  \jt{Phys. Today}  \bvol{70}~(11),  \pg{30--36}.

\bibitem[Lighthill(1950)]{Lighthill1950}
{\sc \au{Lighthill, M.~J.}} \yr{1950}  \at{Contributions to the theory of heat
  transfer through a laminar boundary layer}.  \jt{Proc. Roy. Soc. A. Math.
  Phy.}  \bvol{202}~(1070),  \pg{359--377}.

\bibitem[Loginov {\em et~al.\/}(2006)Loginov, Adams \&
  Zheltovodov]{loginov2006large}
{\sc \au{Loginov, M.~S.}, \au{Adams, N.~A.} \& \au{Zheltovodov, A.~A.}}
  \yr{2006}  \at{{Large-eddy simulation of shock-wave/turbulent-boundary-layer
  interaction}}.  \jt{J. Fluid Mech.}  \bvol{565},  \pg{135--169}.

\bibitem[Lozano-Dur\'an {\em et~al.\/}(2020)Lozano-Dur\'an, Bose \&
  Moin]{lozano2020}
{\sc \au{Lozano-Dur\'an, A.}, \au{Bose, S.~T.} \& \au{Moin, P.}} \yr{2020}
  \at{Prediction of trailing edge separation on the {NASA} {J}uncture {F}low
  using wall-modeled {LES}}.  \jt{AIAA Paper 2020-1776} .

\bibitem[Mack(1984)]{mack1984boundary}
{\sc \au{Mack, L.~M.}} \yr{1984}  \bt{Boundary-layer linear stability theory}.
  {\em Tech. Rep.\/}.  \org{California Inst. of Tech. Pasadena Jet Propulsion
  Lab}.

\bibitem[Marco \& Komives(2018)]{Komives}
{\sc \au{Marco, N.} \& \au{Komives, J.~R.}} \yr{2018}  \at{{Wall-Modeled large
  eddy simulation of a three-dimensional shock-boundary layer interaction}}.
  \jt{AIAA Paper 2018-1298} .

\bibitem[Mettu \& Subbareddy(2018)]{Subbareddy}
{\sc \au{Mettu, B.~R.} \& \au{Subbareddy, P.~K.}} \yr{2018} {Wall modeled LES
  of compressible flows at non-equilibrium conditions,}.  \bt{In {\em 2018 AIAA
  Fluid Dynamics Conference, Atlanta GA\/}}.

\bibitem[Modesti \& Pirozzoli(2016)]{modesti2016reynolds}
{\sc \au{Modesti, D.} \& \au{Pirozzoli, S.}} \yr{2016}  \at{Reynolds and {M}ach
  number effects in compressible turbulent channel flow}.  \jt{Int. J. Heat
  Fluid Fl.}  \bvol{59},  \pg{33--49}.

\bibitem[Morkovin(1962)]{Morkovin1962}
{\sc \au{Morkovin, M.~V.}} \yr{1962} Effects of compressibility on turbulent
  flows.  \bt{In {\em Mecanique de la Turbulence (ed. A. Favre)\/}},  \pg{pp.
  367--380}.

\bibitem[Pirozzoli \& Bernardini(2011)]{pirozzoli_bernardini_2011}
{\sc \au{Pirozzoli, S.} \& \au{Bernardini, M.}} \yr{2011}  \at{Turbulence in
  supersonic boundary layers at moderate {R}eynolds number}.  \jt{J. Fluid
  Mech.}  \bvol{688},  \pg{120–168}.

\bibitem[Pirozzoli {\em et~al.\/}(2010)Pirozzoli, Bernardini \&
  Grasso]{pirozzoli2010direct}
{\sc \au{Pirozzoli, S.}, \au{Bernardini, M.} \& \au{Grasso, F.}} \yr{2010}
  \at{{Direct numerical simulation of transonic shock/boundary layer
  interaction under conditions of incipient separation}}.  \jt{J. Fluid Mech.}
  \bvol{657},  \pg{361--393}.

\bibitem[Pirozzoli \& Grasso(2006)]{pirozzoli2006direct}
{\sc \au{Pirozzoli, S.} \& \au{Grasso, F.}} \yr{2006}  \at{{Direct numerical
  simulation of impinging shock wave/turbulent boundary layer interaction at
  $M$= 2.25}}.  \jt{Phys. Fluids}  \bvol{18}~(6),  \pg{065113}.

\bibitem[Pirozzoli {\em et~al.\/}(2004)Pirozzoli, Grasso \&
  Gatski]{pirozzoli2004direct}
{\sc \au{Pirozzoli, S.}, \au{Grasso, F.} \& \au{Gatski, T.~B.}} \yr{2004}
  \at{{Direct numerical simulation and analysis of a spatially evolving
  supersonic turbulent boundary layer at $M$= 2.25}}.  \jt{Phys. fluids}
  \bvol{16}~(3),  \pg{530--545}.

\bibitem[Polivanov {\em et~al.\/}(2015)Polivanov, Sidorenko \&
  Maslov]{Polivanov}
{\sc \au{Polivanov, P.~A.}, \au{Sidorenko, A.~A.} \& \au{Maslov, A.~A.}}
  \yr{2015}  \at{{Transition effect on shock wave / boundary layer interaction
  at $M=$1.47}}.  \jt{AIAA Paper 2015-1974} .

\bibitem[Robinet(2007)]{robinet2007bifurcations}
{\sc \au{Robinet, J-Ch.}} \yr{2007}  \at{{Bifurcations in
  shock-wave/laminar-boundary-layer interaction: global instability approach}}.
   \jt{J. Fluid Mech.}  \bvol{579},  \pg{85--112}.

\bibitem[Sandham \& L{\"u}deke(2009)]{sandham2009numerical}
{\sc \au{Sandham, N.~D.} \& \au{L{\"u}deke, H.}} \yr{2009}  \at{Numerical study
  of {M}ach 6 boundary-layer stabilization by means of a porous surface}.
  \jt{AIAA J.}  \bvol{47}~(9),  \pg{2243--2252}.

\bibitem[Sandham {\em et~al.\/}(2014)Sandham, Sch{\"u}lein, Wagner, Willems \&
  Steelant]{sandham2014transitional}
{\sc \au{Sandham, N.~D.}, \au{Sch{\"u}lein, E.}, \au{Wagner, A.}, \au{Willems,
  S.} \& \au{Steelant, J.}} \yr{2014}  \at{{Transitional
  shock-wave/boundary-layer interactions in hypersonic flow}}.  \jt{J. Fluid
  Mech.}  \bvol{752},  \pg{349--382}.

\bibitem[Schneider(2008)]{schneider2008development}
{\sc \au{Schneider, Steven~P}} \yr{2008}  \at{{Development of hypersonic quiet
  tunnels}}.  \jt{J. Spacecr. Rockets}  \bvol{45}~(4),  \pg{641--664}.

\bibitem[Sch{\"u}lein(2014)]{schuelein2014effects}
{\sc \au{Sch{\"u}lein, E.}} \yr{2014}  \at{{Effects of laminar-turbulent
  transition on the shock-wave/boundary-layer interaction}}.  \jt{AIAA Paper
  2014-3332} .

\bibitem[Tadmor(2003)]{tadmor2003}
{\sc \au{Tadmor, E.}} \yr{2003}  \at{Entropy stability theory for difference
  approximations of nonlinear conservation laws and related time-dependent
  problems}.  \jt{Acta Numer.}  \bvol{12},  \pg{451--512}.

\bibitem[Thome {\em et~al.\/}(2019)Thome, Knutson \&
  Candler]{thome2019boundary}
{\sc \au{Thome, John}, \au{Knutson, Anthony} \& \au{Candler, Graham~V}}
  \yr{2019} {Boundary layer instabilities on BoLT subscale geometry}.  \bt{In
  {\em AIAA Scitech 2019 Forum\/}},  \pg{p. 0092}.

\bibitem[Touber \& Sandham(2009)]{touber2009large}
{\sc \au{Touber, E.} \& \au{Sandham, N.~D.}} \yr{2009}  \at{{Large-eddy
  simulation of low-frequency unsteadiness in a turbulent shock-induced
  separation bubble}}.  \jt{Theor. Comp. Fluid. Dyn.}  \bvol{23}~(2),
  \pg{79--107}.

\bibitem[Trettel(2019)]{trettel2019transformations}
{\sc \au{Trettel, A.}} \yr{2019}  \at{Transformations for variable-property
  turbulent boundary layers}. PhD thesis, UCLA.

\bibitem[Trettel \& Larsson(2016)]{trettel2016mean}
{\sc \au{Trettel, A.} \& \au{Larsson, J.}} \yr{2016}  \at{{Mean velocity
  scaling for compressible wall turbulence with heat transfer}}.  \jt{Phys.
  Fluids}  \bvol{28}~(2),  \pg{026102}.

\bibitem[Urzay(2018)]{urzay2018supersonic}
{\sc \au{Urzay, J.}} \yr{2018}  \at{Supersonic combustion in air-breathing
  propulsion systems for hypersonic flight}.  \jt{Annu. Rev. Fluid Mech.}
  \bvol{50},  \pg{593--627}.

\bibitem[Vanstone {\em et~al.\/}(2013)Vanstone, Estruch-Samper, Hillier \&
  Ganapathisubramani]{Vanstone}
{\sc \au{Vanstone, L.}, \au{Estruch-Samper, E.}, \au{Hillier, R.} \&
  \au{Ganapathisubramani, B.}} \yr{2013}  \at{{Shock-induced separation of
  transitional hypersonic boundary layers}}.  \jt{AIAA Paper 2013-2736} .

\bibitem[Volpiani {\em et~al.\/}(2018)Volpiani, Bernardini \&
  Larsson]{volpiani2018effects}
{\sc \au{Volpiani, P.~S}, \au{Bernardini, M.} \& \au{Larsson, J.}} \yr{2018}
  \at{{Effects of a nonadiabatic wall on supersonic shock/boundary-layer
  interactions}}.  \jt{Phys. Rev. Fluids}  \bvol{3}~(8),  \pg{083401}.

\bibitem[Vreman(2004)]{vreman2004}
{\sc \au{Vreman, A.~W.}} \yr{2004}  \at{An eddy-viscosity subgrid-scale model
  for turbulent shear flow: {A}lgebraic theory and applications}.  \jt{Phys.
  fluids}  \bvol{16}~(10),  \pg{3670--3681}.

\bibitem[Walz(1962)]{Walz1962}
{\sc \au{Walz, A.}} \yr{1962}  \bt{Compressible turbulent boundary layers}.
  \pg{pp. 299--350}.  \publ{CNRS}.

\bibitem[Walz(1966)]{Walz1966}
{\sc \au{Walz, A.}} \yr{1966} {\em Str{\"o}mungs-und
  Temperaturgrenzschichten\/}.  \publ{Braun}.

\bibitem[Wheaton {\em et~al.\/}(2018)Wheaton, Berridge, Wolf, Stevens \&
  McGrath]{wheaton2018boundary}
{\sc \au{Wheaton, Bradley~M}, \au{Berridge, Dennis~C}, \au{Wolf, Thomas~D},
  \au{Stevens, Ryan~T} \& \au{McGrath, Brian~E}} \yr{2018} {Boundary layer
  transition (BOLT) flight experiment overview}.  \bt{In {\em 2018 Fluid
  Dynamics Conference\/}},  \pg{p. 2892}.

\bibitem[Willems {\em et~al.\/}(2015)Willems, G{\"u}lhan \&
  Steelant]{willems2015experiments}
{\sc \au{Willems, S.}, \au{G{\"u}lhan, A.} \& \au{Steelant, J.}} \yr{2015}
  \at{{Experiments on the effect of laminar--turbulent transition on the SWBLI
  in H2K at Mach 6}}.  \jt{Exp. Fluids}  \bvol{56}~(3),  \pg{49}.

\bibitem[Yang {\em et~al.\/}(2017{\natexlab{{\em a\/}}})Yang, Park \&
  Moin]{yang2017log}
{\sc \au{Yang, X. I.~A.}, \au{Park, G.~I.} \& \au{Moin, P.}}
  \yr{2017{\natexlab{{\em a\/}}}}  \at{{Log-layer mismatch and modeling of the
  fluctuating wall stress in wall-modeled large-eddy simulations}}.  \jt{Phys.
  Rev. Fluids}  \bvol{2}~(10),  \pg{104601}.

\bibitem[Yang {\em et~al.\/}(2017{\natexlab{{\em b\/}}})Yang, Urzay, Bose \&
  Moin]{yang2017aerodynamic}
{\sc \au{Yang, X. I.~A.}, \au{Urzay, J.}, \au{Bose, S.~T.} \& \au{Moin, P.}}
  \yr{2017{\natexlab{{\em b\/}}}}  \at{{Aerodynamic heating in wall-modeled
  large-eddy simulation of high-speed flows}}.  \jt{AIAA J.}  \pg{pp.
  731--742}.

\bibitem[Zhang {\em et~al.\/}(2014)Zhang, Bi, Hussain \&
  She]{zhang2014generalized}
{\sc \au{Zhang, Y.-S.}, \au{Bi, W.-T.}, \au{Hussain, F.} \& \au{She, Z.-S.}}
  \yr{2014}  \at{A generalized {R}eynolds analogy for compressible wall-bounded
  turbulent flows}.  \jt{J. Fluid Mech.}  \bvol{739},  \pg{392--420}.

\end{thebibliography}

\end{document}